\documentclass{article}
\usepackage[utf8]{inputenc}
\usepackage{authblk}
\usepackage[english]{babel}
\usepackage{biblatex}
\usepackage{comment}
\usepackage{biblatex}
\usepackage[export]{adjustbox}
\usepackage{subcaption}
\usepackage{array}
\usepackage{amsmath}
\usepackage{amssymb}
\usepackage{csquotes}

\title{\textit{In silico} model of infection of a CD4(+) T-cell by a human immunodeficiency type 1 virus, and a mini-review on its molecular pathophysiology.}
\author[1,2]{Vivanco-Lira, A. (a.vivancolira@ugto.mx)}
\affil[1]{Department of Medicine and Nutrition, University of Guanajuato, Leon, Mexico.}
\author[1]{Nieto-Saucedo, J.R.}
\affil[2]{Department of Exact Sciences and Engineering, Open and Distance Learning University of Mexico, Mexico City, Mexico.}
\date{Submitted on February 2021}

\begin{document}

\maketitle

\begin{abstract}

\textbf{Introduction}. Infection by the Human Immunodeficiency Virus can be defined as a chronic viral infection which mainly affects T-cells; this virus displays a set of both structural and regulatory proteins that aid in its survival within the host's cell, through these proteins, can the virus alter the host's gene expression pattern and with it, signaling involved in cell cycle control, cytokine response, differentiation, metabolism, and others; therefore by hijacking the host's genetic machinery there exists a promotion in viral fitness, and the ground is cemented for changes in the host's cell differentiation status to occur.  \textbf{Methods}. We will consider two stochastic Markov chain models, one which will describe the T-helper cell differentiation process, and another one describing that process of infection of the T-helper cell by the virus; in these Markov chains, we will consider a set of states $\{X_t \}$ comprised of those proteins involved in each of the processes and their interactions (either differentiation or infection of the cell), such that we will obtain two stochastic transition matrices ($A,B$), one for each process; afterwards, the computation of their eigenvalues shall be performed, in which, should the eigenvalue $\lambda_i=1$ exist, the computation for the equilibrium distribution $\pi^n$ will be obtained for each of the matrices, which will inform us on the trends of interactions amongst the proteins in the long-term. \textbf{Results}. The stochastic processes considered possess an equilibrium distribution, when reaching their equilibrium distribution, there exists an increase in their informational entropy, and their log-rank distributions can be modeled as discrete beta generalized distributions (DGBD). \textbf{Discussion}. The equilibrium distributions of both process can be regarded as states in which the cell is well-differentiated, ergo there exists an induction of a novel HIV-dependent differentiated state in the T-cell; these processes due to their DGBD distribution can be considered complex processes; due to the increasing entropy, the equilibrium states are stable ones. \textbf{Conclusion.} The HIV virus can promote a novel differentiated state in the T-cell, which can give account for clinical features seen in patients (decrease in naïve T cell counts, and the general T-cell count decline); this model, notwithstanding does not give account of YES/NO logical switches involved in the regulatory networks. 
    
\end{abstract}

\tableofcontents

\section{Introduction}

Human Immunodeficiency Virus infection and the spectrum of related disorders are relatively new to the human being, with the first known displays of this disease occurring in the decade of the 1980s, however, efforts were made to determine a more precise temporal origin of the virus, with some estimates computing that it appeared as early as in the 1920s decade \cite{ye}; while it became clearer a bit later that HIV may have originated by means of zoonoses from primates in Africa \cite{sharp}. During the first-ever recorded outbreak in the 1980s, the epidemic was said to be limited to the population of the 4 Hs: homosexuals, haemophiliacs, haitians, heroin addicts \cite{gallo} nevertheless, was this soon proved to be a far too narrow assumption, and that the actual boundaries of the infection were not these, for the set of susceptible population was indeed a larger one. HIV was later discovered to exhibit two basic genotypes: HIV-1 and HIV-2 with differences described both in the clinical outcome and course of the disease \cite{sagoe}. HIV is known to exhibit affinity for the receptors in T cells CD4, CCR5, CXCR4, as well as other cofactors necessary for entry into the cell \cite{woodham}, then using these cells for replication, inducing later apoptosis in these ones, this apoptosis in the CD4 (+) T cells (and as well in CD8(+) T cells) is responsible for the immunodeficiency seen in the HIV(+) patients, in which, when untreated, may result in the development of opportunistic infections, including infection by human herpesvirus type 8, being able this virus to promote the establishment of a malignancy, say Kaposi's sarcoma.

\subsection{Biology of HIV-1.}

Both HIV-1 and HIV-2 are species belonging to the \textit{Lentivirus} genus which contains other notable viruses, such as \textit{Simian immunodeficiency virus}, \textit{Feline immunodeficiency virus} and others; this genus belongs itself to the subfamily \textit{Orthoretrovirinae}, part of the \textit{Retroviridae} family which also contains the subfamily \textit{Spumaretrovirinae} in which we may find those endogenous retroviruses which have incorporated into the host's genome and may take part in the genetic diversification. The \textit{Retroviridae} family is part of those reverse transcribing RNA and DNA viruses \cite{ictv}. The viral particle is about 100 nm in diameter, with an outer membrane as envelope, this envelope contains 72 knobs, each knob comprised by trimers of Env proteins, the Env protein possesses in its structure the gp120 and the gp41 (gp160) proteins in its stead, where the gp41 anchors to the lipid membrane and gp120 binds to gp41 finally protruding from the membrane. In its envelope, the viral particle also contains MHC pertaining to the host. Within the particle, we find another membrane, comprised by the matrix (p17) and the protease (p10) proteins; attached to the matrix protein membrane we find the lateral bodies, which are polarized bodies inside the particle. Finally, the capsid lies in the core of the viral particle, the shell being composed of the p24 protein (capsid protein) and attached to the matrix protein by means of the p6 protein (link protein). In the innards of the capsid lies the genome, where the reverse transcriptase complex is found attached to the genome, along with the nucleic acid binding protein (p7) \cite{german}. The viral particle's bilayer is enriched in specific lipids: aminophospholipids, dyhydrosphingomyelin, plasmenyl phosphoethanolamine, phosphatidylserine, phosphatidylethanolamine and  phosphoinositides; being the lipid composition bearers of importance in HIV fusion and infectivity (chol-chelating compounds inhibit these actions), interaction with T-cell immunoglobulin and mucin domain proteins block release of HIV from infected cells, interaction with bavituximab (which targets phosphatidylserine) suppresses productive HIV infection \cite{huarte}. The clues leading to the thought of HIV being a retrovirus came when in the 1980s outbreak, HIV-2 was shown to also cause AIDS in patients, HIV-2 was related to HIV-1 but also to a simian virus which caused immunodeficiency in macaques. It has been determined that the SIV from the Sooty mangabey originated the HIV-2 (groups A-H), while the SIV from the chimpanzee originated both the HIV-1 (groups M, N and probably O) as well as the SIV from Western gorillas, which in turn gave place to HIV-1 (groups P and probably O) \cite{sharp}; it has been proposed as well that HIV originated from SIV by means of several genetic mutation transitions and that the use of unsterile injections may have promoted the increase in transmission \cite{marx}, other mechanisms have been proposed, such as: population growth, changing sexual practices, migration, increased hunting and deforestation in post-colonial Africa \cite{pela}; however, retroviruses have been computed to appear as far back as 460 to 550 million years ago (during the Palaeozoic Era), along with the radiation of \textit{Vertebrata}, nevertheless it remains unclear whether \textit{Retroviridae} originated from \textit{Metaviridae} or whether \textit{Retroviridae} gave place to \textit{Metaviridae} \cite{haywa}. HIV-1 displays three major groups: major (M) with a transmission event occurring between 1912 and 1941, outlier (O), and nonmajor and nonoutlier (N); the virus may have commenced with the HIV-1 O group in gorillas, then spread among humans from the Congo River into Kinshasa, Zaire, with the earliest documented case of HIV-1 infection (M group) in humans dating from 1959. The M group is the predominant circulating group, and it has been divided into subtypes: A1, A2, A3, A4, B, C, D, F1, F2, F, H, J, K; further recognition of subtypes by means of full-genome sequencing has been made (CRFs and URFs) \cite{taylor}. Meanwhile, HIV-2 displays A-H grups, with the A group possessing two subdivisions: A1 and A2 \cite{german}. The life of those patients infected with HIV gave a very important turn with the advent of the antiretroviral drugs, and has approached the amount of years to be lived at age 20 years two thirds that of the general population \cite{antire}, other measurements have been made estimating expected average ages of death in those patients receiving antiretroviral therapy, finding 67.6 years for men and 67.9 years for women \cite{antiret}. In addition, the mutation rate of the HIV-1 genome in vivo has been shown to be rather high, i.e., $4.1 \pm 1.7 \times 10^{-3}$ per base per cell, this rate is contributed 2\% by the reverse transcriptase and 98\% by the host's cytidine deaminases of the A3 family \cite{cuevas}.

\subsection{Insights into molecular pathophysiology: viral proteins.}

\subsubsection{gp120: organizing the foundations for infection from membrane to nucleus.}

gp120 owes its name to its molecular weight ranging from 110 to 120 kDa \cite{crublet},  being this protein a portion of the gp160 complexes (due to the fact that gp120 matures from a larger, gp160 peptide), the gp120 associates with the gp41 molecule assemblying into a heterodimer, which will trimerize to form the mature Env protein, in its turn bound to the viral bilayer membrane. The protein is comprised by five variable regions and five conserved regions, folding into a globular structure with an inner and an outer domain bound by a bridging sheet. gp120 is glycated mostly by high-mannose glycans, and complex glycans fucosylated and containing multiple antennas, along with a variable amount of sialic acid ; the high-mannose oligosaccharides were found to be $(GlcNAc)_2 (Man)_5$ and $(GlcNAc)_2 (Man)_9$ \cite{raska}; deficiencies in glycosylation (as they have been induced in bacteria-produced gp120) result in unsuccessful binding of gp120 to CD4 \cite{fennie}. gp120 binds to CD4 by means of a pocket found above the bridging sheet \cite{yoon}, this pocket is known to be hydrophobic in gp120 and when bound to CD4 a phenylalanine residue caps this pocket, this residue has proved essential for a successful binding, the whole complex (gp120 hydrophobic pocket + phenylalanine residue) is termed the Phe43 cavity \cite{dey}; moreover, the affinity with which CD4 interacts with gp120 has been determined to be $K_d = 4 \times 10^{-9} M$ \cite{thali} while the entropy of this process has been seen to be $\Delta S= 220 \pm 13 kJ mol^{-1}$ and an enthalpy of $\Delta H= -259 \pm 13 kJ mol^{-1}$ (at 310 K) (in which we find that this binding of CD4 to gp120 is an irreversible (due to $\Delta S>0$) and spontaneous (because $\Delta G <0$, considering $\Delta G$ as the free energy) process \cite{maron}) with an intermolecular hydrogen-bond network of 166 atoms (for CD4) and 130 atoms (for gp120) \cite{hsu}. The binding of gp120 to CD4 induces conformational changes in gp120, and promotes the creation of a high-affinity binding site for CCR5, as well as the exposure of gp41 in order to induce membrane fusion; furthermore, has it been shown that the binding of the soluble form of CD4 (sCD4) to gp120 promotes the dissociation of gp120 from gp41, and some variable loops (V1, V2, V3) change their conformation or become more more exposed; the exposure of these loops (V1 and V2) could be potentially recognized by the monoclonal antibodies 17b and 48d (this nomenclature is due to the epitopes exposed when the conformational changes are induced) \cite{sullivan}, where the monoclonal antibody 17b is an anti-HIV-1 gp120 monoclonal antibody which is obtained from EBV transformation of B cells (let us recall that these transformed B cells are those infected by EBV in which EBNA1, EBER1 and EBER2 are expressed, inducing the translocation of MYC into the immunoglobulin loci, activating permanently this transcription factor \cite{kuppers}) of an asymptomatic HIV-1 infected individual  \cite{robinson}; however, due to steric hindrance and conformational masking, these epitopes remain obscured or unaccesible to the antibodies directed against gp120, therefore, by means of a chimeric protein comprised of a soluble CD4 bound to the monoclonal antibody directed towards 17b, the necessary conformational change occurs and the epitopes are exposed, method through which, the monoclonal antibody may then bind to the epitope \cite{lagenaur}; nevertheless, there exist also entropic barriers in the gp120 core which may decrease the potential binding of antibodies \cite{raja}. The binding of CD4 to the heterohexamer ($(gp120)_3 (gp41)_3$) happens in a stoichiometry of $(CD4)_3 (gp120)_3 (gp41)_3$ this because two CD4 molecules may be as close as $19.8 nm$ in the cell membrane context, that is, the root of the CD4 molecules for these molecules protrude from the cell membrane and be closer in the extracellular space, the maximum distance at which they can be approached one from the other has been measured to be $3.5 nm$ in the extracellular space and without steric interference \cite{kwong}. After the binding of CD4 to gp120, we have described that there exist some conformational changes which promote the interaction of CCR5 with gp120 (used in early infection) and with CXCR4 (in late infection), it has been displayed that in the absence or changes in sequence of the loops V1 and V2, there could be a loss in the ability of gp120 of binding to CD4, then being able to infect CD4(-) cells with CCR5(+) or CXCR4(+); the conformational changes of gp120 by means of CD4 diminish the entropy of the protein \cite{raja}; this binding of gp120 to the chemokines happens through the V3 loop which is predominantly composed of 35 residues, connected by a disulfide bridge between residues 1 and 35, being those residues essential for the binding with the coreceptors those from 13 to 21, additionally, gp120 binds with near equal energetic properties to CCR5 and CXCR4 \cite{tamamis}; the binding of CD4 to gp120 induces a diminishment in the distance of the trimer to the target cell membrane by $2 nm$ \cite{kwong}. We previously commented on the observed fact that the HIV-1 in its early infection phase exhibits tropism towards CCR5, while during late infection, this tropism shifts towards CXCR4 \cite{ribeiro} and there have been some reasons of this shift implicating gp120: loss of N-linked glycosylation site in V2 in late stages, increase in positive-charged residues in V1/V2 and V4/V5, addition of N-linked glycosylation site near the CD4-binding site (N362); however the implications of this change in tropism are viral-fitness-dependent, that is, whether this switch is convenient for the virus to occur or not, in which, given the case, the switch shall not occur \cite{mosier}; this may be challenged (or sustained) by the fact that CCR5 is expressed in memory T cells, while CXCR4 is in change expressed in naïve T cells, in which the memory T cells have shown to possess a higher division rate than naïve T cells, which promotes an advantage in the fitness of the virus during early infection, nevertheless when time passes and CD4(+) T cell depletion occurs, both memory and naïve T cell numbers are consequentially diminished, but the fraction of naïve T cells in division has been seen to divide more rapidly in the late stage of disease than memory T cells, those augmenting the selection of those viruses with CXCR4 tropism or the switch from CCR5 to CXCR4 in late stages of disease \cite{ribeiro}. It has been shown that gp120 folding occurs at the rough endoplasmic reticulum (and not in the Golgi apparatus) and this folding is essential to the binding of gp120 to the CD4 molecule; the folding time of this molecule has been shown to be more lengthy than usual, this due perhaps to the formation of disulfide bonds within gp120, probably aided by disulfide isomerase (disulfide bond-formation are dependent on several enzymes, those thiol-disulphide oxidoreductases, such as disulphide isomerase (this one is yet to know whether it catalyzes de novo formation of disulfide bonds or shuffles pre-existing bonds), as well as other proteins, such as ERO1 which is a glycosylated ER protein which induces oxidizing equivalents into the ER lumen \cite{sevier}); gp120 is constrained to proper folding for its exit from the rough endoplasmic reticulum \cite{fennie}. gp120 not only exerts fusion-related functions, but also contributes to the pathogenesis of HIV by activating complement, inducing polyclonal B cell activation, binding to immune complexes, participates in influencing T cell function (promoting inefficiency in T cell response), augments production of TGFB \cite{yoon}. The gp120-dependent activation of complement occurs via a dettachment of gp120 from the viral envelope either spontaneously or after gp120 binding to the CD4 molecule, then gp120 may circulate in the patient's blood and may attach in a soluble form to CD4(+) T cells, promotion of opsonization and elimination of CD4(+)gp120(+) T cells by means of the reticuloendothelial system; the complement is activated in its classical pathway, and gp120 has been seen to bind to C4, C3, C5, C9 complement proteins \cite{susal}; thus ending with the membrane atttack complex (MAC) formation in the gp120-dependent complement activation directed towards CD4(+) T cells, contributing to the depletion of these cells; this membrane attack complex is assembled by means of the cleavage of C5 to C5b and C5a, C5b recruits C6 (C5b6) through C9 and forms the MAC pore \cite{zewde}; abnormal complement activation has been observed in HIV patients, associating this infection with certain autoimmune disorders, such as: immune thrombocytopenic purpura, inflammatory myositis, Guillain-Barré syndrome, sarcoidosis, myasthenia gravis, Graves' disease, Hashimoto thyroiditis, autoimmune hemolytic anemia, autoimmune hepatitis, rheumatoid arthritis, systemic lupus erythematosus \cite{virot}. Abnormalities of the humoral response have been observed in HIV patients (which relates to abnormal B cell function), such as: polyclonal hypergammaglobulinemia, B-cell hyperplasia, circulating immune complexes, spontaneous Ig secretion, proliferation with elevated autoantibodies; that has been shown to be induced by gp120 due to an increase in TNFA, which increases B cell proliferation, this increase in TNFA not only stimulates cellular proliferation, but also IgM and IgG secretion; furthermore, it has been seen that B cells display binding to the gp120 protein (experimentally, 2.9\% to 4.3\% of the B cell CD20(+) population binds to gp120, therefore we may think of this ratio of binding as a probability p for B cell to bind to gp120 in a range of $p \in [0.029,0.043]$); in B cells, gp120 also promotes cAMP generation \cite{patke} which is known to suppress the activity of NFKB by means of BCR and TLR4 signaling blocking, possibly through a protein kinase A \cite{minguet}, NFKB is also a lymphopoietic regulator in B cells, by maintaining TNFA production at appropriate levels (then we see an increase in TNFA activity, and suppression of NFKB, which may contribute to a malignancy development in HIV patients, even in those patients with antiretroviral treatment, there exists the possibility of developing, for instance, Hodgkin's lymphoma \cite{jacobson}), and plays also a role in the expression of $Ig \lambda$ in immature B cells; mature B cell survival and population restoration is dependent on the NFKB canonical pathway \cite{sasaki}. gp120 also exerts actions down- and upregulating cell cycle- or transcriptional regulation-related genes in T cells (as seen when T cells are treated with the V3 loop of gp120), some of these upregulated genes are: NOC2L (inhibitor of a histone acetyltransferase independent of HDAC, which is seen upregulated) \cite{morou}, SEPT9 (member of a family of GTP-binding proteins, which exhibit roles in cytokinesis, cytoskeleton, cell cycle control; whose hypermethylation is related to several cancers \cite{shen}), IFI6, SPIN, HNRNPM; while some other downregulated genes are: ABCG1, PGPEP1, PTPLA, SPATA21; overall, the V3 loop of gp120 affects the following sets of genes within T cells: cell cycle (62 genes), cellular development (function, maintenance, compromise, morphology) (54 genes), aminoacid or lipid metabolism (36 genes), gene expression (65 genes), DNA metabolism (replication, recombination, repair) (27 genes), cellular assembly and organization (30 genes), cellular movement (25 genes), hematopoiesis (25 genes), immune response (32 genes), cellular death (growth and proliferation) (100 genes), infection mechanism (34 genes); therefore we witness a hijacking of the genetic machinery in T cells which may compromise by all means their further function and proliferation \cite{morou}. Variois gp120 inhibitors have been described, such as: BMS-378806, BMS-488043, BMS-626529, BMS-663068, NBD-556, JRC-II-191, and 18A 
\cite{llu}, BMS-378806 does not display action against HIV-2, SIV, and other viruses, whose metabolism is cythochrome-dependent \cite{zyang} and thus major drug-drug interactions are expected \cite{ywang}. BMS-488043 is an analog of the BMS-378006 compound and has been seen to display in vivo efficacy against HIV-1 in a monotherapy regime \cite{llu}. BMS-626529 or temsavir is administered as a prodrug (BMS-663068), a methyl phosphate prodrug (fostemsavir) which is hydrolized by an esterase, whose metabolism is equally contributed by the cytochromes \cite{lzhu}, the recorded adverse effects include: headache, rash, and micturition urgency \cite{nettles}; additionally, this prodrug has shown a better performance in the inhibition of dually tropic viruses \cite{nowicka}; fostemsavir has very recently been suggested to be used as a drug in multidrug-resistant HIV-1 infection \cite{kozal}.

\subsubsection{gp41: fusion, metabolism, and T-cell function suppression.}

This protein mediates fusion with the host's cell membrane \cite{davidchan}, portion of the env protein pertaining to HIV; this protein displays C-terminal helices which back around N-terminal helices, forming a six-helix bundle \cite{merk}, this structure organizes into an extracellular ectodomain, anchoring gp120 to the cell surface, a transmembrane domain (promoting further anchoring to the lipid bilayer), and a cytoplasmic tail, this tail contains a highly conserved endocytosis motif, and three alpha helical motifs known as lentiviral lytic peptides  \cite{fernandez}; where the transmembrane region of gp41 is followed by a region of high hydrophilicity (the cytoplasmic domain), containing a highly immunogenic region and a C-terminal region with two amphipathic segments (the lentiviral lytic peptides , LLPs), as well as a leucine zipper motif between LLP2 and LLP1; the cytoplasmic domain of gp41 has been suggested to confer conformational stability to the Env protein, but it has been determined that the cytoplasmic domain interacts with host proteins: AP1, AP2, CAM, CTNNA, luman, MA, TP115, perilipin-3, plasma membrane, PRA1, prohibitin 1/2, TAK1. Furthermore, the cytoplasmic domain has been proved essential for both replication and incorporation into several cell lines, such as: monocyte-derived macrophages, peripheral-blood mononuclear cells, B cells, epithelial carcinoma-derived cells; however, the replication and incorporation processes have proved independence of the cytoplasmic domain of gp41 in other cell lines, mostly CD4(+) T cells \cite{postler}. 
Furthermore, gp41 fusion protein works as an inhibitor of T cell activation by different mechanisms. The immunosupressive (ISU) sequence impairs T cell activation through the interaction with the T cell receptor (TCR) complex and the direct inhibition of protein kinase C mediated phosphorylation; gp41 fusion peptide inhibits antigen-specific T cell activation by binding to the TCR$\alpha$ transmembrane domain; gp41 transmembrane domain can enhance the overall immunosupressive effect through an interaction with the fusion peptide; and the recent described immunosupressive loop-associated determinant (ISLAD) is another inhibitor of antigen-specific T cell proliferation and proinflammatory cytokine release by interacting with the TCR$\alpha$ \cite{ashkenazi}. 

Peptides derived from gp41 N-terminal heptad repeat (NHR) and C-terminal heptad repeat (CHR) sequences can inhibit HIV-1 infection by interaction with their counterparts in gp41. Some of these peptide fusion inhibitors are: DP178, later named T20 (generic name: enfuvirtide) was the first fusion inhibitor approved by the U.S. FDA., but the high cost and inconvenience of twice daily injection, prevents it from being considered as a regular drug \cite{cai}; FB006M (generic name: albuvirtide) was approved in 2018 by the Chinese FDA.; SFT (generic name: sifuvirtide) is a novel and potent gp41 inhibitor that has shown promising results; other gp41 fusion inhibitors under development and study are: T1249, 2F5, 4E10, C52L, VIR-576, among others \cite{pu}.

\subsubsection{p7 (nucleoprotein, $NC\in gag$): viral genome package and facilitator.}

This protein derives from the Gag precursor which is cleaved into p17, p24, and p7 proteins \cite{maynard}. Nucleocapsid protein 7 (NCp7) is the major internal component of the HIV virion core, it has been seen to be highly conserved (however not so in spumaretroviruses), displays RNA-binding properties, containing two zinc fingers (ZF) \cite{debaar}, reminding us that these domains are maintained by a zinc ion, coordinating a cysteine and a histidine molecule, these motifs have two $\beta$-sheets and one $\alpha$-helix; it has been seen that these proteins bind to DNA but may as well bind to RNA \cite{cassandri}. p7 possesses functions involved in the selection and packaging of the viral genome, as well as others which play a role in viral replication \cite{maynard}; NCp7 interacts with the viral RNA and is required for its dimerization, encapsidation, and initiation of its reverse transcription, where NCp7 enhances the reverse transcriptase processivity and RNase H activity \cite{levin}; p7 additionally binds to proviral DNA (by means of its basic residues), and protects it from nuclease digestion \cite{lapadat}.
Interestingly, zinc ejection or mutations affecting the zinc finger folding and conformation of the nucleocapsid hydrophobic plateau, lead to non-infectious viral particles \cite{godet}. The importance of its conserved structure is the low probability of mutations found in treatment-resistant strains \cite{klinger}. Thus, NCp7 represents a promising therapeutic target for an effective next-generation antiretroviral therapy. NCp7 inhibitors are divided into covalent and non-covalent inhibitors. Covalent inhibitors, also referred to as irreversible inhibitors or zinc ejectors, which can recognise NCp7 among cellular proteins containing ZFs, some examples of them are 3-Nitrosobenzamides (NOBAs); disulfide-substituted benzamides (DIBAs); thioesters and pyridinioalkanoyl derivatized thioesters (PATEs); benzisothiazolone (BITAs); azodicarbonamide (ADA); thiocarbamates (TICAs); S-acyl 2-mercaptobenzamides (SAMTs); transition metal complexes; diselenobisbenzamides (DISeBAs); and more recently, thioether prodrugs. All of these covalent inhibitors are structurally characterised by a weak electrophilic group that is attached by the distal ZF domain, after this covalent complex has been formed, the zinc is ejected, causing the loss of the tertiary protein structure and consequently, all of its functions \cite{turpin, sancineto}. Non-covalent inhibitors and nucleic acid (NA) binders are another therapeutic target against NCp7. These compounds have weaker antiviral potency when compared with covalent binders, and they have not been approved for clinical trials. Several interactions in the NA/NCp7 complexes are involved with the W37 hydrophobic plateau residue, offering a chance to develop competitive inhibitors. Some of these are pseudodinucleotides, HTS-derived small molecules, thiadiazoles, and thiazolidinones. All of these molecules are characterized by a $\pi$-rich area capable of interacting with W37, seeking to avoid the interaction of NCp7 with NAs. On the other hand, NA-binding NCp7 inhibitors include stem loop structure-binders and anthraquinones, which block the NCp7-RNA-DNA complex formation, but most of them are unable to disrupt a preformed complex \cite{iraci}.

\subsubsection{p6 ($p6\in gag$): incorporating vpr into salient viral particles.}

This protein, also a byproduct of the Gag precursor cleavage, promotes virus particle budding, and the incorporation of the vpr protein into the viral particles \cite{jenkins}; inclusively, p6 is a factor involved in the capsid maturation and virus core formation processes, this by means of its phosphoprotein features; these functions are performed due to the close Euclidean distance which is encountered between p6 and the plasma membrane of the cells, this protein may be adsorbed onto the inner surface of the plasma membrane and promote, for instance, the incorporation of vpr into the viral particles. This protein possesses two $\alpha$-helical domains which are connected by a flexible region, this structure is more pronounced in hydrophobic conditions; its C-terminal region contains vpr-binding residues; the Ser40 residue has been seen to be a potential protein kinase C phosphorylation site \cite{oie}. TSG101 (tumor susceptibility gene 101) is a key cellular protein as part of the endosomal sorting complexes required for transport (ESCRT), which is recruited to viral assembly sites via p6, where the Pro-Thr-Ala-Pro (PTAP) motif in p6 acts as a docking site for TSG101. This process is critical for HIV release. The duplication of this PTAP motif has shown an enhanced replication advantage of HIV-1 subtype C by engaging TSG101 with a higher affinity \cite{sharma}. Furthermore, the deletion of the YPx\textsubscript{n}L motif, which binds to ESCRT component ALG-2-interacting protein X (ALIX), is associated with a decrease in virus release from infected cells (as seen in HIV-1 subtype C), and conversely, PYxE motif insertion can reconstitute the p6 binding to ALIX and consequently, viral budding mediated through the ESCRT pathway \cite{vanDomselaar}. Depletion of the ESCRT components have shown a powerful block to HIV particle release \cite{spearman}, for this reason, therapeutic targets are principally focused on disrupting the p6-TSG101 interface, such as peptoid hydrazones, cyclic peptide 11, F15 (esomeprazole) and N16 (tenatoprazole), these last two are able to join to the ubiquitin E2 variant domain of TSG101, highlighting the possibility of interfering with previously unknown therapeutic targets and expanding the future perspectives of TSG101 inhibitors \cite{dick}.

\subsubsection{p24 (capsid protein, $CA\in gag$): protector of the viral genome.}

Capsid protein (CA) is member of the subset of proteins derived from the Gag polyprotein cleavage, a $[24,25]kDa$ protein which may be detected before seroconversion, this protein may assemble into a protective shell around the viral RNA \cite{graybain} which is a spontaneous process \cite{campbell}, this has been described as a "fullerene cone", hexamers of p24 link into an hexagonal surface lattice, with 12 capsid-protein pentamers, finally possessing around 1500 p24 monomers. This protein is comprised by seven $\alpha$-helices, a $\beta$-hairpin (N-terminal region), a C-terminal region with four $\alpha$-helices, and a flexible linker \cite{perilla}, when assembled into the cone, the N-terminal domain is located on the outer surface of the cone and the C-terminal domain is oriented towards its interior; inside this cone, the RNA genome, and the POL proteins are located (that is the integrase, protease, reverse transcriptase and others); the cone may act in order to shield the genetic content from a host response \cite{campbell}, such as those induced by STING, DDX14 or IFI16 acting as foreign DNA sensors \cite{wallach}. Several models have been proposed in regards to the viral uncoating, that is, the dissociation of the capsid cone into monomers or simpler polymers: immediate uncoating, cytoplasmic uncoating and nuclear pore complex uncoating; there exist studies which support each of these models, in the first case, the uncoating occurs rapidly after the entry into the host's cell has been performed, however, this model has lost its predictive power due to the fact that the core provides protection against the host's foreign-DNA sensors, furthermore, the capsid possesses a pocket to which the reverse transcriptase complex binds to and allows it to be imported into the nucleus. The second model, the one of cytoplasmic uncoating proposes the disassembly of the capsid after a certain time interval has been spanned within the host's cell, this model may be nevertheless challenged as well by the factors which we have already commented (foreign DNA host's response), moreover some additional cellular factors may protect the viral's genome from the host's response, it may occur by means of the HMGA1 protein (high mobility group AT-hook, also called HMGIY), which has been recollected from the preintegration complexes, that is the reverse transcriptase complex before they integrate into the host's genome \cite{farnet}. The third model refers to the nuclear pore complex (NPC) uncoating, in which when the intact capsid reaches the nuclear pores, and disassembles in situ, while the reverse transcriptase complex is imported into the nucleus; this model displays another range of problems which are not compatible entirely with the experimental data, ergo a more suitable model should be provided in following years. Both viral and host's factors take part in the uncoating process: PPIA (or CYPA, a peptidylpropyl isomerase A, with a native function of accelerating the folding of proteins and isomerize the proline imidic bonds \cite{ppia}), dynein (cytoplasmic trafficking), CPSF6 (nuclear import, which is a cleavage and polyadenylation specific factor, interacts with RNA \cite{cpsf6}), TNPO3 (nuclear import, this protein is member of the importin-$\beta$ family of proteins, has been seen to bind to the viral integrase, and as well interacts with the capsid protein), NUP358, NUP153 (nuclear import) \cite{campbell}. Maturation inhibitors are a novel class of antiretroviral drugs targeting the cleavage site between the C-terminal portion of CA and the spacer peptide 1 (SP1), this cleavage site usually triggers a conformational switch that destabilizes the immature Gag and the mature core formation. Maturation inhibitors cause an accumulation of CA-SP1 precursor, which eventually leads to the loss of viral infectivity \cite{dick}. The first reported maturation inhibitor was Bevirimat (BVM), which causes an abnormal virion morphology and inhibition of viral replication, but failed in the phase IIb trial due to resistance mutations in CA-SP1. A second compound named PF-46396 was identified, it shows structural differences compared with BVM, but it induced resistance mutations at different locations, this lead to the identification of second-generation maturation inhibitors, such as GSK3532795, which successfully overcame the inconvenience of drug resistance, but showed a high rate of adverse gastrointestinal events and frequency of treatment-emergent nucleoside reverse transcriptase inhibitor (NRTI) resistance, reason why its evaluation was interrupted \cite{morales-ramirez}. However, the promising results obtained support the continued development of drugs against this therapeutic target. Small molecules and peptide-based antivirals designed to disrupt CA-CA interactions in the immature Gag lattice, the mature core, or both have been studied during the last years. CAP-1 was the first small molecule developed to target the CA protein, it produced abnormal core morphologies, and consequently noninfectious particles. Besides small molecules compounds, 12-mer peptide CA inhibitor (CAI), binds in a hydrophobic CA dimerization interface, but it can not penetrate cell membranes, limiting its clinical use. Some more stable $\alpha$-helical peptides, such as NYAD-1 and NYAD-13, showed a stronger affinity for the binding site than CAI. Other classes of CA inhibitors are benzodiazepienes and benzimidazoles. PF74 and the pyrrolopyrazolones BI-1 and BI-2, seem to compete with CPSF6 and Nup153 for CA binding, disrupting nuclear import. The main problem of the compounds described so far, is the low clinical relevance due to their pharmacological characteristics. Recently, a new type of CA inhibitors has been described, this group includes GS-CA1 and its derivative, GS-6207, are promising drugs that have showed higher potency than PF74 \cite{dick}.

\subsubsection{p17 (matrix protein, $MA\in gag$): multifunctional protein.}

\begin{figure}
    \centering
    \includegraphics[width=0.9\textwidth]{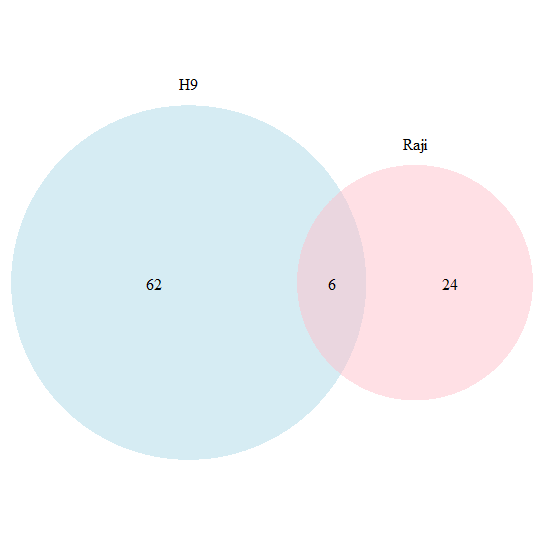}
    \caption{Venn diagram of the sets representing the receptors expressed in Raji cells (Raji) and H9 cells (H9), where we regard at their cardinalities, that is, $|R|=28$, $|H|=68$, and $R\cap H=6$; in this case the p17R protein lies in the $R-H$ set. }
    \label{fig:figset}
\end{figure}

The HIV-1 p17 protein (matrix protein, MA) is associated with the inner surface of the viral envelope, and may primarily function as an anchor of the gp41 protein on the virion surface. When in solution it is mainly encountered in a monomeric form, while in a solid state it trimerizes \cite{massiah}; it displays five $\alpha$-helices, and a three-strand $\beta$-sheet, whereby the C-terminal domain, exposes carboxyl-terminal residues which aid in the early states of HIV infection, and basic residues promote membrane binding and nuclear localization (the residues which take part in this process are located in a cationic loop connecting $\beta$-strands one and two). This protein additionally functions in RNA targeting to the plasma membrane, incorporation of the envelope into virions and particle assembly, and aids in the transport of the reverse transcriptase complex through the nuclear pores \cite{starich}, it also acts as a viral cytokine, by binding to a cellular receptor, namely p17R \cite{fiorentini}, this receptor has been seen to be expressed in Raji B cells, and activates AP-1, as well as ERK1/2 and downregulates AKT, this by means of maintaining PTEN in its active state through the serine/threonin kinase ROCK \cite{ciagulli}; the identity of this p17R protein is a bit obscured in the literature, nonetheless through a non-exhaustive search in regards to Raji cells' receptors we came up with a list of potential proteins which may play such role, bearing in mind as well that the H9 cell line lacks this p17R protein in its surface, if we consider that the surface proteins of the Raji cells are members of the set $R=\{p_1,...,p_n \}$ and those of the H9 cell line pertain to the set $H=\{r_1,...,r_m \}$, therefore $\exists p_i\vee r_i : p_i \in R,p_i \notin H$ \cite{harmonizome};  those proteins in the set $R\cap H$ are: PVRL1, LILRB1, DRD4, CRLF3, and ADGRD2; such that these receptors may not be the p17R protein; now, those proteins which are exclusively expressed in Raji cells and take part in PI3K/AKT/PTEN/AP1/GPC (G-protein coupling) are: NTSR2, TACR3, FOLR2, MC5R, OR2J3, TNFRS13B, PTAFR, CCR8, GPR173; this sets are shown in Figure \ref{fig:figset}; the activation of this pathway (by means of this p17R) may lead to a promotion in proliferation and release of proinflammatory cytokines from T-cells \cite{fiorentini}; but if p17R is also expressed in B cells, it ought indeed to promote B cell growth and tumorigenesis \cite{ciagulli}. MA protein has a fundamental role in virion assembly, because of its highly conserved PI[4,5]P\textsubscript{2}/nucleic acid binding site, it has become an attractive site for the development of new antiretroviral drugs, nevertheless, other drugs have been described targeting the nuclear localization signal of MA or the MA-RNA interaction. Thiadiazolane based compounds where first described, they target the MA-RNA interaction, but showed significant levels of toxicity. In contrast, PI[4,5]P\textsubscript{2} binding site inhibitors were not associated with cytotoxic effects, but work is actively ongoing in optimizing the affinity/potency of this type of chemotypes. New targeting sites are highly desirable, one of them could be the involvement of MA in Env incorporation, where MA trimerization is important for the recognition of Env cytoplasmic tail (gp41-CT) and virus assembly \cite{dick}.

\subsubsection{p10 (protease, $PR\in POL$): catalyzer of viral and host's proteins breakdown.}

This protease is encoded by the \textit{pol} gene being an aspartic protease, and as in other retroviruses, displays its function when in its homodimeric form; each monomer is an aspartic peptidase with four elements: two hairpin loops, wide catalytic aspartic acid loop, and an $\alpha$ helix \cite{dunn}, the dimeric interface comprises eight N- and C-terminal residues of each chain, some of these are exposed to the solvent, and some others (the hydrophobic ones) are oriented towards the interior of the enzyme \cite{tropsha}. This protein displays two states (when in the homodimer form), either open or closed, depending on the presence of a ligand (open when in the free state, and closed when in the bound state) \cite{nkeze} where the open form is more stable than the closed one (considering the free energy of the process $\Delta G$) \cite{tropsha}. The HIV protease cleaves the polyproteins Gag and Gag-Pol creating protein subunits \cite{miaohuang}; all the same the viral protease displays proteolytic abilities of the host's proteins, such as those belonging to: the cytoskeleton: vimentin, desmin, GFAP \cite{shoeman}, actin, troponin \cite{tomasselli}, laminin \cite{niebren}; immune system: proIL1B; and transmembrane proteins: APP \cite{tomasselli}; cytosolic proteins: BCL2, CASP8 \cite{niebren}. Vimentin plays a role in the diapedesis process of T cells, whereby this protein, T cells migrate across the endothelium, and in genotypes $vim^{-/-}$, there is a diminished capacity of T cells to home to mesenteric lymph nodes and spleen, this may be mediated by changes in expression or distribution of ICAM1 and VCAM1 \cite{nieminen}; while actin polymerizes or depolymerizes when the T cell becomes active, this T cell activation leads towards the formation of a distal-pole complex which is an actin-rich structure; disruptions in the cytoskeleton might as well induce changes in the organization of the supramolecular activation clusters in T cells (SMACs) \cite{billadeau} which takes a central part in the immunological synapse, being the SMAC a nanocluster comprised of an actin mesh, transmembrane proteins (members of the TCR complex), and cholesterol-enriched nanodomains \cite{parajo}; these disruption in the cytoskeletal proteins might be added to the one induced by the gp120 protein, as this protein increases ICAM1 expression \cite{renyao,tardif}, inducing the formation of a type of SMAC but in this case gp120-dependent (and likely protease-aided), in which gp120 clusters in the centre, and LFA1 and ICAM1 concentrate in the periphery, this type of SMAC is termed the virological synapse, throughout which the virus may spread between T cells without the need of being externalized from the cell (no need of viral budding) \cite{martin}; this SMAC recruits the host's proteins which we may as well find in the immunological synapse, such as: TCRZ, ZAP70, LAT, SLP76, ITK, PLCG, and only a weak recruitment of: CD3E, and ABTCR; however when this protein nanocluster was assembled, there was an abnormal signaling (distinct to the one induced by the immunological synapse), where no PKC recruitment, no calcium mobilization nor CD69 upregulation were observed \cite{vasiliver}. By means of the cleavage of both BCL2 and CASP8 can the protease induce cell death \cite{niebren, leibowitz}. HIV-1 protease inhibitors (PI) play a key role in the antiretroviral treatment. First generation of PIs were based on hydroxyethylene and hydroxyethylamine isosteresstarted. Saquinavir was the first PI approved by the FDA in 1995, since then many other PIs have been developed. Ritonavir was found to be a potent inhibitor of cytochrome P450 3A, a major metabolic enzyme for PIs \cite{kumar}, due to this finding, ritonavir is more frequently used as a PI pharmacokinetic booster than as a PI itself. Other first generation antiretrovirals are indinavir, nelfinavir, and amprenavir. Due to the major problems of first-generation PIs (high metabolic clearance, low half-life, poor oral bioavailability, gastrointestinal distress, and the emergence of drug-resistance strains), second-generation PIs were developed. The balance of hydrophobicity and hydrophilicity allowed for a longer half-life. Some of these drugs are lopinavir, atazanavir, tipranavir (reserved as a salvage therapy), and darunavir (high potency against multidrug-resistance strains) \cite{ghosh}. Despite the progress in PIs therapy, some drug-resistance variants have emerged, they are classified in two groups: primary mutations, which involve changes in residues directly involved in substrate binding and manifest themselves in the active site of the enzyme; while secondary mutations are located away from the active site and are usually compensatory mutations to mitigate the deleterious effects of primary mutations \cite{mitsuya}. Even, secondary mutations may not manifest in the protease itself but instead in the protein cleavage site on the Gag-Pol and Gag substrates \cite{perno}. One of the main strategies to avoid drug resistance, is the design of PIs by promoting hydrogen bonding interactions with the backbone atoms in the HIV-1 protease active site, since mutations that cause drug resistance cannot significantly alter protease active site backbone conformation, as an example we have darunavir, that showed extensive binding interactions with backbone atoms and maintained a potent antiviral activity against panels of clinically relevant multidrug-resistant HIV-1 variants \cite{ghosh,ghosh2}. Recently, a new pharmacokinetic enhancer, cobicistat, which does not have anti-HIV activity, has been developed, offering the advantage over ritonavir, where cobicistat will not contribute toward the emergence of drug-resistant HIV-1 variants \cite{wu}. Many other new classes of PIs with innovative ligands are under preclinical development for the next generation, molecular design efforts have focused on the synthesis of P2-ligands promoting enhanced backbone binding and non-peptide PIs containing different structural scaffolds distinct from hydroxyethylsulfonamide isosteres \cite{ghosh}. 

\subsubsection{p51 (reverse transcriptase, $RT\in POL$): loader of nucleic acids and structural support.}

The reverse transcriptase's main function is to convert the viral RNA which has just entered the host's cell after membrane fusion into double-stranded DNA, this occurs in the host's cell cytoplasm, this DNA will later be translocated towards the nucleus in order for it to be finally integrated into the host's genome. It is a heterodimer comprised of p66 (or p15 or RNase H) and p51, both of which derive from a common Gag-Pol polyprotein, cleaved by the viral protease (p10, PR); these subunits (p66 and p51) share a common amino terminus; the details of p15 will be addressed in the following section \ref{p15}. p51 possesses 4 subdomains (which are homonymous to those of p15 or p66): fingers, palm, thumb, and connection  \cite{sarafianos}. The heterodimerization process is dependent on factors such as: mutations (e.g., the L289K mutation, the L289 residue in the p51 subunit makes various hydrophobic contacts with p66 but the mutation does not yield towards an ill-heterodimerization or no heterodimerization at all; contrarywise, the same mutation but occurring at the protein p66 yields towards a lack of dimerization), nucleotide substrates, temperature, magnesium, and solution conditions \cite{zheng}. It has been noted that the catalytic site of the reverse transcriptase lies in the p15 (p66) domain, however p51 may take part in: structural support and facilitating the loading of nucleic acids onto p66, which would explain why the N348I mutation to p51 may confer resistance to reverse transcriptase inhibitors (both nucleoside and non-nucleoside), as well, p51 has been seen to contribute to the architecture of the RNase H primer grip/phosphate binding pocket, finally, deletions to the C-terminal sequence of p51 may alter changes in the RNase H activity \cite{chung}. The p51 subunit in itself may undergo homodimerization under certain conditions, this was observed by small-angle X-ray scattering (SAXS), this phenomenon occurs dependent on the concentration of the p51 monomer, and may coexist with the homodimer p51/p51 in equilibrium, furthermore it has been noted that p51 may exist in two forms: p51E and p51C, the latter is the form in which it may be encountered when bound to the p66 in the heterodimer, and the former is an extended form in which it may resemble the structure of p66, the homodimerization may occur with one p51E species and one p51C species assemblying into p51E/p51C, this could happen due to the stabilizing effect of p51C on the structure of p51E. p51 interacts with non-nucleoside reverse transcriptase inhibitors (NNRTIs) forming two different species: p51E-NNRTI and p51E/p51C-NNRTI, this due to the fact that p51E is in possession of a p66-like structure \cite{zheng}; the monomeric form of p51 is favored by low concentration, low salt, and low temperature conditions \cite{zheng2}. Reverse transcriptase inhibitors are divided into nucleoside/nucleotide and non-nucleoside reverse transcriptase inhibitors, NRTIs and NNRTIs, respectively. NRTIs inhibit viral replication through competition with purines and pyrimidines, avoiding adding new nucleotides and consequently, finishing viral DNA replication. In contrast, NNRTIs bind to an hydrophobic pocket in p66, this binding leads to a stereo-chemical change in the protein, preventing the addition of new nucleosides and blocks the cDNA elongation. Typically, NRTIs constitute the backbone of the antiretroviral therapy, among them we can find tenofovir disoproxil fumarate (TDF) and tenofovir alafenamide (TAF), which are adenosine-derivated; emtricitabine (FTC) and lamivudine as cytosine analogs; abacavir as guanosine analog; zidovudine and stavudine as thymidine analogs; and didanosine as an inosine derived. An advantage of the current NRTIs is the low clinically significant drug-drug interactions but they have the disadvantage of significant side effect profiles and problems with resistant HIV-1 variants  \cite{rai}. All of them can be affected by selected resistance mutations, either by mutations in the N-terminal polymerase domain of the enzyme, where the most common are K65R, L74V, Q151M, and M184V; or by thymidine analog mutations, where are included M41L, D67N, K70R, L210W, T215Y/F, and K219Q/E \cite{rai,marcelin,asahchop}. Among NNRTIs, we can find the first generation (efavirenz and nevirapine) and second generation (rilpivirine and etravirine). Resistance mutations against NNRTIs are based on the inhibited drug interaction with the NNRTI binding pocket (as seen with K103N and K101E), disruption between the drug and NNRTI binding domain residues (Y181C and Y188L), or changes in the conformation or size of the NNRTI binding pocket (Y188L and G190E)\cite{adams,das,das2,ren}. Unlike NRTIs, NNRTIs have a higher plasma half-life, so they can act as perpetrators of drug interactions additionally to their significant side-effect profile \cite{rai}. New NRTIs and NNRTIs are focused on overcoming resistances and reduce side effects \cite{gulick}. Islatravir (MK-8591) is a nucleoside nonobligate chain terminator that inhibits the enzyme by preventing its translocation, and therefore it has been categorized as a nucleoside reverse transcriptase translocation inhibitor (NRTTI), demostrating a robust antiviral activity \textit{in vivo} against HIV-1 \cite{markowitz}. Doravirine (MK-1439), a novel NNRTI, has showed its potency at low doses against common resistant strains and has less potential drug interactions than its counterparts  because it does not have appreciable inductive or inhibitory effects on CYP enzymes \cite{blevins}. Elsulfavirine is the prodrug of the active compound VM-1500A, it has shown a high genetic barrier to the development of resistant drug mutations \cite{al-salama}. A study showed that elsulfavirine was not inferior to efavirenz combined with TDF/FTC in virologic control, and in addition, elsulfavirine was better tolerated \cite{murphy}.

\subsubsection{p15 (RNase H, $(p15\vee p66 \in pol$): converter of RNA into dsDNA.}\label{p15}

We have previously commented that the reverse transcriptase protein is comprised of two subunits: p51 and p66 (or p15), where the enzymatic activity of the protein is performed by the p15 subunit, this enzymatic activity is  performed by both the polymerase and by the RNase H, the polymerase domain contains 4 subdomains: fingers, palm, thumb, and connection \cite{sarafianos}. The RNase H belongs to a nucleotidyl-transferase superfamily, this family includes enzymes such as: transposase, retroviral integrase, Holliday junction resolvase, and RISC nuclease Argonaute. In retroviruses, this RNase H converts an ssRNA into dsDNA, removes the template RNA after the DNA synthesis and produces a polypurine primer for the second-DNA-strand synthesis. In reactions catalyzed by the RNase H, the nucleophile is derived from a water molecule or a 3'-OH from the nucleic acid; the active site of these enzymes is dependent on magnesium or manganese to bind to the substrate and catalyze the reactions, these reactions occur in a one-step bimolecular nucleophilic substitution; two metallic ions are required for the reactions, in which one metal ion activates the hydroxyl nucleophile and the other one stabilizes the pentacovalent intermediate. Various RNases display a conserved $\sim 100$ residue core structure: five stranded $\beta$ sheet and three $\alpha$ helices (A,B,D). Specificity for the RNA strand is determined by contacts between the RNase H and five consecutive 2'-OH groups, and when the RNA complexes (around 6 bp) with the enzyme, around 100 nm of the enzymatic surface becomes buried, a groove which protrudes near the site of the buried surface is the active site, where catalytic carboxylates (E109 and D132) make hydrogen bonds with 2'-OH groups. The DNA binding site is located in the alpha helices (polymerase), which fold into a groove, a phosphate binding pocket accompanies this binding site (comprised of the amino acids T104, N106, S147, T148); the pocket when not in enzymatic activity is occupied by a sulfate ion, which is replaced by the DNA's phosphate \cite{nowotny}. It has been noted that the manganese may specifically promote a specific hydrolytic activity (hydrolysis of dsRNA) to the enzyme, rather than the more general endonuclease and directional processing activities, divalent metal binding may be mediated by the p66 residue E478 \cite{cirino}. The initial RNAse H inhibitors were described almost 30 years ago, however, the first inhibitor with a relevant effect was \textit{N}-(4-\textit{tert}-butylbenzoyl)-2-hydroxy-1-naphthaldehyde hydrazine (BBNH), although it was also an inhibitor of the RT DNA polymerase \cite{borkow}. After, a derivative,  (\textit{E})-3,4-dihydroxy-\textit{N}'-((2-methoxynaphthalen-1-yl)methylene)benzohydrazide (DHBNH), which has its binding site between the RT polymerase active site and the polymerase primer grip, alters the trajectory of the template-primer, so the RNAse H cannot cleave the RNA strand in RNA/DNA complexes \cite{himmel}; metalchelating RNase H active-site inhibitors, such as 1-hydroxy-pyridopyrimidinone, pyridopyrimidinone, dihydroxycoumarin, diketo acids, $\beta$-keto acids, 3-hydroxypyrimidine-2,4-diones and hydroxypyridonecarboxylic acids, might sequester Mg\textsuperscript{+2} ions required for RNAse H activity \cite{tramontano,wang}. An alternative to metalchelating inhibitors are allosteric inhibitors that interfere with the RNA-DNA binding and induce changes in the RNAse H active site, examples of these are the cycloheptathiophene-3-carboxamide (cHTC) derivatives \cite{wang}. Several other inhibitors have been proposed and developed such as the dual inhibitors RDS1643 (co-targeting HIV-1 IN and RNAse H), EMAC2005 (co-targeting HIV-1 RNA-dependent DNA polymerase and RNAse H), and RMNC6 (co-targeting RT and RNAse H) \cite{wang}. However, despite extensive research and improvement in RNAse H inhibitors, the efficacy of these molecules remains low \cite{tramontano}.

\subsubsection{p32 (Integrase, $IN\in pol$): immersing viral DNA into the host's genome.}

After the dsDNA has been produced in the reverse transcriptase enzyme, this DNA is part of a large nucleoprotein complex, the reverse transcription complex \cite{craigie}, this DNA shall later transite from the reverse transcriptase towards the pre-integration complex (PIC) which contains the integrase enzyme, as well as the viral proteins: vpr, capsid, and matrix, and the host proteins: INI1 and PML \cite{iordanski}, moreover, it has been noted that vpr-defective viral particles fail for their PICs to be effectively translocated to the nucleus, whereas matrix-defective viral particles are only partially effective in the transport process, this could be related to the finding that vpr binds to KPNA (karyopherin alpha), the ideal concentration of the vpr protein in order for this to occur is $<5 nM$, for when concentrations of the vpr protein such as $>42.8 nM$ happen, the transport mechanism is inhibited \cite{popov}; this pre-integration complex will afterwards be transported towards the nucleus by active mechanisms (due to the size impediment of the PIC, being this complex around 28 nm wide, versus 9 nm as the limit of passive transport towards the nucleus), the active transport mechanism is dependent on the capsid protein (CA) from the HIV-1 particle, as well as various nuclear transport proteins: NUP153, NUP358, CPSF6, TNPO3, and ADAM10 \cite{endsley}. The integration of the viral genome into that of the host's seems partially non-random, due to the fact that the viral genome has been seen to prefer transcriptionally active regions of the host's genome; the integrase enzyme also displays preference towards chromatinised DNA substrates rather than naked ones, i.e., the viral genome integrating process can be described by means of a stochastic process $\{S_x\}$, where the x represents the host's DNA fragment, the probability of the viral DNA being bound to the genome in a position near the nucleosome increases 4-fold, and when this position shifts towards naked DNA, the probability distribution becomes a uniform one. The efficiency of integration has also been computed to increase when the chromatin is found in an open fashion (euchromatin) and not in its closed one (heterochromatin) \cite{michieletto}; furthermore, not only does the viral DNA prefers transcriptionally active sites, but as well as intronic genetic ones. The viral DNA integration is aided partially by the host, through a protein product of the PSIP1 gene, say the LEDGF/p75 protein which belongs to the hepatoma-derived growth factor-related proteins, which contains a conserved N-terminal PWWP (this domain is a member of the Royal superfamily, functions as a chromatin methylation reader, by recognizing methylated DNA or histones, through its high content of basic residues, thus raising its isoelectric point and facilitating DNA interactions \cite{rona}) and A/T hook domains (this is a DNA binding peptide motif, with a Gly-Arg-Pro tripeptide, where we may find in its sides basic amino acids, these DNA binding motifs, that might as well bind to RNA \cite{filarsky}) \cite{agosto}. 
Available integrase inhibitors include bictegravir, cabotegravir, dolutegravir, elvitegravir, and raltegravir, that may be used in combination with other antiretrovirals (cobicistat, emtricitabine, tenofovir, rilpivirine, and others) \cite{scarsi}; bictegravir tends to be used along with emtricitabine and tenofovir alafenamide, (BET) in which case bictegravir displays a high genetic barrier in HIV-1 resistance development, this combination of antiretrovirals was similar or superior to the one of dolutegravir, abacavir, and lamivudine (DAL); where in naïve adults, BET established virological suppression in treatment during 96 weeks \cite{deeks}; bictegravir inhibits the integrase with an $IC_{50}=7.5 \pm 0.3 nM$ \cite{tsiang}. Cabotegravir, dolutegravir, and bictegravir have been seen to most efficiently inhibit B HIV-1 subtypes, compared to non-B subtypes \cite{neogi}.

\subsubsection{vpr (virus protein r): nuclear importer of viral genome.}

This protein is a 96 amino acid, 14 kDa protein, which is conserved among other retroviruses \cite{kogan}. The structure of the vpr protein is featured by three $\alpha$-helices, with the N- and C-terminal domains surrounding these helices \cite{gonzalez}, this protein has been seen to exists as an oligomer being the amino acids 36-76 of essence for this polymerization to occur \cite{zhaowang}, \textit{in vivo} both dimers and hexamers have been observed, where in the dimeric state only helices II and III of the protein's structure are involved, while in the hexameric state, helix I also takes part \cite{venkatachari}. Its functions include: the nuclear import of the preintegration complex \cite{tsurutani}, induction of G2 cell cycle arrest, modulation of T-cell apoptosis, transcriptional activation of both viral and host genes, regulation of NFKB activity \cite{kogan}; these functions are regulated by distinct domains of the vpr protein, the N-terminal domain may take part in nuclear localization and virion packaging, the LR domain is involved in the nuclear localization of vpr, and the C-terminal domain regulates cell cycle arrest \cite{mahalingam}. Cell cycle arrest (in G2-M phase) is described as one of the most dramatic functions of vpr, process occurring with the aid of some host's proteins, for instance: DCAF1, CRL4, which includes CUL4A, RBX1, and DDB1 \cite{yingwu}; DCAF1 can promote ubiquitylation of proteins through CRL4 and another E3 ligase, assemblying later with another protein into a complex, DCAF1-DDB1-DYRK2-EDD; it has been determined that DCAF1 regulates cell-cycle progression and apoptosis either with TP53 or without it, where DCAF1 negatively regulates TP53, as well as some TP53 target genes (TP21, TP27), while another E3 ligase, MDM2 can promote TP53's ubiquitylation, inducing then its degradation; DCAF1 not only acts directly on TP53 but can additionally alter its epigenetic landscape through an interaction with HDAC1, then recruiting this deacetylase towards TP53-dependent promoters \cite{schabla}; nonetheless TP53 is not the only protein whose expression is affected by DCAF1, for MUS81 and MCM10 are also affected by it \cite{yingwu}, where MUS81 is a Holliday junction resolvase, regulating cell growth and meiotic recombination \cite{haber}, and MCM10 is a protein essential for the initiation of chromosome replication containing a single-stranded DNA binding domain, this protein complexes with CDC45 and GINS proteins which togetherly display helicase activity \cite{watase} (ATP-dependent DNA unwinding enzyme \cite{brosh}). Interestingly, vpr can alter the host's transcriptome, this done through ubiquitination of HDAC1, HDAC2, HDAC3, and HDAC8, being HDAC3 the most affected one; this has a two-fold response: enhance the viral genome expression and that of the host \cite{romani}, some of the host's proteins whose expression has been seen to become altered by vpr are: ANT, VDAC, PMCA, GLUD2 (mitochondrial), HK1 (glucose metabolism), G6PDH (pentose phosphate pathway), GAPDH (glycolysis), ATR (stress response), DCAF1 (proteasome), NFAT, NFKB, CEBP, AP1, HIF1A, HDAC1, GR, SP1 (transcription), SLX4, HLTF, and UNG2 (DNA metabolism) \cite{gonzalez}; therefore, vpr possesses the potential to radically modify the host's T-cell transcriptome. 
Some naturally occurring compounds have been recorded to exhibit inhibitory activity against vpr, such as isopimarane diterpenoids and picrasane quassinoids \cite{nwetwin}; while coumarin compounds in \textit{in silico} experiments have yielded information that these compounds may bind to the hydrophobic pocket in vpr and could inhibit the protein \cite{choi}; a novel hematoxylin derivative was developed, this compound binds to vpr and can inhibit HIV-1 replication, this hematoxylin derivative can be produced by reacting the hematoxylin molecule with 2,2-dimethoxypropane in acetone with p-toluene sulphonic acid and phosphorus pentoxide, with an intermediate molecule being produced, the acetonide-protected hematoxylin, this is reacted with allyl bromide with $K_2 CO_3$ in dimethylformamide, where we might then obtain the stable hematoxylin derivative \cite{hagiwara}.  

\subsubsection{vpu (virus protein unique, p16): CD4 downregulator.}

This protein is an 81-amino acid type I transmembrane protein, pertains to the viroporin family of proteins; this protein displays certain functions: downregulates CD4 receptor expression, promotes release of virions from infected cells; this protein is translated from the  gene which also encodes for the \textit{env} complex \cite{eugenia}, and it has been suggested that the expression of \textit{env} and \textit{vpu} are coordinated; the \textit{vpu} gene lies between the first exon of the \textit{tat} and \textit{env} genes, this gene is absent in HIV-2. Vpu may assemble into homooligomers, and the monomers exhibits an N-terminal domain (luminal), a single transmembrane domain, and  C-terminal hydrophilic domain (into the cytoplasm); where this transmembrane domain appears essential for the oligomerization of Vpu, and a stoichiometry of $(Vpu)_n,n=5$ is required for the formation of the pore \cite{dube}; there have been experimental suggestions in regards to the ion channel activity of this protein, such as that result of a selective conductance for cations of this protein (where vpu has been expressed in frog oocytes), this protein then has been compared with the M2 protein of the influenza virus, because vpu possesses a similar transmembrane topology, an uncleaved transmembrane domain, similar length, and casein kinase II phosphorylation sites; features whereby the M2 protein has been seen to transport protons, in this protein a motif within the transmembrane domain has been proved essential for its ion channel activity, the His-X-X-X-Trp motif, in this case, His is the proton sensor and Trp is the pore; when drugs which targeted the transmembrane domain of vpu, the release of virions could be decreased, additionally, the vpu protein displays an Ala-X-X-X-Trp motif in a very similar position as the previously mentioned motif does in the M2 protein; nonetheless, these discoveries while they indeed proof the necessity of this protein by the virus in releasing the virions mainly through its transmembrane domain, its role as an ion channel has not yet been fully proved \cite{guatelli}, but this protein may alter potassium transport, with a possible role as an ion channel and additionally with an increased TASK1/KCNK3 protein degradation (KCNK3 is a member of the superfamily of potassium channel proteins, it is activated by halothane and isoflurane \cite{kcnk3}) which can lead to membrane depolarization and viral release \cite{eugenia}. The mechanism through which vpu promotes CD4 downregulation is by means of targeting newly synthesized CD4 molecules for proteasomal degradation, promoting the binding of CD4 to the SKP1-CUL1 E3 ubiquitin ligase which allows for the ubiquitination of CD4 on lysine, serine, and threonine residues, then enabling the retention of the CD4 protein in the endoplasmic reticulum, being afterwards recruited the VCP-UFD1-NPL4 dislocase complex (NPL4 tends to recognize Lys48-linked polyubiquitinylated substrates \cite{sato}, VCP extracts ubiquitinated proteins from lipid membranes \cite{vcp}), finally leading to degradation by means of the proteasome. Downregulation of CD4 is not accomplished purely by the action of the vpu protein, but instead it is a constructive result of vpu, nef, and env proteins \cite{pacini}. If the viral particle uses CD4 as a main receptor for its entry into the host's cell, why would it downregulate this molecule? Because the unlimited binding of the viral particle to the CD4 molecules would lead to superinfection (acquisition of different viral strains infecting a single T-cell \cite{benson}) of a T-cell, and superinfection of T-cells leads to increased apoptotic rate, by diminishing the available CD4 molecules for the viral particle to bind to, this superinfection is in its stead decreased, and consequently, the likelihood of increased T-cell apoptosis, which can lead to a more everlasting viral infection \cite{wildum}. Downregulation of the CD4 molecule expression on the membrane surface could also serve the virus in promoting the virion release from the cell surface (due to the finding that CD4 may bind to the \textit{env} protein in virions, as well as CD4 prevents the \textit{env} protein from inserting itself into the virion surface) \cite{ueno}. In vpu-defective virions, where the viral release was still allowed, when treated with type 1 interferons, these viral particles were no longer released, and were tethered onto the host's cell surface; it was later observed that tetherin (BST2) was one of the proteins involved in this process \cite{dube}, whose expression is promoted by type I interferons \cite{prevost}. Vpu aids virion release and avoidance of virion tethering onto the cellular surface by ubiquitinating BST2 at the cytoplasmic lysine residues 18 and 21, likely promoting subsequently proteasomal degradation, nevertheless, lysosomal degradation is also possible and has been docummmented (because lysosomal inhibitors have been seen to inhibit vpu-mediated BST2 degradation) \cite{dube}. Tetherin displays this restraining activity in regards to viral release from the host's cell not only in HIV1, but as well in: HIV2, SIV (simian immunodefficiency virus), alpha-, beta-, and gammaretroviruses; filoviruses (Marburg and ebola viruses), arenaviruses (Lassa virus), herpesvirus (type 8 or associated with Kaposi's sarcoma); thereof can tetherin be regarded as an antiviral molecule \cite{kuhl}. 
A molecule BIT225 (2-naphthalenecarboxamide, N-(aminoiminomethyl)-5-(1-methyl-1H-pyrazol-4-yl)) has been described as a vpu inhibitor regarding its viroporin function, and can also work as a hepatitis C virus protein 7 inhibitor \cite{kuhl2}, SM111 (1-(2-(azepan-1-yl))nicotinoyl)guanidine) has also been described as a potential vpu inhibitor, for HIV-1 replication diminishes when SM111 comes into play even with protease, reverse transcriptase, and integrase inhibitors-resistant strains \cite{mwimanzi}; another class of molecules which show promise in inhibiting vpu function, are phlorotannins (brown-seaweed derived molecules, with a structural unit of polyphenols \cite{venkatesan}), where these molecules have been tested in silico, one advantage of these molecules is the finding that they can as well inhibit the reverse transcriptase and the protease proteins, reducing the likelihood of HIV replication within the cell, one point, nonetheless, to be considered is the hydrophyllicity of these molecules which may reduce oral bioavailability \cite{langarizadeh}.

\subsubsection{nef (negative regulating factor, p27): downregulator of host's proteins.}

Nef is a 27 to 35 kDa N-myristoylated protein, in which the myristoylation process is mediated by the N-myristoyl transferase (NM1), and is further cleaved by the viral protease, in a site near a hydrophobic groove, which stabilizes a PxxP loop \cite{geyer}; structurally, this protein is highly plastic, and many conformations have been determined \cite{xiaofei}. Through its N-myristoylation, is nef able to latch onto the inner surface of the host's cell membrane and can therefore interact with the clathrin-coated vesicle machinery, redirecting some transmembrane proteins from the cell surface. Nef can interact with AP2 (heterotetramer comprising the monomers $\alpha$, $\beta2$, $\mu2$, and $\sigma2$) which is a bridge between a membrane protein substrate (which will be later cargoed onto the clathrin-mediated endocytosis process towards its degradation) and clathrin, throughout this AP2-Nef interaction the latter protein can downregulate the host's proteins: CD4, CD8, CD28, CD3, SERINC3, SERINC5, CXCR4, CCR5; Nef also interacts with AP1 (a heterotetrameric adaptin, which mediates the recruitment of clathrin and recognition of sorting signals of transmembrane receptors \cite{ap1b1}) being able subsequently to downregulate both tetherin and MHCI \cite{buffalo}. We can witness the impact of the Nef-induced downregulation of proteins, such as CD3, essential for TCR signaling; and CD28 with vital roles in T-cell proliferation, and Th2 cell development \cite{cd28}. The small molecule 2c (2,4-dihydroxy-5-(1-methoxy-2-methylpropyl)benzene-1,3-dialdehyde \cite{dikeakos}) has been seen to inhibit the downregulation of MHC-1 mediated by the action of nef through disrupting the interaction of nef with a Src family kinase \cite{dekaban}; 4-amino substituted diphenylfuropyrimidines have also been described as nef inhibitors, and subsequently inhibiting the nef-dependent HCK activation, the values of the inhibitory concentration $IC_{50}$ range the micromolar order; one of these compounds relates to the tyrosine kinase type of inhibitors (which are built around a 5,6-biarylsulfo[2,3-d]pyrimidine pharmacophore \cite{emert}. 

\subsubsection{tat (transactivator protein, p14): promoting viral genome expression.}

This protein has a weight which may be found within the interval of $w\in[14,16] kDa$ \cite{pugliese}, which uses the alternative splicing machinery of the host in producing this set of possible tat translated isoforms \cite{sertznig}. It has been observed that the tat protein displays what could be considered as antagonistic properties to other viral proteins, such as vpu or nef, in regards to the fact that tat promotes CD4 surface expression, as well as it promotes cellular survival (through BCL2) \cite{pugliese}, however it has been observed that the interplay between the proteins tat and nef, which would seem initially antagonistic, instead provide optimal viral infection, which by binding nef to tat, induction of transcription occurs, leading to a specific cellular fate determined by the viral fitness \cite{joseph}. This protein promotes gene expression (both viral and those of the host cell) by binding to a nascent RNA structure termed transactivator response region (TAR) which results in the transactivation of the long terminal repeat (LTR) promoter; tat can also recruit histone acetyltransferases, inducing histone acetylation in those genes whose expression is mediated by the LTR; these LTR-induced expressed genes may not only be promoted by means of TAR, but as well tat may induce these genes' expression by binding to the NFKB enhancer \cite{dandekar}. Furthermore, tat regulates a set of cellular genes by various mechanisms: by means of binding to TAR-like sequences, binding to promoter region or interaction with transcription factors; these mechanisms regulate then the expression of genes such as: IL6, TNFB, MAP2K6, IRF7, PTEN, PPP2R1B, PPP2R5E, REL, IL2, IL2RA, OGG1, LMP2, CD69, VAV3, ADCYAP1, and FAM46C \cite{clark}; as it is possible for us to regard, these genes are related to cytokine signaling, cell cycle, TCR activation, amongst other processes. Tat can be inhibited by various compounds: peptide-based, oligonucleotide-based, and small molecules. Peptide-based compounds are arginine-enriched and can assemble into complexes with the Transactivation response element RNA (TAR RNA), both of which can compete for binding to the tat protein; peptoids are another type of peptide-based compounds which have been seen to inhibit the assembly of the tat-TAR RNA complex, peptoids are peptide isomers, in which the branching amino acids are all bound to nitrogens in the backbone. In regards to oligonucleotide-based compounds we can find TAR decoys, which resemble the TAR RNA sequence and can bind to the tat protein, however with a limited impact on HIV-1 replication; antisense sequences to that of the TAR RNA one can also inhibit tat activity; small interfering RNAs can also be used in inhibiting tat activity by promoting the degradation of the target mRNA sequence. Interestingly, in regards to the small molecule compounds, the quinolones have displayed anti-tat activity at the nanomolar range by means of chelating $Mg^{+2}$ in the TAR RNA sequence, one specific quinolone compound, WM5 has been tested (6-amino-1-methyl-4-oxo-7-[4-(2-pyridyl)piperazin-1-yl]quinoline-3-carboxylic acid) \cite{richter}, the use of quinolones in tat inhibition can aid in the treatment of bacterial coinfections, although by all means can also may give rise to quinolone-resistant bacteria, which poses a challenge in its clinical use. 

\subsubsection{rev (RNA splicing regulator, p19): exporter of viral RNA.}

This protein is a 18 kDa, 116 amino acid phosphoprotein, and binds to rev-responsive elements, assemblying into multimers \cite{pollard}; rev is phosphorylated by the kinases CSNK2A1 and MAPK, and it has been seen that the residues Ser8 as well as Ser 5 are phosphorylated by CSNK2A1, and CSNK2A1-induced phosphorylation may induce downregulation of the rev protein \cite{meggio}. These multimers, when in solution, display a hollow fiber-like configuration (with around 20 nm in diameter), and when interacting with RNA, the RNA molecule might be engulfed into the hollow volume in order to protect it from nucleic acid metabolism, promoting its cytoplasmic translation \cite{wingfield}. HIV-1 mRNA nuclear export has been seen to be one of the main functions of the rev protein, but it may as well promote the translation of rev-responsive mRNAs (RRE), and is able also to downregulate its own expression as well as that of other viral genes \cite{pollard}; the RRE is found within \textit{env}-coding region, it spans around $\sim 350$ nucleotides in length, and when transcribed, this region binds to rev, where the complex rev-mRNA (ribonucleoproteic in nature) then promotes the recruitment of XPO1 and RAN \cite{fernandes}, where this multimeric ribonucleoproteic complex shall subsequently be exported from the nucleus, XPO1 is an exportin (also termed CRM1 or chromosome region maintenance 1), member of the importin beta superfamily of nuclear transport receptors, required for the export of many RNAs (but is enabled to export proteins by all means), CRM1 interacts with a nuclear export signal (NES) which is leucine-enriched (rev possesses a NES domain \cite{behrens}), but can additionally be a hydrophobic domain; thus, the rev-mRNA-CRM1-RAN-GTP complex found in the nucleus binds to the RAN binding protein 3 (RANBP3) which will then attach to the nucleoporin complex comprised by the NUP214 and NUP88 proteins (found in the nuclear membrane), hatched to the NUP358 protein with its cytoplasmic ending bound to the RAN GTPase activating protein 1; resulting in the export of the rev-mRNA-CRM1-RAN-GDP complex and release onto the cytoplasm \cite{hutten}. Cellular splicing tends to occur within the nucleus \cite{fica,casolari} (however there do exist reports of cytoplasmic intronic unspliced mRNA \cite{talhouarne}, some of these unspliced mRNAs are: APP, ATF4, CACNA1H, CAMK2B, FMR1, HP, IL1B, and others which may promote in increasing the variability of the transcribed message \cite{buckley}), and export of mRNAs tends to be that of spliced mRNAs, nevertheless, rev promotes the export towards the cytoplasm of both spliced and unspliced viral RNA \cite{favaro}. A small molecule PKF050-638 (Ethyl (Z)-3-[5-(2-amino-5-chlorophenyl)-1,2,4-triazol-1-yl]prop-2-enoate) is recorded to disrupt the XPO1-NES interaction, thereof being able to inhibit rev-mediated nuclear export, this compounds binds to the Cys539 residue of XPO1, and exhibits similar effects to those induced by the XPO1 inhibitor, leptomycin B \cite{daelemans}. Three alternate compounds were found to suppress HIV-1 gene expression of tat and rev, these compounds are a pyrimidin-7-amine, a diphenylcarboxamide, and a benzohydrazide \cite{balachandran}.

\subsubsection{vif (viral infectivity protein, p23): evading APOBEC3G response.}

This is a basic $\sim 23 kDa$ phosphoprotein, required by the HIV-1 particle in order to replicate in nonpermissive cells, such as lymphocytes, macrophages and leukemic T-cells; but vif is redundant in permissive cells (T-cells) \cite{kozak}. Vif's phosphorylation occurs within the C-terminus: Ser144, Thr155, and Thr188, which may occur through protein kinase C or cGMP-dependent kinases \cite{goncalves} (serine-threonine protein kinase family of proteins). Interestingly, vif-defective viral particles when infecting a permissive cell, may continue to replicate, and produce infectios vif-defective viral particles which may then infect nonpermissive cells; when these vif-defective viral particles derived from nonpermissive cells infect either a permissive cell or a nonpermissive cell, viral replication ceases, this is due to negative imprinting mechanisms applied onto vif \cite{kozak}. It has been observed that the permisiveness feature of the cell relies on the expression pattern of the APOBEC3G protein which can be found in T-cells, B-cells, macrophages, and myeloid cells; in T-cells, APOBEC3G's expression is promoted by T-cell activation \cite{koning}, this protein is member of a family of cytosine deaminases which play a role in innate immunity by restricting viral replication through inducing deamination and mutation of viral genes, these enzymes display one or two zinc-binding motifs, one of the members of this family is essential for antigen-driven B-cell differentiation (AICDA); the APOBEC3 genes are clustered in  chromosome 22; and the expression of the gene APOBEC3G in monocytes has been seen to be linked to IFNA, IFNB, IFNG, TNFA, and IL4 \cite{covino}. APOBEC3G is able to deaminate ssDNA cytosines to uracils, whose ability is not constrained merely to HIV1, but as well to other retroviruses, and inclusively retrotransposons which effectively reduces the viral capability of replicating in the absence of vif \cite{iwatani}, due to vif being able to induce ubiquitination and degradation ov APOBEC3G through complexing with CUL5, and KITLG (vif-CUL5-KITLG) \cite{xianghui}. How can vif-defective viral particles derived from permissive cells infect nonpermissive cells which indeed contain APOBEC3G? This could be evaded by action of both the viral core and the nucleocapsid proteins which can then esterically protect reverse transcription from the deleterious actions of APOBEC3G \cite{kozak,buckman}. The compound RN-18 which is a small molecule containing three benzene rings, has been reported to diminish vif activity, and increasing APOBEC3G levels \cite{nathans}; another small molecule compound (IMB-301) inhibits vif-mediated APOBEC3G by binding to APOBEC3G and inhibiting the binding of vif to APOBEC3G \cite{mazhang}, another study reported that the small molecule compound, Redoxal, inhibits the DHODH enzyme (dihydroorotate dehydrogenase), thereof diminishing the pool of available pyrimidine molecules, which has been seen to increase APOBEC3G protein stability \cite{pery}; these three molecules, RN-18, IMB-301, and Redoxal, aid then in decreasing HIV-1 replication rate.  

\subsubsection{tev (tat/env/rev protein, p26): fusion protein.}

This is one of the fusion proteins present in the HIV1 proteome, comprised of portions from three genes: env, rev, and tat, displaying a molecular weight of $26 kDa$; this protein is able to transactivate viral transcription (through its tat portion) at an activity of 70\% relative to that of the tat protein, however its rev activity is weak or nonexistant. This protein localizes to the nucleolus, and may be phosphorylated \cite{langer}.

\subsection{T-cell differentiation.}

Based on the one cell multiple fate theory, several T cell subsets can be differentiated from a common T cell precursor (either a naïve T-cell or already primed T-cell) \cite{gagliani}. Each subset of T helper cells is involved in immunity against specific pathogens. It is well known that the cytokine-mediated signal traducers and activators of transcription (STATs) activation is critical for Th-cell-fate determination \cite{zhu}.

For Th1 differentiation, T-cell receptors (TCRs) are stimulated by IL12 through the STAT4 pathway, promoting TBX21 expression, which promotes the transcription of the IFNG gene, and its own via STAT1 activation. TBX21 inhibits GATA3, the principal regulator of Th2 cells. Th1 cells mainly express CXCR3, INFG, IL2, and TNFA \cite{gagliani, zhu, schmitt}.

Th2 cells are essential in the control of helminth infections. Their effector cytokines are IL4, IL5, and IL13, while IL2 and IL4 mediate their differentiation. These cells express CCR4 and STAT6 (mediated by IL4), which induce GATA3 expression, the master transcriptional regulator of Th2 cells differentiation, it directly binds to the \textit{Il4/Il13} gene locus and through this binding, it induces IL5 and IL13 transcription. Retroviral expression of GATA3 is sufficient to induce endogenous GATA3 \cite{zhu}. In addition to IL4, IL2 activates STAT5, critical for Th2 differentiation, inducing IL4RA and maintaining GATA3 expression. Other proposed transcriptional regulators are NLRP3 and IL25 \cite{gagliani, schmitt}.

Th17 cells are principally located in the gastrointestinal tract. Their primary function is to respond to extracellular bacteria or fungi infections. Their differentiation is mediated by TGFB, IL1B, IL6, and IL23. Through IL6-mediated STAT3 activation, TGFB induces IL23R and RORC (the principal regulator of Th17 differentiation), leading to the production of both IL17A, and IL17F. IL21 and IL23 have a similar function as IL6, inducing STAT3 activation. Th17 cells also can produce IL22 and TNFA. IL2 activates STAT5, which inhibits STAT3 and consequently reduces IL17A expression. FOSL2, MINA, FAS, and POU2AF1 promote Th17 differentiation, against TSC22D3, which negatively regulates it. NFIL3 (induced by melatonin) blocks RORC \cite{gagliani, zhu}.

FOXP3(+) Treg cells are essential to regulate auto-reactive T cells. They are characterized by the constitutive expression of the IL2 receptor alpha chain (CD25). These cells can be generated within the thymus (tTreg) or in peripheral lymphoid organs (pTreg). Their master transcriptional regulator is FOXP3. IL2 and TGFB1 are their main differentiation factors, through the IL2-mediated STAT5 activation, which also suppresses Th17-cell production. In the absence of IL2, IL2RA, or IL2RB, Treg cells are downregulated \cite{gagliani, zhu}.

Transcriptional factors usually cross-regulate the expression of factors involved in the development of other lineages. For example, TBX21 suppresses GATA3 through direct protein-protein interaction; GATA3 suppresses the expression of STAT4 and the production of IFNG; RORC and FOXP3 antagonize each other through a protein-protein interaction \cite{zhu}.
Interestingly, all T-cell subsets have demonstrated some degree of plasticity, adapting their phenotype to the other T-cell subsets under certain conditions \cite{gagliani}.

\subsection{Role of TOX in thymocyte differentiation}

Thymocyte selection-associated high-mobility group box (TOX) is a DNA-binding factor that regulates transcription by producing specific changes in DNA structure and allowing the formation of multi-protein complexes \cite{maestre}.

TOX was initially identified as a thymic transcript; it is transiently upregulated during $\beta$-selection and positive selection of developing thymocytes. This upregulation is mediated by TCR-mediated calcineurin signaling. Expression of TOX induces the factor RUNX3, resulting in CD4 downregulation and CD8 single positive cell formation, but TOX stimulus alone is insufficient to substitute the complete TCR signaling during positive selection. Mice deficient in TOX revealed a requirement for TOX in CD4 T cell lineage development. RUNX3 and ZBTB7B are key nuclear factors for CD8 and CD4 T cells differentiation in the thymus, respectively. It has been suggested that TOX is an upstream regulator of ZBTB7B. TOX is required also for the development of NK, LTi and CD4(+) cells, probably by modulating E protein activity thought the upregulation of ID2, an E protein inhibitor \cite{aliahmad}.

\section{Materials and methods.}
\subsection{Computational requirements.}

The stochastic model, further computations, and graphs were performed by means of RStudio: Integrated Development Environment for R \cite{rstudio}. 

\subsection{Data acquisition.}

The data was acquired from two main databases: The BioGrid \cite{biogrid}, and STRING \cite{string}, as well as from various papers: \cite{jäger,postler,cai, yakovian,zhou,ivanusic,wyma,thielens,chen,qi,sanhadji,checkley,henrick,takeshita,stoiber,planelles,zhao,fabryova,pereira,buffalo,pyeon,marrero,arhel,hu,wang2,salamango,wang3,huttenhain,azimi,spector,jean,mahmoudi,gautier,remoli,cujec,li,solbak,sharma2,solbak2,popov2,müller,ott,yu,anton,karnati,rumlova,wagner,impens,sheng,wangw,yuXG,caccuri,bugatti,lu,tang,couturier,fu,campbell,bejarano,summers,schur,carnes,novikova,yuant,achutan,yamashita,larsen,liD,bradyS,warren,wangY,toro,taniguchi,landi,hammond,suhasini,yedavalli,cochrane,kiss,langer,ronsard,liL,arizala,arizala2,benko,hauber,meggio,gonzalezME,guatelli2,guo}.

\subsection{Stochastic model.}

\begin{figure}[hbt!]

\begin{subfigure}{1\textwidth}
\includegraphics[width=1\linewidth,height=7cm]{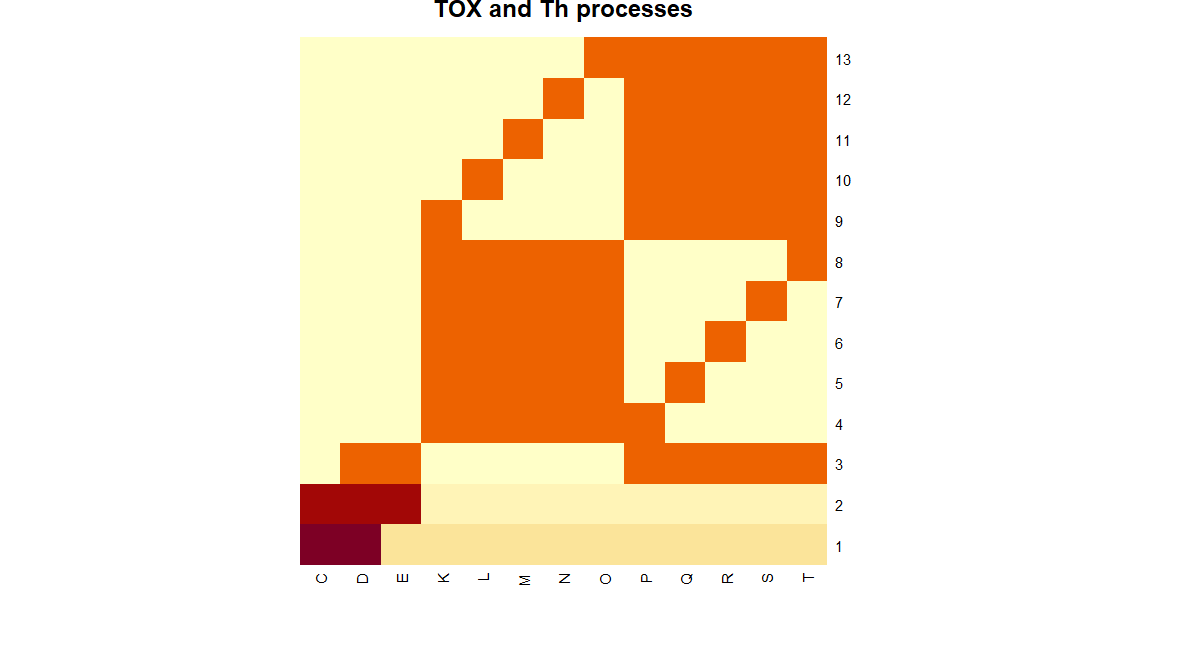}
\caption{The subsets belonging to the space of states, and their interactions are shown by means of their adjacency matrix, colour key: shell/beige: nonexistence of interaction between networks, brown/dark orange: interaction between networks.}
\label{fig:fig1up}
\end{subfigure}

\begin{subfigure}{1\textwidth}
\includegraphics[width=1\linewidth,height=7cm]{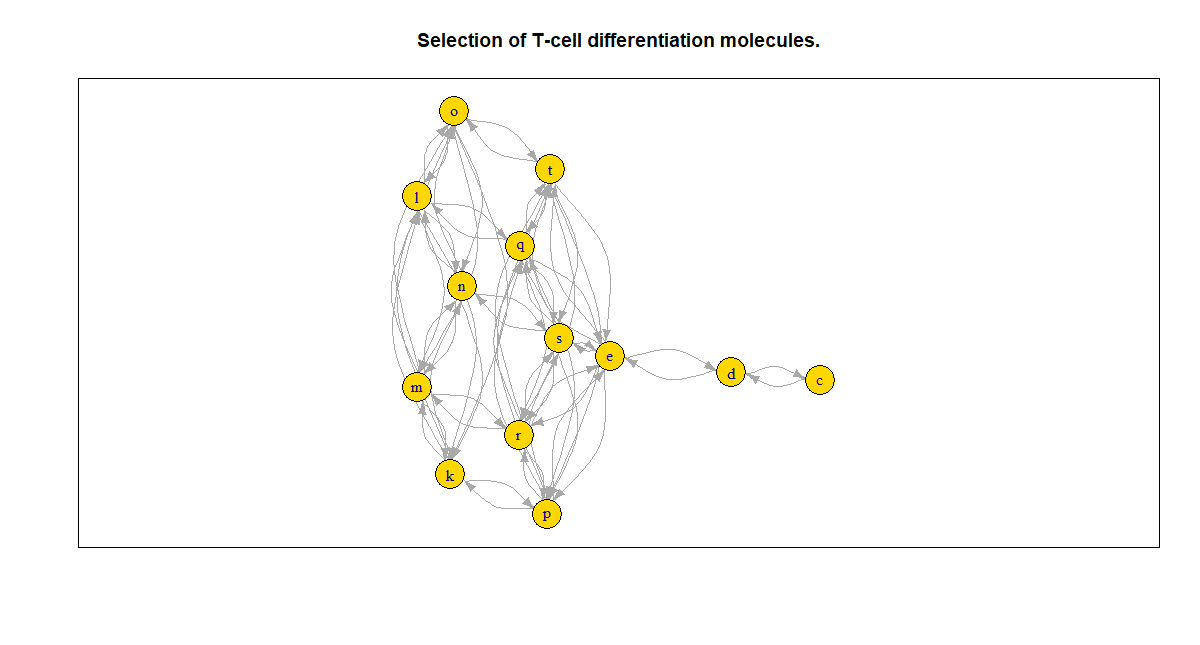}
\caption{Graph displaying the adjacency matrix interactions between the TOX and Th-related networks. C=TOX root (TOX), D=TOX secondary network, E=TOX tertiary network, K=Tfh root network, L=Th1 root network, M=Th2 root network, E=TOX tertiary network, O=Treg root network, P=Tfh secondary network, Q=Th1 secondary network, R=Th2 secondary network, S=Th17 secondary network, T=Treg secondary network.}
\label{fig:fig1low}
\end{subfigure}

\caption{Adjacency matrix and corresponding graph depicting the interactions between the TOX and the Th networks.}
\label{fig:fig1}
\end{figure}

The interactions pertaining to the T-cell differentiation molecular network were assessed, 562 proteins were included, a Markov chain was then associated to these interactions, the stochastic process $\{X_t\}$ displays a space of states $S$, such that $S=\{p_1 , p_2 ,..., p_n\}$ where $n=562$ and $p_i$ stands for the protein or gene that the system encounters itself in; the space of the temporal parameter is $T=\{1,2,...m\}$, the units of this temporal parameter shall be defined subsequently. The space of states of the stochastic process $\{X_t\}$ may be divided, in order to analyse the selection of the molecules taking part of it, into two subsets, the TOX-related process subset $A\subset S$, and the Th-cell-related process subset $B \subset S$, such that, $A+B-(A\cap B)=S$ due to the fact that some molecules in each subset are repeated, $A\cap B \neq \emptyset$. The TOX-related process is comprised of three major networks, $A=\{C,D,E\}$, where the C network stands for one of the roots of the process: the TOX protein; the D network stands for those proteins which interact with TOX and amongst themselves (KRTAP12-1, KRTAP26-1, DMRTB1, ZDHHC17, TOX2, BANP, GRAP, FUS); the E network stands for those proteins which interact with TOX-related proteins and amongst themselves. The Th-cell-related process subset is in its turn comprised of ten networks, $B=\{K,L,M,N,O,P,Q,R,S,T\}$, the K network stands for those proteins involved in the Tfh-differentiation process, the L network stands for the proteins involved in the Th1-differentiation process, the M network stands for those proteins involved in the Th2-differentiation process, the N network stands for those proteins involved in the Th17-differentiation process, the O network is composed of those proteins involved in the Treg-differentiation process; we may regards that the set $U\subset B$, $U=\{K,L,M,N,O\}$ is a set of roots of the process (TBX21, GATA3, STAT4, STAT5, STAT6, RUNX3, IRF1, HLX, EOMES, EST1, FI1, ASCL2, TCF7, BCL6, IRF4, CMAF, JUNB, DEC2, IKAROS, RORC, FOXP3, BATF1, RORA, RUNX1, HELIOS, FOXO). The P network stands for those proteins interacting with the Tfh-related ones, the Q network is comprised of proteins interacting with the Th1-related ones, the R network stands for those proteins interacting with the Th2-related ones, the S network stands for those proteins interacting with the Th17-related ones, and finally the T network is comprised of proteins which interact with the Treg-related ones; the set $v=\{P,Q,R,S,T\}$ is thus a secondary network. The adjacency matrix and corresponding graph summarising this data is shown in \textbf{\ref{fig:fig1}}. The states comprising the processes forecommented are not mutually exclusive, i.e., one protein may be involved in various networks. 

We may describe the probabilities of going from one state to another in a Markov chain by means of a transition probability matrix $A=(a_{ij})$ where $a_{ij}$ is the probability of transiting from the i-th state to the j-th state, the existence of a probability between two molecules in this model is given by the following condition, 

\begin{equation} \label{eq:1}
\exists R \mid p_i R p_j \rightarrow \exists a_{ij}
\end{equation}

Should the relationship R between two elements of the space of states ($p_i$ and $p_j$) exist, the probability of transiting from the i-th state to the j-th state exists ($a_{ij}$), the probability is computed by the amount of states the process can transite to from a given initial state; say we start our process at the $p_1$ molecule, this molecule interacts with $q$ molecules, these molecules in the vector $Q=(p_2,...p_q)$, $Q\subset S$, then the probability of transiting from $p_1$ to $p_q$ shall be, 

\begin{equation} \label{eq:2}
(a_{1q})=\frac{1}{q}
\end{equation}

in this case, we have considered that all molecules involved interact with themselves, i.e., $a_{ii}>0$. And the condition for the stochastic matrix is thus fulfilled, 

\begin{equation} \label{eq:3}
\sum_{i=1}^k (a_{ik})=1
\end{equation}

\begin{figure}[hbt!]
    \centering
    \includegraphics[width=1\textwidth]{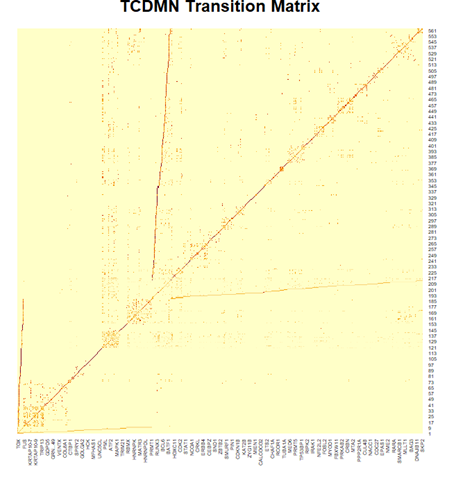}
    \caption{T-cell differentiation molecular network (TCDMN) transition matrix, represented as a heatplot, the minimum value of this matrix is $a_{ij}\in P, min\{a_{ij}\}=0$, that is where there exists no interaction between two molecules, and the maximum possible value is $a_{ij}\in P, max\{a_{ij}\}=\frac{1}{2}$, due to the fact that every molecule interacts at least with itself (as we can see that all the elements of the main diagonal of the matrix, $\forall a_{ii}\in P, a_{ii}>0$ and the "root" molecule.}
    \label{fig:heatplo1}
\end{figure}

The row is normalised. The transition matrix corresponding to the molecules interacting in the T-cell differentiation process is shown in \ref{fig:heatplo1}. \\

The eigenvalues were also computed in order to obtain a matrix $\Lambda$, where $\lambda \in \Lambda$ is an eigenvalue of P; since $\lambda_1 =1 \in \Lambda$, we may intuit that the equilibrium distribution $\pi^n$ exists, such that, 

\begin{equation}
\pi^n = \pi P^n , n\rightarrow \infty
\end{equation}

The equilibrium distribution has been computed taking in consideration the following linear system, 

\begin{equation}
\begin{split}
\pi(1)(p_{11}-1)+p_{21} \pi(2)+...+p_{n1}\pi (n)=0 \\
p_{12} \pi(1)+\pi(2)(p_{22}-1)+...+p_{n2}\pi(n)=0 \\
\vdots \\
p_{1n} \pi(1)+p_{2n}\pi(2)+...+\pi(n)(p_{nn}-1)=0 \\
\sum_{i=1}^n \pi_i (n)=1
\end{split}
\end{equation}

Where $\pi(i)$ stands for the i-th state, $p_{ii}$ stands for the probability of transiting from the i-th state to the i-th state when the transition matrix is $P^1$. The eigenvalues' matrix $\Lambda$ is shown in section \ref{results}, as is as well displayed the equilibrium distribution.  

\begin{figure}[hbt!]
    \centering
    \includegraphics[scale=0.8]{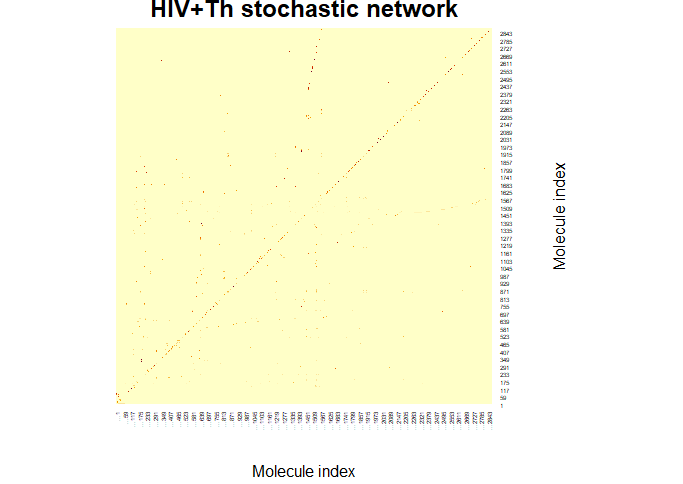}
    \caption{Transition matrix corresponding to the interaction of one viral particle with a T-cell, 2874 molecules are included, displaying thus 8'259,876 interactions in total.}
    \label{fig:heatplot2}
\end{figure}

In regards to the T-cell differentiation network when the HIV particle enters the cell, the network was extended to 2874 molecules (including those that were previously in the T-cell differentiation network). The most connected node in this network was the one of the TP53 molecule with an adjacency $k_{TP53}$ of $k_{TP53}=169$, the transition matrix is shown in figure \ref{fig:heatplot2}.

\section{Results.} \label{results}

\subsection{Equilibrium distribution.}

\begin{figure}[hbt!]
\centering

\begin{subfigure}{0.45\textwidth}
\centering
\includegraphics[width=0.9\linewidth,height=3.5cm]{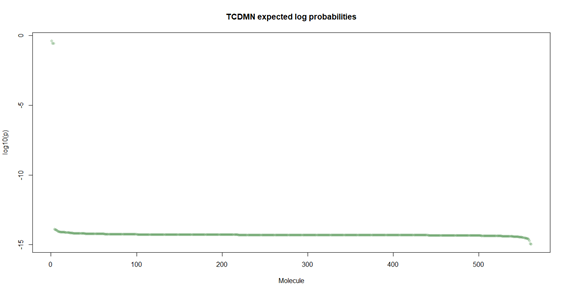}
\caption{{\footnotesize Equilibrium distribution of the T-cell differentiation molecular network process, the x-axis represents the index of the molecule, where there exists a bijection between the space of states $S$ and the set of the index; the y-axis represents the base-10 logarithm of the corrected probability.}}
\label{fig:logprob}
\end{subfigure}
\hfill
\begin{subfigure}{0.45\textwidth}
\centering
\includegraphics[width=0.9\linewidth,height=3.5cm]{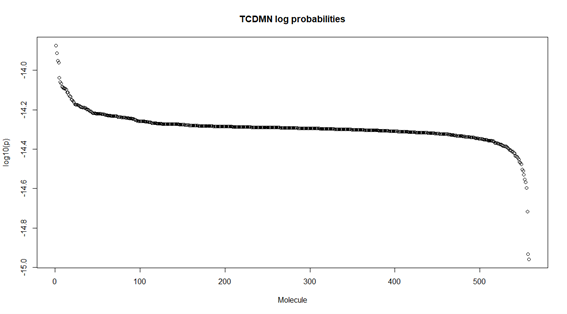}
\caption{{\footnotesize Equilibrium distribution of the T-cell differentiation molecular network process, in this case, we have excluded the extreme cases in which the probability was accumulated, here we display those molecules in which $p=2.588919e^{-12}$ of the probability is condensed.}}
\label{fig:logprob2}
\end{subfigure}
\vspace{0.3cm}

\begin{subfigure}{0.45\textwidth}
\centering
\includegraphics[width=0.9\linewidth,height=3.5cm]{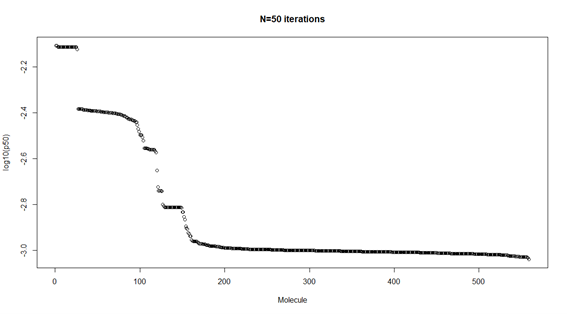}
\caption{{\footnotesize Probability vector when 50 iterations have been performed, $\pi^{50}=\pi P^{50}$.}}
\label{fig:50iter}
\end{subfigure}
\hfill
\begin{subfigure}{0.45\textwidth}
\centering
\includegraphics[width=0.9\linewidth,height=3.5cm]{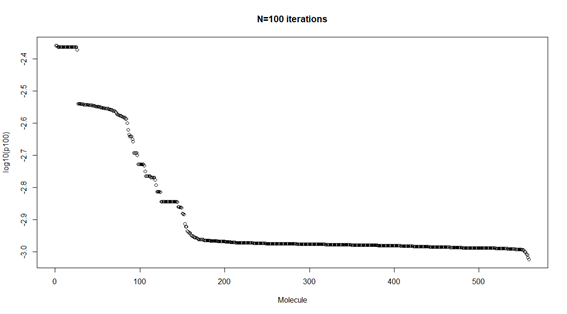}
\caption{{\footnotesize Probability vector when 100 iterations have been performed, $\pi^{100}=\pi P^{100}$.}}
\label{fig:100iter}
\end{subfigure}
\vspace{0.3cm}

\begin{subfigure}{0.45\textwidth}
\centering
\includegraphics[width=0.9\linewidth,height=3.5cm]{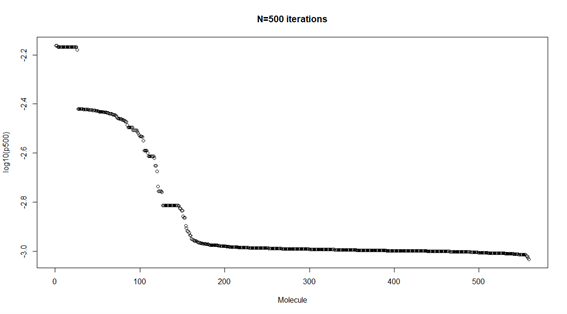}
\caption{{\footnotesize Probability vector when 500 iterations have been performed, $\pi^{500}=\pi P^{500}$.}}
\label{fig:500iter}
\end{subfigure}
\hfill
\begin{subfigure}{0.45\textwidth}
\centering
\includegraphics[width=0.9\linewidth,height=3.5cm]{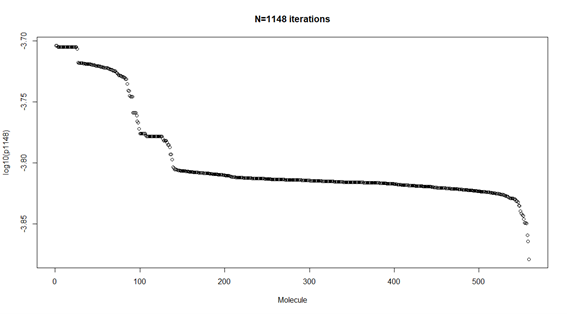}
\caption{{\footnotesize Probability vector when 1148 iterations have been performed, $\pi^{1148}=\pi P^{1148}$, this number has not been randomly chosen, it is $f(\lfloor T \rfloor)$, that is the floor function of the rate towards convergence, $T$.}}
\label{fig:1148iter}
\end{subfigure}

\caption{Rank-logarithmic graphs of the equilibrium distribution pertaining to the T-cell differentiation molecular network process.}
\label{fig:log}
\end{figure}

\begin{figure}[hbt!]
\centering
\includegraphics[width=0.9\linewidth,height=6cm]{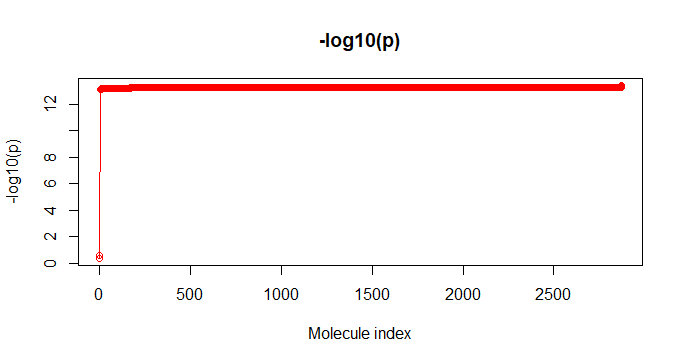}
\caption{Equilibrium distribution of the HIV molecular network process, the x-axis represents the index of the molecule, where there exists a bijection between the space of states $S$ and the set of the index; the y-axis represents the base-10 logarithm of the corrected probability.}
\label{fig:loghiv}
\end{figure}

The eigenvalues for the T-cell differentiation process matrix were obtained, further assembled into a set $\Lambda$, this set contained two eigenvalues of special interest, the presence of $\lambda_1=1$ which, as has been commented, permits us to intuit the existence of an equilibrium distribution $\pi^n$ \cite{rincon}, which has been computed; the computation of this equilibrium distribution has been performed by means of a numerical method yielding results $\hat{f}(\pi)$ where $\hat{f}$ is the approximation method which differ from the theoretical ones by a factor $\epsilon_a=|f(\pi)-\hat{f}(\pi)|$ where $\epsilon_a$ is the absolute error from the process; considering that the events are subject to a $\sigma$-algebra $\mathcal{F}$ and that there exists a measurement $P$ of the probability over $\mathcal{F}$, we may consider that $\sum \pi =1$. In this case, $\epsilon_a =5.32039697\times 10^{-15} $ and we are able to correct our probabilities, therefore we performed a mapping $g:\hat{f}(\pi)\rightarrow f(\pi)$. These corrected probabilities were then the argument of the base-10 logarithm (in order to more vividly represent the computed probabilities, where some probabilities were $p<<0$) that is, $h(\pi)=\log_{10}(p)$, this is shown in \ref{fig:logprob}; from this we can regard that $2.588919\times 10^{-12}$ of the probability is contained in 555 molecules, while only $3$ molecules span $p=1-2.588919\times 10^{-12}$ of the probability, should we exclude for a moment these 3 molecules, the logarithmic probability would be seen as is displayed in \ref{fig:logprob2}. The corresponding eigenvalues for the HIV+Th-cell differentiation network were computed, and poured into a matrix $\Lambda_1$, where we come to find $\lambda_i \in \Lambda_1 : \lambda_i=1$, therefore should we start intuiting about the existence of an equilibrium distribution for our compound network; the second largest eigenvalue in the network is $\lambda_j\neq \lambda_i: \mod(\lambda_j)=0.999654150505886$; our intuition serves well for the HIV process indeed possesses an equilibrium distribution $\pi_{HIV}^n$, this is shown in figure \ref{fig:loghiv}.\\

\subsection{The logarithmic equilibrium distribution follows a discrete beta generalized distribution.}

Graph \ref{fig:logprob2} may be considered a rank-ordered relation, where the rank is the molecule and the probabilities are logarithmic frequencies, Zipf showed that when plotted, the logarithmic frequencies fall on a slope $m=-1$, which may suggest a power law, we may try and fit the data by means of the following function, 

\begin{equation}
\begin{split}
x(r)=C\frac{(N+1-r)^b}{r^a} \\
\equiv log_{10}(p)
\end{split}
\end{equation}

Where $r$ is the rank value, $N$ is the maximum value, $C$ is a normalization constant and $a,b \in \mathbb{R}$ are two constants to be determined \cite{cocho}, furthermore, it has been observed that many types of data follow this distribution: ranks of articles and citations in journals, linguistic data, citation profile, English, and Spanish letter frequency distribution, and more; this distribution, depending on the value of the constants $a,b$, may give rise to other common probability distributions, that is, when $a=b=0$, it yields a constant random variable, when $a=0,b=1$ we may obtain a uniform distribution, when $b=0$ it becomes a Pareto distribution, when $a=b$ it yields a Lavalette distribution \cite{fontanelli}. Here we will determine the constants $a,b$ for the equilibrium distributions of both the T-cell differentiation and the HIV infection processes, as well as the $R^2$ fitting values for said functions. This we may do by means of firstly obtaining the logarithm of the function $X(r)$ (in order to ease the computations), 

\begin{equation}
\ln(x(r))=\ln(C)+b \ln (N+1-r)-a\ln(r)
\end{equation}

Then, the first moment of the logarithmic function is ($E[\ln(x(r)] \equiv \break E[\ln(\log_{10}(p))]$), 

\begin{equation*}
E[g(r)] = \int^{\infty}_{-\infty} g(r) dF_X(x)
\end{equation*}
\begin{equation}
 \rightarrow E[\ln(x(r))]  = \int^1_0 \ln(x(r))\cdot \frac{dr}{dx} dx 
\end{equation}
\begin{equation*}
=\int^1_0 \ln(x(r))dr    
\end{equation*}

We can substitute the function $\ln(x(r))$, 

\begin{equation*}
E[\ln(x(r))]=\int^1_0 [\ln(C)+b\ln(N+1-r)-a\ln(r)]dr
\end{equation*}
\begin{equation}
=[r\ln (C)+b(N+1-r)(1-\ln(N+1-r))-ar(\ln(r)-r)]^1_0
\end{equation}
\begin{equation*}
=\ln(C)+a+b[N(1-\ln(N))-(N+1)(1-\ln(N+1))]
\end{equation*}

When $N=0$, 

\begin{equation}
E[\ln(x(r))]=\ln(C)+a-b
\end{equation}

Which goes in line with \cite{fontanelli}; now, when we plug in the values from the simulation ($N=2.70546008=\max\{\ln(\log_{10}(p))\}$), we obtain, 

\begin{equation}
    E[\ln(x(r))]=\ln(C)+a+1.60703188b
\end{equation}

Let us compute the variance now, 

\begin{equation}
Var [\ln(x(r))]=E\big[[\ln(x(r))]^2] \big]-E^2 [\ln(x(r))]
\end{equation}

We have already computed the expression for the first moment $E[\ln(x(r))]$, we are missing now the computation of $E[[\ln(x(r))]^2]$, 

\begin{equation*}
\ln^2(x(r))=(\ln(C)+b \ln(N+1-r)-a\ln(r))^2
\end{equation*}
\begin{equation}
=\ln^2(C)+2\ln(C)(\ln(N+1-r)-a\ln(r))+(b\ln(N+1-r)-a\ln(r))^2
\end{equation}

The integral that we are looking for then is, 

\begin{multline}
I=\ln^2 (C) \int^1_0 1\cdot dr+2\ln(C)\int^1_0 \ln(N+1-r)dr-2a\ln(C) \int^1_0 \ln(r) dr+\\
b^2 \int^1_0 \ln^2(N+1-r)dr-2ab\int^1_0 \ln(N+1-r)\ln(r)dr+a^2\int^1_0 \ln^2(r) dr
\end{multline}

Which equals, 

\begin{multline}
I=\ln^2(C)+2\ln(C)[(N+1)\ln(N+1)-N(\ln(N)-1)]+2a\ln(C)+ \\ b^2[(N+1)\ln(N+1)(\ln(N+1)-2)-N\ln(N)(\ln(N)-2)+2]- \\
2ab\bigg[(N+1)\bigg(\ln(-N)-\ln(-N-1)-Li_2\bigg(\frac{1}{N+1} \bigg)\bigg)-\ln(N)+2 \bigg]+2a^2
\end{multline}

Where $Li_2(.)$ is the polylogarithm function. The variance equals (we make $\ln(C)=\phi$), 

\begin{multline}
   Var(\ln(x(r)))= \phi^2 +2\phi[\ln(N+1)(N+1)-N(l\ln(N)-1)]+2a\phi +b^2[(N+1) \\
   \ln(N+1)(\ln(N+1)-2)-N\ln(N)(\ln(N)-2)+2]-2ab\bigg[N \bigg( \ln(-N) \\
   -\ln(-N-1)-Li_2\bigg(\frac{1}{N+1}\bigg)-\ln(N)+2\bigg)+1\bigg]+2a^2-(c+a+ \\
   b[N(1-\ln(N))-N(1-\ln(N+1))-1])^2
\end{multline}

In the case of $N=2.70546008$, 

\begin{multline}
    Var[\ln(\log_{10}(x(r)))]=a^2-9.475027627b^2-0.025582926ab+7.385337236b\phi \\
    -2a\phi +9.732461206\phi
\end{multline} 

The median is where $r=0.5$, 

\begin{equation}
    Med(\ln(x(r)))=\phi+b\ln(N+1/2)-a\ln(1/2)
\end{equation}
\begin{equation}
    = \phi+1.164855631b+0.6931471806a
\end{equation}

the values of the first moment, variance, and median are, 

\begin{equation*}
E[\ln(\log_{10}(p))]=\overline{Z}=2.64148234\times 10^0
\end{equation*}
\begin{equation}
Var[\ln(\log_{10}(p))]=Var(Z)=6.1653831\times 10^{-2}
\end{equation}
\begin{equation*}
    Med(\ln(x(r)))=Med(Z)=2.659718516\times 10^0
\end{equation*}

Therefore we possess a system of three variables and three equations, 

\begin{equation*}
    \overline{Z}=\phi+a+1.60703188b = 2.64148234
\end{equation*}
\begin{multline}
     Var(Z)=a^2-9.475027627b^2-0.025582926ab+7.385337236b\phi \\
    -2a\phi +9.732461206\phi
\end{multline}
\begin{equation*}
    Med(Z)=\phi+1.164855631b+0.06931471806a=2.659718516
\end{equation*}

We have consequently an underdetermined linear system, 

\begin{equation}
   \Xi = \begin{pmatrix}
        1 & 1 & 1.60703188 & 2.64148234 \\
        1 & 0.6931471806 & 1.164855631 & 2.659718516 
    \end{pmatrix}
\end{equation}

Which yields the partial solution, 

\begin{equation*}
    a= 3.22596 - 1.2164 \phi
\end{equation*}
\begin{equation}
   b= 0.134657 \phi - 0.363698
\end{equation}

By plugging these solutions into the quadratic equation, we obtain the expression for $\phi$,

\begin{equation}
  2.306500645\phi^2+3.323241286\phi-23.48052236=0
\end{equation}

the corresponding solutions are then, $\Phi = (\phi_1, \phi_2)= (2.550545919,-3.991361162)$, and the constants for each $\phi_i\in \Phi $, 

$$
\phi_1, \begin{pmatrix} \hat{a}_1 \\ \hat{b}_1 \end{pmatrix} = \begin{pmatrix} 8.0810517 \\ -0.9011627 \end{pmatrix}
$$
\begin{equation}
    \phi_2 , \begin{pmatrix} \hat{a}_2 \\ \hat{b}_2 \end{pmatrix} = \begin{pmatrix} 3.2034876 \\ -0.3612103 \end{pmatrix}
\end{equation}

\begin{figure}
    \centering
    \includegraphics[width=0.9\textwidth]{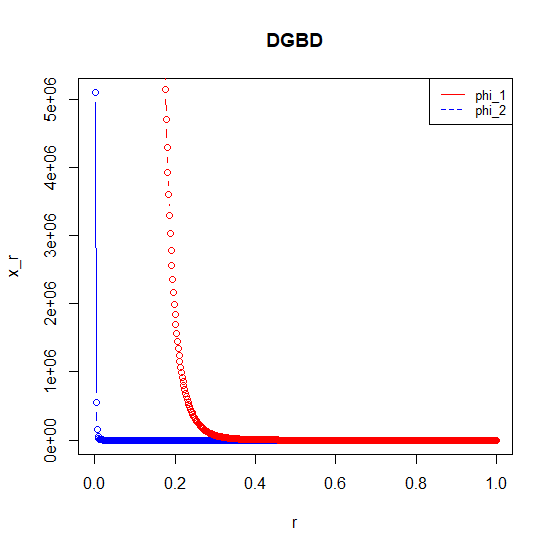}
    \caption{Discrete generalized beta distributions modeling the logarithmic equilibrium distribution corresponding to the Th-cell differentiation process, where $\hat{a}_1=8.0810517,\hat{b}_1=-0.9011627$, $\hat{a}_2=3.2034876,\hat{b}_2=-0.3612103$. }
    \label{fig:dgbd}
\end{figure}

Whose graphs are shown in figure \ref{fig:dgbd}. 

\subsection{The stochastic processes can be coupled with both \textit{in vivo} processes of T-cell differentiation, and HIV-cellular infection.}

The second eigenvalue of interest is one which allows us to compute the rate of convergence, should it exist, and in this case, we have already commented on its existence; this rate is dependent on the second maximum eigenvalue $\mu$ \cite{stanford}, 

\begin{equation}
    \begin{split}
\mu=max\{\lambda_i \in \Lambda\ :\mu \neq \lambda_1=1 \} \\
\mu=0.99799714        
    \end{split}
\end{equation}

We thus can compute the rate to convergence $T$, that is, the necessary steps for the process to achieve equilibrium, 

\begin{equation}
    \begin{split}
T_1=\frac{1}{log(\mu^{-1})}\\
=1148.49753 u        
    \end{split}
\end{equation}

The process, in this case a T-cell, may reach equilibrium after $1148.49753$ arbitrary temporal units of time, these units may be then converted to an international system unit, taking in account the \textit{in vivo} or \textit{in vitro} data collected in the past, this achievement of equilibrium may be regarded as the complete lifespan of a T-cell, and has been reported to be as high as 17 years \cite{imamichi} and as low as 6 days \cite{borghans, gossel}, thus this unit may range in the following interval,

\begin{equation}
1 u \in [4.51372324\times 10 ^4, 4.66985088\times 10^5]s
\end{equation}

In regards to the HIV infection process, we may regard that the coupled set of viral particle and T-cell reach equilibrium when there begins to exist a decline in the Th-cell count, moment in which we can start to perceive that the genetic/proteic machinery of the virus has effectively hijacked that of the cell towards another more deleterious, that of more massive apoptosis, this is more generally seen after five  years of infection \cite{feasey}, then the time to reach equilibrium $T_2$, will depend on the second maximum eigenvalue $\nu$, 

\begin{equation}
    \begin{split}
T_2=\frac{1}{log(\nu^{-1})}\\
=6656.62441 u
    \end{split}
\end{equation}

Now, $6656.62441 u \equiv 5y\rightarrow 1331.32488 u/y$. 

\subsection{Evolution of distributions until equilibrium.}

\begin{figure}[hbt!]
    \centering
    \includegraphics[width=1\textwidth]{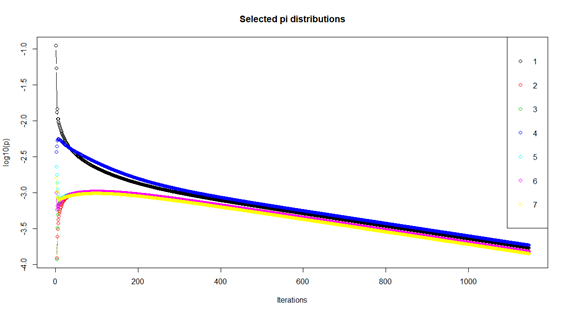}
    \caption{Selected probability distributions of seven molecules pertaining to the Th-cell differentiation process, where the probability for every time step in the interval $[1,T]$ is shown, in this plot, 1=TOX, 2=EOMES, 3=PFDN1, 4=NOTCH2NL, 5=TP53, 6=HDAC1, 7=EP300.}
    \label{fig:some}
\end{figure}

The probability distributions $\pi_i$ for $i\in [1,T]$ where $T$ is the time when the process has converged towards equilibrium $\pi^n $ have been computed and are shown in \ref{fig:some}.

\subsection{The Th-cell differentiation and the HIV-infection processes tend towards an increasing entropy, $S_1 \leq S_2 \leq ... \leq S_n$ when reaching its equilibrium distribution. } \label{subsection:entr}

Entropy in a thermodynamic setting may relate to the concepts of reversibility or order in a system, more specifically, should we have a protein with a thermodynamic state $\pi_0(x) =exp\big[-\frac{\beta \mathcal{H}_0(x)}{Z_0} \big]$ where $\mathcal{H}_0(x)$ is the Hamiltonian ($\mathcal{H}=T+V$, T is the kinetic energy and V is the potential energy), $\beta=\frac{1}{kT}$ and $Z_0$ is the partition function, we can thus compute the entropy $S(\pi)$ of the system by means of, 

\begin{equation} \label{eq:19}
H(\pi)=-k\sum_{i=1}^N \pi(x) \ln \pi(x)
\end{equation}

\cite{parrondo} this could be therefore considered as the summation over all the particles of each of their stochastic entropies $s_t$,

\begin{equation} \label{eq:20}
s_t=-\ln \pi(x_t , t)
\end{equation}

The probability is to be computed by means of solving the Fokker-Planck equation \cite{seifert}, which should we recall, describes the evolution of the probabilities $\pi$ from an ensemble of particles when this particles start their trajectories in $t=0$, we would call this distribution the initial vector $\pi^0$ \cite{baron} that we have previously commented on, thus we may regard the solution to this Fokker-Planck equation in our case as a discrete one (which comes as no surprise, due to the fact that the continuous probability considered in the Fokker-Planck equation is a continuous Markovian one). Should we divide the expression \ref{eq:19} by $1/k$, we would obtain, 

\begin{equation}
S(\pi)=-\sum_{i=1}^N \pi(x) \ln \pi(x)
\end{equation}

\begin{figure}[hbt!]
    \centering
    \includegraphics[width=1\textwidth]{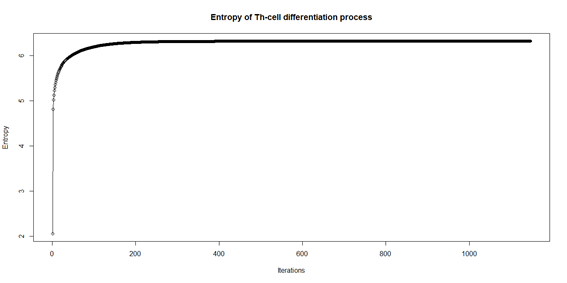}
    \caption{Entropy computations for each time step in the Th-cell differentiation process, until this process reaches its equilibrium distribution in $t=1148 u$, we may regard the increasing values of entropy with time, $S_1 < S_2 <...<S_n$.}
    \label{fig:entropyth}
\end{figure}

\begin{figure}
    \centering
    \includegraphics[width=1\textwidth]{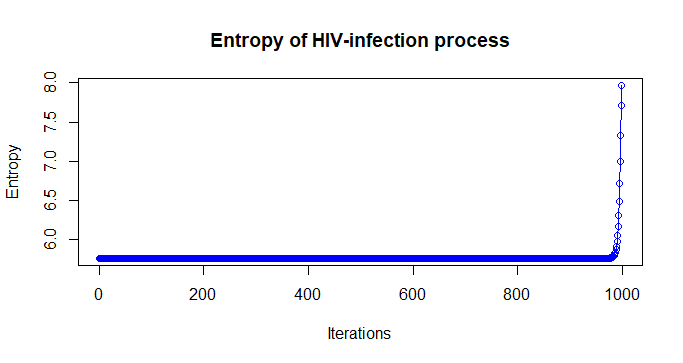}
    \caption{Shannon's entropy for the HIV-infection process, in which it is as clear as for the previous process that the values of entropy $S_i$ display increases, $S_1\leq S_2\leq ...\leq S_n$ for $n\rightarrow \infty$.}
    \label{fig:entrophiv}
\end{figure}

This is the Shannon entropy of the system \cite{parrondo}, which will help us in the understanding of the changes of the microscopical states $\pi$ in our system, these states are then the thermodynamic states pertaining to each particle in the regarded system. We have thereof computed $S(\pi_i)$ for $i\in [1,N]$ where $N=1148$ in this case, due to this being the time of convergence for the Th-differentiation process, which may be seen in \ref{fig:entropyth}, the entropy $S(\pi_{HIV})$ was also computed for the HIV-infection process, and is seen in \ref{fig:entrophiv}

We may regard that there exists a feature for each $S_i$ computed, 

\begin{equation} \label{eq:24}
    S_1 < S_2 < ... <S_n
\end{equation}

Therefore, the process tends towards an increasing entropy value $S_n$, which according to Boltzmann, systems tend towards increases in entropy when they reach equilibrium due to this being the state in which there exist the most possible configurations \cite{attard}. However we may notice two aspects in this entropy computation: 

\begin{enumerate}
    \item $\Delta S_1 > \Delta S_2 > ... > \Delta S_{n-1} \rightarrow (\Delta S_{n}\rightarrow 0) $, therefore the system tends towards steadiness in its equilibrium distribution. \\
    
Interestingly, considering the series in \ref{eq:24}, if we regard the entropy $S_t$ as a measurement of the information or the number of states in the system at any time $t\in T$, therefore, the set of states $Z_s^t$ in the system accounted for by the entropy are defined by,

\begin{equation}
t=1,Z^1_s =\{Z_1^1, Z_2^1, ..., Z_m^1 \}
\end{equation}

Then $Z_s$ is a measurement of events $\pi \in \Omega$ (some molecules may have a repeated value of probability) over $P=[0,1]$, so $Z_s$ is a $\sigma$-algebra; now, let us establish a relationship between the cardinalities of the sets of states, 

\begin{equation}
    |Z_s^1| < |Z_s^2|<...< |Z_s^n|
\end{equation}

And, 

\begin{equation}
    \begin{split}
        \forall Z^1_m \in Z^1_s , Z^1_m \in Z^2_s \\
        \forall Z^2_{m+i} \in Z^2_s , Z^2_{m+i} \notin Z_s^1 \\ 
        \rightarrow Z^1_s \subseteq Z^2_s \wedge Z^2_s \supset Z^1_s
    \end{split}
\end{equation}

In our simulation we can state thus, 

\begin{equation}
    Z_s^1 \subseteq Z_s^2 \subseteq ... \subseteq Z_s^n
\end{equation}

And we may say that there exists a collection of $\sigma$-algebras for every step $t=1,2,...,N$ in the process, this collection is then a filtration $\{Z_s^n\}_{n \geq 1}$ such that the following is fulfilled, 

\begin{equation}
    Z^n_s \subseteq Z^{n+1}_s
\end{equation}

The process $\{X_n\}$ can be $Z$-measurable, and we even may couple the entropy with the filtration $\{Z_s^n\} _{n\geq 1}$ and, the acquisition of information throughout the Th-cell differentiation process may be considered a martingale. 
    
    \item The Euclidean distance $\delta_1 \in \mathbb{R}$ between the initial entropy $S_1$ and the final entropy $S_f$ is $\delta(S)=S_f -S_1 > 0$, therefore this model describes a possible process, and moreover an irreversible one: When reaching equilibrium the change in entropy, $\delta (S)>0$ tells us the differentiated status is a committed one in the cell. It has been previously commented that natural systems tend to display an increase in entropy in time \cite{wilson, buchel}, therefore this simulation is a phenomenologically accurate one.
    
\end{enumerate}

\section{Discussion.}

HIV-1 transmission results from viral exposure at mucosal surfaces or from percutaneous inoculation. Since direct analysis of these mechanisms is inaccessible, understanding HIV-1 infection has been done by indirect methods \cite{shaw}.

The initial period between the moment when the first cell is infected and the moment when the virus can be detected in blood, is called the eclipse phase (estimated at 7-21 days). During this phase, clinically silent, HIV-1 is propagated. Once virus becomes detectable in blood plasma,  it increases exponentially \cite{shaw, ribeiro2}.

Most dynamic models of HIV infection do not usually take into account the earliest stage of infection, where stochastic models play a key role \cite{pearson}.

\subsection{Th-cell fully differentiated state as an equilibrium distribution, $\pi^n$.}

\begin{figure}
    \centering
    \includegraphics[width=0.9\textwidth]{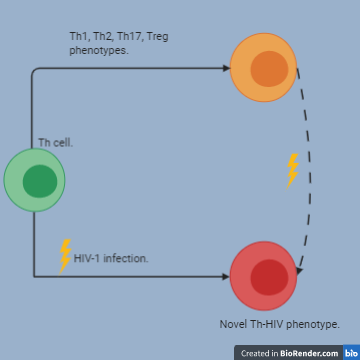}
    \caption{Commitment and differentiation of T-helper cells towards their known phenotypes Th1, Th2, Th17, and Treg, where we may find that the stochastic molecular network which is involved in the regulation of the differentiation process displays an equilibrium distribution, then can we intuit these states are stable ones, where the cell should necessitate an amount of energy in order to leave such state, this perturbation in the energy distribution might be induced by chronic infection of HIV-1, and then promote the differentiation of a T-cell towards a particular Th-HIV cell, where the genetic machinery has been hijacked by the virus, and can also be regarded as a stable cellular state. Created with BioRender.com}
    \label{fig:thhiv}
\end{figure}

One of the motivating questions for this paper was the question: can the HIV change the differentiation status of a T-cell? In order to answer such conundrum one would ask oneself how would the modeling for the differentiation process in the T-cell ever occur, we have addressed this problem in a two-fold fashion, one in a biological manner, i.e., a differentiated state in a cell may be understood as a committed state in the cell, one in which the cell cannot return towards a totipotent or pluripotent cell and has a narrower, more defined set of functions \cite{mitalipov}; and another in a more physical manner, i.e., this differentiated state as a state which is in equilibrium, or that it would require energy from its environment in order to \textit{de}differentiate and return to a less differentiated state (as it may occur in certain processes, such as neoplasms \cite{mercardouribe}), this energy $E$ is not available readily for such endeavours for any cell, requiring a special effort for it to happen, this equilibrium state can be regarded as the equilibrium reached for a Markov process, a state which is independent of any initial state of the system for it will be reached notwithstanding the initial conditions; a Markov process can model as well the behaviour of the differentiating cell due to the stochastic nature of interactions between signaling, structural, and energy-related molecules. Through this simulation we have observed that in both stochastic processes (Th-cell differentiation, and HIV-infection) there exist equilibrium distributions, providing us with \textit{in silico} evidence that these processes indeed lead to differentiated states of the cell, the Th-cell differentiation may lead to the various phenotypes (Th1, Th2, Th17, Treg, etc), and the HIV-infection-dependent differentiation may lead to a novel differentiated state, $Th_{HIV}$ \ref{fig:thhiv}, this state may be featured by an altered immunological synapse, and promotion of induction of the virological synapse, expression of antiapoptotic factors in early phases of infection, but lately would it be overrun with promotion of proapoptotic factors leading to the decline of T-cell count in a chronic phase of the infection. This has been seen to occur in an \textit{in vivo} setting, where CD45RA(+) (PTPRC(+)) T cells decline over time, with a more rapid decrease in the naïve phenotype type of T cells than those which bear a T-cell memory phenotype \cite{connors}, this molecule was included in our simulation, and displayed $-\log_{10}(p)=13.24022$ interacting with other 6 molecules, here can we witness ways in which this simulations can be embedded onto the \textit{in vivo} or clinical settings, for we can track the decline in the CD45RA(+) T-cells in the course of the HIV-1 infection, and establish a relation with the adimensional temporal units in this model. 

\subsection{Th-differentiation, and HIV-infection processes as martingales.}

Here, we have commented on a method through which one may be able to prove an stochastic process $\{X_t\}$ is a martingale by a non-conventional path, for we have proposed that a martingale can be regarded as such set of random variables which are coupled with filtrations, these filtrations are steps in the Markov chain in which by each step, the process gains information, therefore increases its informational Shannon entropy. Both stochastic processes considered here have been proved to gain entropy when $t\rightarrow \infty$, this increase in entropy can also be seen as a way in which the systems tend towards equilibrium, both the T-cell differentiation process, and the HIV-infection process. A connection  between Shannon's entropy and the more thermodynamic sense of it can be performed should we remind ourselves that the increase in the states in which the system may be encountered (the increase in information) are molecules within such system, and probabilities the system may find itself in given molecule for a given amount of time. Interestingly, our simulation has yielded results, in which the probabilities of the system are very unequally divided between the members of the system, that is, there is a small set of molecules $\{S\} \subset \{X\}: |S|<<|X|$, while $\sum_i p^S_i >> \sum_j p^X_j $, these molecules contain most of the probability within a very small set of them, this has been observed in other simulations which do not pertain to the biological realm, but to the economical one, in which in simulations of continuous bids within a given population, a very small amount of people possess most of the wealth, and the majority of the population end up with a very small amount of wealth \cite{fontanelli2,boghosian}, in the yard-sale model trying to give account for wealth distribution inequality and the formation of oligarchies, it is also of interest to note that should we consider $\{X_n\}$ as the wealth of a given player in the model  in the n-th moment, the conditional expectancy equals, 

\begin{equation}
    E[(X_{n+1}^i-X_n^i)^2|X_n^i]=a^2 \min {(X_n^i,1-X_n^i)^2}
\end{equation}

where $X_{\infty}\sim Ber(X_0)$, that is the stochastic process $\{X_n\}$ is also a non-negative martingale which converges to the random variable $X^i_{\infty}$ \cite{chorro}. In our model, we can consider that there exists the formation of molecular "oligarchies" which possess most of the wealth (time spent on those molecules) which can play central roles in both stochastic processes considered. 

\subsection{Th-cell differentiation, and HIV-infection equilibrium distributions as discrete generalized beta ones.}

One additional result which astounded the authors of this paper is that of the log-rank equilibrium distribution of the stochastic processes, for they display the morphology of the discrete generalized beta distributions, this distribution has been encountered in various processes: distribution of impact factor in journals \cite{egghe}, letter frequency distribution in political speeches \cite{miramontes}, k-mer distributions in the human genome \cite{freudenberg}, traffic networks \cite{nguyen}, however it extends to art, and genetic regulatory networks \cite{cocho}; all these phenomena might be regarded as complex networks, which is the case we display here, and its importance of an HIV-infection towards a single cell (as well as the T-cell differentiation process) as complex ones, that is, that the gross phenomena which arise from it might differ from the individual products of its constituents, i.e., while we might not be able to explain the clinical course of the HIV infection by means of tracking the course of a single molecule included in the process, should we regard the network and the behaviour it displays as a whole, then can we establish those mappings from the molecular realm towards the clinical and more macroscopic one. 

\section{Conclusions \& Limitations.}

Three main findings may be exposed: the T-cell differentiation process tends towards an equilibrium state which can ensure its differentiated state (for it could be required an energetic stimulus in order to exit from it), and the HIV-infection of a T-cell leads to another, novel T-cell differentiated state, which can be antiapoptotic at first, but when time passes by and equilibrium is reached, there is a tendency towards a proapoptototic state. These processes lead toward "oligarchic" states, that is, states in which only a few molecules possess the wealth of the process (time spent on those molecules), this can be seen as how some molecules can play central roles in the process. Finally, these two processes can be considered complex ones, their phenomenology corresponds to that in which the behavior in whole cannot be directly explained by the individual behaviour of its constituents. This model can serve as scaffold for future simulations of pathogen/cell infections, the interaction of this systems with additional pathogens can be considered, for instance, what occurs when an opportunistic virus such as herpes virus type 8 infects the body? What is the nature of the interaction of this virus with the T-cells and their novel HIV-induced differentiated state? Other questions which can be addressed are those which correspond to the treatment: what happens when we block gp120? Can the path towards equilibrium continue? If so, how does this path look? 

One of the clearest limitations of this study is the lack of the logical gates YES and NO, that is, whether a protein with a higher probability to be interacted with represents an inhibition or a promotion of its function, this can be addressed in further studies; another limitation is the one of the heterogeneity in the nature of the information treated with, that is, the levels of evidence included in the interactions, The BioGrid database already displays a system of levels of evidence regarding the interactions of every protein, however in the data-acquisition process, we have not only considered interactions derived from this database, for we have considered interactions recorded in individual papers, as well as in the STRING database, all of them with distinct levels of evidence. Additionally, the interactions considered do not consider the very nature of the interaction, it does not distinguish whether the interaction is genetic, physical, or else. Finally, this model does not take into account mutations which may lead to dysfunctional protein products of both the host and the virus, as well as it does not consider the compound effect of the infection in various cells, but merely in one cell.

\section{Acknowledgements.}

To my friend and colleague Luz Fernanda Jiménez Muñoz for her input in regards to section \ref{subsection:entr}.

\printbibliography

@article{ye,
    author = {Ye, J. and Lu, H. and Su, X. and Xin, R. and Xu, K. and Bai, L. and Yu, S. and Feng, X. and Yan, H. and He, X. and Zeng, Yi.},
    title = {Phylogenetic and temporal dynamics of Human Immunodeficiency Virus type 1B in China: four types of B strains circulate in China.},
    journal = {AIDS Research and Human retroviruses. },
    volume = {30},
    number = {9},
    pages = {920--927},
    year = {2014},
    doi = {10.1089/aid.2014.0074},
}

@article{sharp,
    author = {Sharp, P.M. and Hahn, B.H.},
    title = {Origins of HIV and the AIDS pandemic.},
    journal = {Cold Spring Harbor Perspectives in Medicine.  },
    volume = {1},
    pages = {1--22},
    year = {2011},
    doi = {10.1101/cshperspect.a006841},
}

@article{gallo,
    author = {Gallo, R.C.},
    title = {A reflection on HIV/AIDS research after 25 years. },
    journal = {Retrovirology.  },
    volume = {3},
    number = {72},
    pages = {1--7},
    year = {2006},
    doi = {10.1186/1742-4690-3-72},
}

@article{sagoe,
    author = {Sagoe, K.W.C. and Mingle, J.A.A. and Afrram, R.K. and Britton, S. and Dzokoto, A. and Sonnerborg, A.},
    title = {Virological characterization of dual HIV-1/HIV-2 seropositivity and infections in Southern Africa.  },
    journal = {Ghana Medical Journal.},
    volume = {42},
    number = {1},
    pages = {16--22},
    year = {2008},
    url = {https://www.ncbi.nlm.nih.gov/pmc/articles/PMC2423332/pdf/GMJ4201_0016.pdf},
}

@article{tamamis,
    author = {Tamamis, P. and Floudas, C.A.},
    title = {Molecular recognition of CCR5 by an HIV-1 gp120 V3 loop.},
    journal = {PLoS One.},
    volume = {9},
    number = {4},
    pages = {e95767},
    year = {2014},
    doi = {10.1371/journal.pone.0095767},
}

@article{woodham,
    author = {Woodham, A.W. and Skeate, J.G. and Sanna, A.M. and Taylor, J.R. and Da Silva, D.M. and Cannon, P.M. and Kast, W.M.},
    title = {Human immunodeficiency virus immune cell receptors, coreceptors and cofactors: implications for prevention and treatment.},
    journal = {AIDS Patient Care and STDs},
    volume = {30},
    number = {7},
    pages = {291--307},
    year = {2016},
    doi = {10.1089/apc.2016.0100},
}

@online{ictv,
    author = {Stoye, J.P. and Blomberg, J. and Coffin, J.M. and Fan, H. and Hahn, B. and Neil, J. and Quackenbush, S. and Rethwilm, A. and Tristem, M.},
    title = {Retroviridae},
    url  = {talk.ictvonline.org/ictv-reports/ictv_9th_report/reverse-transcribing-dna-and-rna-viruses-2011/w/rt_viruses/161/retroviridae},
    addendum = {(accessed: 16.04.2020)},
}

@article{german,
    author = {German Advisory Commitee Blood (Arbitskreis Blut) subgroup "Assessment of Pathogens Transmissible by Blood"},
    title = {Human immunodeficiency virus (HIV)},
    journal = {Transfusion Medicine and Hemotherapy.},
    volume = {43},
    pages = {203--222},
    year = {2016},
    doi = {10.1159/000445852},
}

@article{huarte,
    author = {Huarte, N. and Carravilla, P. and Cruz, A. and Lorizate, M. and Nieto-Garai, J.A. and Kräusslich, H.-G. and Pérez-Gil, J. and Requejo-Isidro, J. and Nieva, J.L.},
    title = {Functional organization of the HIV lipid envelope.},
    journal = {Scientific Reports. },
    volume = {6},
    number = {34190},
    pages = {1--14},
    year = {2016},
    doi = {10.1038/srep34190},
}

@article{marx,
    author = {Marx, P.A. and Alcabes, P.G. and Drucker, E.},
    title = {Serial human passage of simian immunodeficiency virus by unsterile injections and the emergence of epidemic human immunodeficiency virus in Africa.},
    journal = {Phil. Trans. R. Soc. Lond. B},
    volume = {356},
    pages = {911--920},
    year = {2001},
    doi = {10.1098/rstb.2001.0867},
}

@article{pela,
    author = {Ona Pela, A. and Platt, J.J.},
    title = {AIDS in Africa: emerging trends. },
    journal = {Soc. Sci. Med.},
    volume = {28},
    number = {1},
    pages = {1--8},
    year = {1989},
    doi = {10.1016/0277-9536(89)90298-0},
}

@article{haywa,
    author = {Hayward, A.},
    title = {Origin of the retroviruses: when, where, and how?},
    journal = {Current Opinion in Virology. },
    volume = {25},
    pages = {23--27},
    year = {2017},
    doi = {10.1016/j.coviro.2017.06.006},
}

@article{taylor,
    author = {Taylor, B.S. and Sobieszczyk, M.E. and McCutchan, F.E. and Hammer, S.M.},
    title = {The Challenge of HIV-1 subtype diversity.},
    journal = {N. Engl. J. Med.},
    volume = {358},
    number = {15},
    pages = {1590--1602},
    year = {2008},
    doi = {10.1056/NEJMra0706737},
}

@article{antire,
    author = {The Antiretroviral Therapy Cohort Collaboration.},
    title = {Life expectancy of individuals on combination antiretroviral therapy in high-income countries: a collaborative analysis of 14 cohort studies.},
    journal = {Lancet.},
    volume = {372},
    number = {9635},
    pages = {293--299},
    year = {2008},
    doi = {10.1016/S0140-6736(08)61113-7},
}

@article{antiret,
    author = {The Antiretroviral Therapy Cohort Collaboration.},
    title = {Survival of HIV-positive patients starting antiretroviral therapy between 1996 and 2013: a collaborative analysis of cohort studies.},
    journal = {Lancet.},
    volume = {4},
    pages = {e349--e357},
    year = {2017},
    doi = {10.1016/S2352-3018(17)30066-8},
}

@article{cuevas,
    author = {Cuevas, J.M. and Geller, R. and Garijo, R. and López-Aldeguer, J. and Sanjuán, R.},
    title = {Extremely high mutation rate of HIV-1 in vivo.},
    journal = {PLoS Biology.},
    volume = {13},
    number = {9},
    pages = {1--19},
    year = {2015},
    doi = {10.1371/journal.pbio.1002251},
}

@Manual{rstudio,
title = {RStudio: Integrated Development Environment for R},
author = {{RStudio Team}},
organization = {RStudio, PBC.},
address = {Boston, MA},
year = {2020},
url = {http://www.rstudio.com/},
}

@online{eomes,
    author = {NIH},
    title = {EOMES eomesodermin [ Homo sapiens (human) ]},
    url  = {https://www.ncbi.nlm.nih.gov/gene?Db=gene&Cmd=DetailsSearch&Term=8320},
    addendum = {(accessed: 02.08.2020)},
}

@article{gagliani,
    author = {Gagliani, N. and Huber, S.},
    title = {Basic Aspects of T Helper Cell Differentiation.},
    journal = {Methods Mol. Biol.},
    volume = {1514},
    pages = {19--30},
    year = {2017},
    doi = {10.1007/978-1-4939-6548-9_2},
}

@article{zhu,
    author = {Zhu, J.},
    title = {T Helper Cell Differentiation, Heterogeneity, and Plasticity.},
    journal = {Cold Spring Harb. Perspect. Biol.},
    volume = {10},
    number = {10},
    pages = {a030338},
    year = {2018},
    doi = {10.1101/cshperspect.a030338},
}

@article{schmitt,
    author = {Schmitt, N. and Ueno, H.},
    title = {Regulation of human helper T cell subset differentiation by cytokines.},
    journal = {Curr. Opin. Immunol.},
    volume = {34},
    pages = {130--136},
    year = {2015},
    doi = {10.1016/j.coi.2015.03.007}
}

@article{maestre,
    author = {Maestre, L. and García-García, JF. and Jímenez, S. and Reyes-García, AI. and García-González, A. and Montes-Moreno, S. and Arribas, AJ. and González-García, P. and Caleiras, E. and Banham, AH. and Piris, MA. and Roncador, G.},
    title = {High-mobility group box (TOX) antibody a useful tool for the identification of B and T cell subpopulations.},
    journal = {PLoS ONE.},
    volume = {15},
    number = {2},
    pages = {e0229743},
    year = {2020},
    doi = {10.1371/journal.pone.0229743}
}

@article{aliahmad,
    author = {Aliahmad, P. and Seksenyan, A. and Kaye, J.},
    title = {The many roles of TOX in the immune system.},
    journal = {Curr. Opin. Immunol.},
    volume = {24},
    number = {2},
    pages = {173-177},
    year = {2012},
    doi = {10.1016/j.coi.2011.12.001}
}

@book{rincon,
    title = {Introducción a los Procoesos Estocásticos.},
    author = {Rincón, L.},
    isbn = {9786070230448},
    year = {2013},
    publisher = {UNAM}
}

@online{stanford,
    author = "Stanford University",
    title = "Lecture 17. Perron Frobenius Theory.",
    url  = "https://stanford.edu/class/ee363/lectures/pf.pdf",
}

@article{imamichi,
    author = {Imamichi, H. and Natarajan, V. and Adelsberger, J. and Rehm, C. and Lempicki, R. and Das, B. and Lane, H.},
    title = {Lifespan of effector memory CD4+ T cells determined by replication-incompetent integrated HIV-1 provirus.},
    journal = {AIDS},
    volume = {28},
    pages = {1091--1099},
    year = {2014},
    doi = {10.1097/QAD.0000000000000223}
}

@article{borghans,
    author = {Borghans, J. and Ribeiro, R.},
    title = {The maths of memory.},
    journal = {eLife},
    volume = {6},
    number = {e23013},
    pages = {1--3},
    year = {2017},
    doi = {10.7554/eLife.23013}
}

@article{gossel,
    author = {Gossel, G. and Hogan, T. and Cowden, D. and Seddon, B. and Yates, A.},
    title = {Memory CD4 T cell subsets are kinetically heterogenous and replenished from naïve T cells at high levels.},
    journal = {eLife},
    volume = {6},
    number = {e23013},
    pages = {1--30},
    year = {2017},
    doi = {10.7554/eLife.23013}
}

@article{cocho,
    author = {Martínez-Mekler, G. and Álvarez Martínez, R. and Beltrán del Río, M. and Mansilla, R. and Miramontes, P. and Cocho, G.},
    title = {Universality of rank-ordering distributions in the arts and sciences.},
    journal = {PLoS One.},
    volume = {4},
    number = {3},
    pages = {e4791},
    year = {2009},
    doi = {10.1371/journal.pone.0004791}
}

@article{fontanelli,
    author = {Fontanelli, O. and Miramontes, P. and Mansilla, R. and Cocho, G. and Li, W.},
    title = {Beta-rank function: a smooth double-Pareto-like distribution.},
    journal = {arXiv},
    pages = {1--33},
    year = {2019},
    url = {https://arxiv.org/abs/1910.05364v1}
}

@article{crublet,
    author = {Crublet, E. and Andrieu, J.-P. and Vivès, R.R. and Lortat-Jacob, H.},
    title = {The HIV-1 envelope glycoprotein gp120 features four heparan sulfate binding domains, including the co-receptor binding site.},
    journal = {The Journal of Biological Chemistry.},
    volume = {283},
    number = {22},
    pages = {15193--15200},
    year = {2008},
    doi = {10.1074/jbc.M800066200},
}

@article{raska,
    author = {Raska, M. and Takahashi, K. and Czernekova, L. and Zachova, K. and Hall, S. and Moldoveanu, Z. and Elliott, M.C. and Wilson, L. and Brown, R. and Jancova, D. and Barnes, S. and Vrbkova, J. and Tomana, M. and Smith, P.D. and Mestecky, J. and Renfrow, M.B. and Novak, J.},
    title = {Glycosylation patterns of HIV-1 gp120 depend on the type of expressing cells and affect antibody recognition. },
    journal = {The Journal of Biological Chemistry.},
    volume = {285},
    number = {27},
    pages = {20860--20869},
    year = {2010},
    doi = {10.1074/jbc.M109.085472},
}

@article{fennie,
    author = {Fennie, C. and Laski, L.A.},
    title = {Model for intracellular folding of the human immunodeficiency virus type 1 gp120.},
    journal = {Journal of Virology.},
    volume = {63},
    number = {2},
    pages = {639--646},
    year = {1989},
    url = {ncbi.nlm.nih.gov/pmc/articles/PMC247734/pdf/jvirol00069-0181.pdf},
}

@article{yoon,
    author = {Yoon, V. and Fridkis-Hareli, M. and Lee, J. and Anastasiades, D. and Stevceva, L.},
    title = {The gp120 molecule of HIV-1 and its interaction with T cells.},
    journal = {Current Medicinal Chemistry.},
    volume = {17},
    pages = {741--749},
    year = {2010},
    doi = {10.2174/092986710790514499},
}

@article{dey,
    author = {Dey, B. and Berger, E.A.},
    title = {Blocking HIV-1 gp120 at the Phe43 cavity: if the extension fits.},
    journal = {Structure.},
    volume = {21},
    number = {6},
    pages = {871--872},
    year = {2014},
    doi = {10.1016/j.str.2013.05.004},
}

@article{thali,
    author = {Thali, M. and Olshevsky, U. and Furman, C. and Gabuzda, D. and Li, J. and Sodroski, J.},
    title = {Effects of changes in gp120-CD4 binding affinity on human immunodeficiency virus type 1 envelope glycoprotein function and soluble CD4 sensitivity.},
    journal = {Journal of Virology.},
    volume = {65},
    number = {9},
    pages = {5007--5012},
    year = {1991},
    url = {https://www.ncbi.nlm.nih.gov/pmc/articles/PMC248964/pdf/jvirol00052-0471.pdf},
}

@book{maron,
    title = {Fundamentos de Fisicoquímica.},
    author = {Maron, S.H. and Prutton, C.F.},
    year = {1975},
    publisher = {Limusa Editorial}
}

@article{hsu,
    author = {Hsu, S.-T.D. and Peter, C. and van Gunsteren, W.F. and Bonvin, A.M.J.J.},
    title = {Entropy calculation of HIV-1 Env gp120, its receptor CD4, and their complex: an analysis of configurational entropy changes upon complexation.},
    journal = {Biophysical Journal.},
    volume = {88},
    pages = {15--24},
    year = {2005},
    doi = {10.1529/biophysj.104.044933},
}

@article{sullivan,
    author = {Sullivan, N. and Sun, Y. and Sattentau, Q. and Thali, M. and Wu, D. and Denisova, G. and Gershoni, J. and Robinson, J. and Moore, J. and Sodroski, J.},
    title = {CD4-induced conformational changes in the human immunodeficiency virus type 1 gp120 glycoprotein: consequences for virus entry and neutralization.},
    journal = {Journal of Virology.},
    volume = {72},
    number = {6},
    pages = {4694--4703},
    year = {1998},
    url = {https://www.ncbi.nlm.nih.gov/pmc/articles/PMC109994/pdf/jv004694.pdf},
}

@article{kuppers,
    author = {Küppers, R.},
    title = {B cells under influence: transformation of B cells by Epstein-Barr virus.},
    journal = {Nature Reviews | Immunology.},
    volume = {3},
    pages = {801--813},
    year = {2003},
    doi = {10.1038/nri1201},
}

@article{robinson,
    author = {Robinson, J.E. and Holton, D. and Pacheco-Morell, S. and Liu, J. and McMurdo, H.},
    title = {Identification of conserved and variant epitopes of human immunodeficiency virus type 1 (HIV-1) gp120 by human monoclonal antibodies produced by EBV-transformed cell lines.},
    journal = {AIDS Research and Human Retroviruses.},
    volume = {6},
    number = {5},
    pages = {567--580},
    year = {1990},
    doi = {10.1089/aid.1990.6.567},
}

@article{lagenaur,
    author = {Lagenaur, L.A. and Villaroel, V.A. and Bundoc, V. and Dey, B. and Berger, E.A.},
    title = {sCD4-17b binfunctional protein: extremely broad and potent neutralization of HIV-1 Env pseudotyped viruses from genetically diverse primary isolates.},
    journal = {Retrovirology.},
    volume = {7},
    number = {11},
    pages = {1--13},
    year = {2010},
    doi = {10.1186/1742-4690-7-11},
}

@article{raja,
    author = {Raja, A. and Venturi, M. and Kwong, P. and Sodroski, J.},
    title = {CD4 binding site antibodies inhibit human immunodeficiency virus gp120 envelope glycoprotein interaction with CCR5.},
    journal = {Journal of Virology.},
    volume = {77},
    number = {1},
    pages = {713--718},
    year = {2003},
    doi = {10.1128/JVI.77.1.713–718.2003},
}

@article{kwong,
    author = {Kwong, P.D. and Wyatt, R. and Sattentau, Q.J. and Sodroski, J. and Hendrickson, W.A.},
    title = {Oligomeric modeling and electrostatic analysis of the gp120 envelope glycoprotein of Human Immunodeficiency Virus.},
    journal = {Journal of Virology.},
    volume = {74},
    number = {4},
    pages = {1961--1972},
    year = {2000},
    doi = {10.1128/jvi.74.4.1961-1972.2000},
}

@article{ribeiro,
    author = {Ribeiro, R.M. and Hazenberg, M.D. and Perelson, A.S. and Davenport, M.P.},
    title = {Naïve and memory cell turnover as divers of CCR5-to-CXCR4 tropism switch in immunodeficiency virus type 1: implications for therapy.},
    journal = {Journal of Virology. },
    volume = {80},
    number = {2},
    pages = {802--809},
    year = {2006},
    doi = {10.1128/JVI.80.2.802–809.2006},
}

@article{mosier,
    author = {Mosier, D.E.},
    title = {How HIV changes its tropism: evolution and adaptation?},
    journal = {Curr. Opin. HIV AIDS.},
    volume = {4},
    number = {2},
    pages = {125--130},
    year = {2010},
    doi = {10.1097/COH.0b013e3283223d61},
}

@article{sevier,
    author = {Sevier, C.S. and Kaiser, C.A.},
    title = {Formation and transfer of disulphide bonds in living cells.},
    journal = {Nature Reviews | Molecular and Cellular Biology.},
    volume = {3},
    pages = {836--848},
    year = {2002},
    doi = {10.1038/nrm954},
}

@article{susal,
    author = {Süsal, C. and Kirschfink, M. and Kröpelin, M. and Daniel, V. and Opelz, G.},
    title = {Identification of complement activabtion sites in human immunodeficiency virus type-1 glycoprotein gp120.},
    journal = {Blood.},
    volume = {87},
    number = {6},
    pages = {2329--36},
    year = {1996},
    url = {https://www.ncbi.nlm.nih.gov/pubmed/8630395},
}

@article{zewde,
    author = {Zewde, N. and Mohan, R.R. and Morikis, D.},
    title = {Immunophysical evaluation of the initiation step in the formation of the membrane attack complex.},
    journal = {Frontiers in Physics.},
    volume = {6},
    number = {130},
    pages = {1--17},
    year = {2018},
    doi = {10.3389/fphy.2018.00130},
}

@article{virot,
    author = {Virot, E. and Duclos, A. and Adelaide, L. and Miailhes, P. and Hot, A. and Ferry, T. and Seve, P.},
    title = {Autoimmune diseases and HIV infection. },
    journal = {Medicine.},
    volume = {96},
    number = {4},
    pages = {e5769},
    year = {2017},
    doi = {10.1097/MD.0000000000005769},
}

@article{patke,
    author = {Patke, C.L. and Shearer, W.T.},
    title = {gp120- and TNFA-induced modulation of human B cell function: proliferation, cyclic AMP generation, Ig production, and B-cell receptor expression.},
    journal = {J. Allergy Clin. Immunol.},
    volume = {105},
    pages = {975--82},
    year = {2000},
    doi = {10.1067/mai.2000.105315},
}

@article{minguet,
    author = {Minguet, S. and Huber, M. and Rosenkranz, L. and Schamel, W.W.A. and Reth, M. and Brummer, T.},
    title = {Adenosine and cAMP are potent inhibitors of the NF-KB pathway downstream of immunoreceptors.},
    journal = {European Journal of Immunology.},
    volume = {35},
    pages = {31--41},
    year = {2005},
    doi = {10.1002/eji.200425524},
}

@article{jacobson,
    author = {Jacobson, C.A. and Abramson, J.S.},
    title = {HIV-associated Hodgkin's lymphoma: prognosis and therapy in the era of cART.},
    journal = {Advances in Hematology.},
    volume = {2012},
    number = {507257},
    pages = {1--8},
    year = {2012},
    doi = {10.1155/2012/507257},
}

@article{sasaki,
    author = {Sasaki, Y. and Iwai, K.},
    title = {Roles of the NF-KB pathway in B-lymphocyte biology.},
    journal = {Curr. Top Microbiol. Immunol.},
    volume = {393},
    pages = {177--209},
    year = {2016},
    doi = {10.1007/82_2015_479},
}

@article{morou,
    author = {Morou, A.K. and Porichis, F. and Kramovitis, E. and Sourvinos, G. and Spandidos, D.A. and Zafiropoulos, A.},
    title = {The HIV-1 gp120/V3 modifies the response of uninfected CD4 T cells to antigen presentation: mapping of the specific transcriptional signature. },
    journal = {Journal of Translational Medicine.},
    volume = {9},
    number = {160},
    pages = {1--16},
    year = {2011},
    doi = {10.1186/1479-5876-9-160},
}

@article{shen,
    author = {Shen, N. and Wang, T. and Li, D. and Zhu, Y. and Xie, H. and Lu, Y.},
    title = {Hypermethylation of the SEPT9 gene suggests significantly poor prognosis in cancer patients: a systematic review and meta-analysis. },
    journal = {Frontiers in Genetics.},
    volume = {10},
    number = {887},
    pages = {1--8},
    year = {2019},
    doi = {10.3389/fgene.2019.00887},
}

@article{davidchan,
    author = {Chan, D.C. and Fass, D. and Berger, J.M. and Kim, P.S.},
    title = {Core structure of gp41 from the HIV-1 envelope glycoprotein.},
    journal = {Cell.},
    volume = {89},
    pages = {263--273},
    year = {1997},
    doi = {10.1016/S0092-8674(00)80205-6},
}

@article{merk,
    author = {Merk, A. and Subramaniam, S.},
    title = {HIV-1 envelope glycoprotein structure.},
    journal = {Curr. Opin. Struct. Biol.},
    volume = {23},
    number = {2},
    pages = {268--276},
    year = {2013},
    doi = {10.1016/j.sbi.2013.03.007},
}

@article{fernandez,
    author = {Fernandez, M.V. and Freed, E.O.},
    title = {Meeting review: 2018 international workshop on structure and function of the lentiviral gp41 cytoplasmic tail.},
    journal = {Viruses.},
    volume = {10},
    number = {613},
    pages = {1--11},
    year = {2018},
    doi = {10.3390/v10110613},
}

@article{postler,
    author = {Postler, T.S. and Desrosiers, R.C.},
    title = {The tale of the long tail: the cytoplasmic domain of HIV-1 gp41.},
    journal = {Journal of Virology.},
    volume = {87},
    number = {1},
    pages = {2--15},
    year = {2013},
    doi = {10.1128/JVI.02053-12},
}

@article{sarafianos,
    author = {Sarafianos, S.G. and Marchand, B. and Das, K. and Himmel, D. and Parniak, M.A. and Hughes, S.H. and Arnold, E.},
    title = {Structure and function of HIV-reverse transcriptase: molecular mechanisms of polymerization and inhibition.},
    journal = {J. Mol. Biol.},
    volume = {385},
    number = {3},
    pages = {693--713},
    year = {2009},
    doi = {10.1016/j.jmb.2008.10.071},
}

@article{chung,
    author = {Chung, S. and Miller, J.T. and Lapkouski, M. and Tian, L. and Yang, W. and Le Grice, S.F.J.},
    title = {Examining the role of the HIV-1 reverse transcriptase p51 subunit in positioning and hydrolysis of RNA/DNA hybrids.},
    journal = {The Journal of Biological Chemistry.},
    volume = {288},
    number = {22},
    pages = {16177--16184},
    year = {2013},
    doi = {10.1074/jbc.M113.465641},
}

@article{zheng,
    author = {Zheng, X. and Mueller, G.A. and Cuneo, M.J. and DeRose, E.F. and London, R.E.},
    title = {Homodimerization of the p51 subunit of HIV-1 reverse transcriptase.},
    journal = {Biochemistry.},
    volume = {49},
    number = {2821},
    pages = {2821--2833},
    year = {2010},
    doi = {10.1021/bi902116z},
}

@article{zheng2,
    author = {Zheng, X. and Pedersen, L.C. and Gabel, S.A. and Mueller, G.A. and Cuneo, M.J. and DeRose, E.F. and Krahn, J.M. and London, R.E.},
    title = {Selective unfolding of one ribonuclease H domain of HIV reverse transcriptase is linked to homodimer formation.},
    journal = {Nucleic Acids Research.},
    volume = {42},
    number = {8},
    pages = {5361--5377},
    year = {2014},
    doi = {10.1093/nar/gku143}
}

@article{nowotny,
    author = {Nowotny, M. and Gaidamakov, S.A. and Crouch, R.J. and Yang, W.},
    title = {Crystal structures of RNase H bound to an RNA/DNA hybrid: substrate specificity and metal-dependent catalysis.},
    journal = {Cell.},
    volume = {121},
    pages = {1005--1016},
    year = {2005},
    doi = {10.1016/j.cell.2005.04.024}
}

@article{cirino,
    author = {Cirino, N.M. and Cameron, C.E. and Smith, J.S. and Rausch, J.W. and Roth, M.J. and Benkovic, S.J. and Le Grice, S.F.J.},
    title = {Divalent cation modulation of the ribonuclease functions of human immunodeficiency virus reverse transcriptase.},
    journal = {Biochemistry.},
    volume = {34},
    pages = {9936--9943},
    year = {1995},
    doi = {10.1021/bi00031a016}
}

@article{craigie,
    author = {Craigie, R.},
    title = {The molecular biology of HIV integrase.},
    journal = {Future virology.},
    volume = {7},
    number = {7},
    pages = {679--686},
    year = {2012},
    doi = {10.2217/FVL.12.56 }
}

@article{endsley,
    author = {Endsley, M.A. and Somasunderam, A.D. and Li, G. and Oezguen, N. and Thiviyanathan, V. and Murray, J.L. and Rubin, D.H. and Hodge, T.W. and O'Brien, W.A. and Lewis, B. and Ferguson, M.R.},
    title = {Nuclear trafficking of the HIV-1 pre-integration complex depends on the ADAM10 intracellular domain.},
    journal = {Virology.},
    volume = {454},
    pages = {60--66},
    year = {2014},
    doi = {10.1016/j.virol.2014.02.006}
}

@article{iordanski,
    author = {Iordanskiy, S. and Berro, R. and Altieri, M. and Kasanchi, F. and Bukrinsky, M.},
    title = {Intracytoplasmic maturation of the human immunodeficiency virus type 1 reverse transcription complexes determines their capacity to integrate into chromatin.},
    journal = {Retrovirology.},
    volume = {3},
    number = {4},
    pages = {1--12},
    year = {2006},
    doi = {10.1186/1742-4690-3-4}
}

@article{popov,
    author = {Popov, S. and Rexach, M. and Zybarth, G. and Reiling, N. and Lee, M.-A. and Ratner, L. and Lane, C.M. and Moore, S. and Blobel, G. and Bukrinsky, M.},
    title = {Viral protein R regulates nuclear import of the HIV-1 pre-integration complex.},
    journal = {The EMBO Journal.},
    volume = {17},
    number = {4},
    pages = {909--917},
    year = {1998},
    doi = {10.1093/emboj/17.4.909}
}

@article{michieletto,
    author = {Michieletto, D. and Lusic, M. and Marenduzzo, D. and Orlandini, E.},
    title = {Physical principles of retroviral integration in the human genome.},
    journal = {Nature Communications. },
    volume = {10},
    number = {575},
    pages = {1--11},
    year = {2019},
    doi = {10.1038/s41467-019-08333-8}
}

@article{agosto,
    author = {Agosto, L.M. and Gagne, M. and Henderson, A.J.},
    title = {Impact of chromatin on HIV replication.},
    journal = {Genes.},
    volume = {6},
    pages = {957--976},
    year = {2015},
    doi = {10.3390/genes6040957}
}

@article{rona,
    author = {Rona, G.B. and Eleutherio, E.C.A. and  Pinheiro, A.S.},
    title = {PWWP domains and their modes of sensing DNA and histone methylated lysines.},
    journal = {Biophys. Rev.},
    volume = {8},
    pages = {63--74},
    year = {2016},
    doi = {10.1007/s12551-015-0190-6}
}

@article{shaw,
    author = {Shaw, G.M. and Hunter, E.},
    title = {HIV Transmission.},
    journal = {Cold Spring Harb. Perspect. Med.},
    volume = {2},
    number = {11},
    pages = {a006965},
    year = {2012},
    doi = {10.1101/cshperspect.a006965}
}

@article{pearson,
    author = {Pearson, J.E. and Krapivsky, P. and Perelson, A.S.},
    title = {Stochastic Theory of Early Viral Infection: Continuous versus Burst Production of Virions.},
    journal = {PLoS Comput. Biol.},
    volume = {7},
    number = {2},
    pages = {e1001058},
    year = {2011},
    doi = {10.1371/journal.pcbi.1001058}
}

@article{ribeiro2,
    author = {Ribeiro, R.M. and Qin, L. and Chavez, L.L. and Li, D. and Self, S.G. and Perelson, A.S.},
    title = {Estimation of the initial viral growth rate and basic reproductive number during acute HIV-1 infection.},
    journal = {J. Virol.},
    volume = {84},
    pages = {6096--6102},
    year = {2010},
    doi = {10.1128/JVI.00127-10}
}

@article{parrondo,
    author = {Parrondo, J.M.R. and Horowitz, J.M. and Sagawa, T.},
    title = {Thermodynamics of information.},
    journal = {Nature Physics.},
    volume = {11},
    pages = {131--140},
    year = {2015},
    doi = {10.1038/NPHYS3230}
}

@article{seifert,
    author = {Seifert, U.},
    title = {Stochastic thermodynamics, fluctuation theorems and molecular machines.},
    journal = {Reports of Progress in Physics.},
    volume = {75},
    pages = {1--59},
    year = {2012},
    doi = {10.1088/0034-4885/75/12/126001}
}

@inbook{baron,
    author = "Baron, P.",
    title = "Continuous stochastic variables.",
    publisher = "Elsevier.",
    year = "2017",
    chapter = "15"
}

@inbook{attard,
    author = "Attard, P.",
    title = "Thermodynamics and Statistical Mechanics.",
    publisher = "Elsevier.",
    year = "2002",
    chapter = "Prologue."
}

@article{wilson,
    author = {Wilson, J.A.},
    title = {Increasing entropy of biological systems.},
    journal = {Nature.},
    volume = {219},
    pages = {534--535},
    year = {1968},
    doi = {10.1038/219534a0}
}

@article{buchel,
    author = {Büchel, W.},
    title = {Entropy and information in the Universe.},
    journal = {Nature.},
    volume = {213},
    pages = {319--320},
    year = {1967},
    doi = {10.1038/213319a0}
}

@article{debaar,
    author = {De Baar, M.P. and van der Horn, K.H.M. and Goudsmit, J. and De Ronde, A. and De Wolf, F.},
    title = {Detection of Human Immunodeficiency Virus type 1 nucleocapsid protein p7 in vitro and in vivo.},
    journal = {Journal of Clinical Microbiology.},
    volume = {37},
    number = {1},
    pages = {63--67},
    year = {1999},
    url = {https://www.ncbi.nlm.nih.gov/pmc/articles/PMC84168/pdf/jm000063.pdf}
}

@article{cassandri,
    author = {Cassandri, M. and Smirnov, A. and Novelli, F. and Pitolli, C. and Agostini, M. and Malewicz, M. and Melino, G. and Raschellà, G.},
    title = {Zinc-finger proteins in health and disease.},
    journal = {Cell Death Discovery.},
    volume = {3},
    number = {17071},
    pages = {1--12},
    year = {2017},
    doi = {10.1038/cddiscovery.2017.71}
}

@article{maynard,
    author = {Maynard, A.T. and Huang, M. and Rice, W.G. and Covell, D.G.},
    title = {Reactivity of the HIV-1 nucleocapsid protein p7 zinc finger domains from the perspective of density-functional theory.},
    journal = {Proc. Natl. Acad. Sci. USA.},
    volume = {95},
    pages = {11578--11583},
    year = {1998},
    doi = {10.1073/pnas.95.20.11578}
}

@article{lapadat,
    author = {Lapadat-Tapolsky, M. and De Rocquigny, H. and Van Gent, D. and Rocques, B. and Plasterk, R. and Darlix, J.L.},
    title = {Interactions between HIV-1 nucleocapsid protein and viral DNA may have important functions in the viral life cycle.},
    journal = {Nucleic Acids Res.},
    volume = {21},
    number = {4},
    pages = {831--839},
    year = {1993},
    doi = {10.1093/nar/21.4.831}
}

@article{oie,
    author = {Öie Solbak, S.M. and Ragna Reksten, T. and Hahn, F. and Wray, V. and Henklein, P. and Henklein, P. and Halksau, Ö. and Schubert, U. and Fossen, T.},
    title = {HIV-1 p6 -- a structured to flexible multifunctional membrane-interacting proteins.},
    journal = {Biochimica et Biophysica Acta.},
    number = {1828},
    pages = {816--823},
    year = {2013},
    doi = {10.1016/j.bbamem.2012.11.010}
}

@article{jenkins,
    author = {Jenkins, Y. and Pornillos, O. and Rich, R.L. and Myszka, D.G. and Sundqist, W.I. and Malim. M.H.},
    title = {Biochemical analyses of the interactions between human immunodeficiency virus type 1 Vpr and $p6^{Gag}$.},
    journal = {Journal of Virology.},
    volume = {75},
    number = {21},
    pages = {10537--10542},
    year = {2001},
    doi = {10.1128/JVI.75.21.10537-10542.2001}
}

@article{graybain,
    author = {Gray, E.R. and Bain, R. and Varsaneux, O. and Peeling, R.W. and Stevens, M.M. and McKendry, R.A.},
    title = {p24 revisited: a landscape review of antigen detection for early HIV diagnosis.},
    journal = {AIDS},
    volume = {32},
    number = {15},
    pages = {2089--2103},
    year = {2018},
    doi = {10.1097/QAD.0000000000001982}
}

@article{perilla,
    author = {Zhao, g. and Perilla, J.R. and Yufenyuy, E.L. and Meng, X. and Chen, B. and Ning, J. and Ahn, J. and Gronenborn, A.M. and Schulten, K. and Aiken, C. and Zhang, P.},
    title = {Mature HIV-1 capsid structure by cryo-electron microscopy and all-atom molecular dynamics.},
    journal = {Nature.},
    volume = {497},
    pages = {643--647},
    year = {2013},
    doi = {10.1038/nature12162}
}

@article{campbell,
    author = {Campbell, E.M. and Hope, T.J.},
    title = {HIV-1 capsid: the multifaceted key player in HIV-1 infection.},
    journal = {Nature Reviews | Microbiology.},
    volume = {13},
    pages = {471--484},
    year = {2015},
    doi = {10.1038/nrmicro3503}
}

@article{wallach,
    author = {Wallach, D. and Kovalenko, A.},
    title = {How do cells sense foreign DNA? A new outlook on the function of STING.},
    journal = {Molecular Cell.},
    volume = {50},
    pages = {1--2},
    year = {2013},
    doi = {10.1016/j.molcel.2013.03.024}
}

@article{farnet,
    author = {Farnet, C.M. and Bushman, F.D.},
    title = {HIV-1 cDNA integration: requirement of HMG I(Y) protein for function of preintegration complexes in vitro.},
    journal = {Cell.},
    volume = {88},
    pages = {483--492},
    year = {1997},
    doi = {10.1016/S0092-8674(00)81888-7}
}

@online{ppia,
    author = "NIH",
    title = "PPIA peptidylprolyl isomerase A [ Homo sapiens (human) ]",
    url  = "https://www.ncbi.nlm.nih.gov/gene?Db=gene&Cmd=DetailsSearch&Term=5478",
    addendum = "(accessed: 27.11.2020)"
}

@online{cpsf6,
    author = "NIH",
    title = "CPSF6 cleavage and polyadenylation specific factor 6 [ Homo sapiens (human) ]",
    url  = "https://www.ncbi.nlm.nih.gov/gene?Db=gene&Cmd=DetailsSearch&Term=11052",
    addendum = "(accessed: 27.11.2020)"
}

@article{fiorentini,
    author = {Fiorentini, S. and Marini, E. and Caracciolo, S. and Caruso, A.},
    title = {Functions of the HIV-1 matrix protein p17.},
    journal = {New Microbiol.},
    volume = {29},
    number = {1},
    pages = {1--10},
    year = {2006},
    pmid = {16608119}
}

@article{ciagulli,
    author = {Ciagulli, C. and Marsico, S. and Magiera, A.K. and Bruno, R. and Caccuri, F. and Barone, I. and Fiorentini, S. and Andò, S. and Caruso, A.},
    title = {Opposite effects of HIV-1 p17 variants on PTEN activation and cell growth in B cells.},
    journal = {PLoS One.},
    volume = {6},
    number = {3},
    pages = {e17831},
    year = {2011},
    doi = {10.1371/journal.pone/0017831}
}

@online{harmonizome,
    author = "Rouillard, A.D. and Gundersen, G.W. and Fernandez, N.F. and Wang, Z. and Monteiro, C.D. and McDermott, M.G. and Ma'ayan, A.",
    title = "The harmonizome: a collection of processed datasets gathered to serve and mine knowledge about genes and proteins",
    url  = "https://maayanlab.cloud/Harmonizome/gene_set/Raji/GDSC+Cell+Line+Gene+Expression+Profiles",
    addendum = "(accessed: 28.11.2020)"
}

@article{massiah,
    author = {Massiah, M.A. and Worthylake, D. and Christensen, A.M. and Sundqist, W.I. and Hill, C.P. and Summers, M.F.},
    title = {Comparison of the NMR and X-ray structures of the HIV-1 matrix protein: evidence for conformational changes during viral assembly},
    journal = {Protein Science.},
    volume = {5},
    number = {12},
    pages = {2391--2398},
    year = {1996},
    doi = {110.1002/pro.5560051202}
}

@article{starich,
    author = {Massiah, M. A. and Starich, M.R. and Paschall, C. and Summers, M.F. and Christensen, A.M. and Sundqist, W.I.},
    title = {Three-dimensional structure of the human immunodeficiency virus type 1 matrix protein.},
    journal = {J. Mol. Biol.},
    volume = {244},
    number = {2},
    pages = {198--223},
    year = {1994},
    doi = {10.1006/jmbi.1994.1719}
}

@article{dunn,
    author = {Dunn, B.M. and Goodenow, M.M. and Gustchina, A. and Wlodawer, A.},
    title = {Retroviral proteases.},
    journal = {Genome biology.},
    volume = {3},
    number = {4},
    pages = {3006.1--3006.7},
    year = {2002},
    doi = {10.1186/gb-2002-3-4-reviews3006}
}

@article{tropsha,
    author = {Zhang, S. and Kaplan, A.H. and Tropsha, A.},
    title = {HIV-1 protease function and structure studies with the simplicial neighborhood analysis of protein packing method.},
    journal = {Proteins.},
    volume = {73},
    number = {3},
    pages = {742--753},
    year = {2008},
    doi = {10.1002/prot.22094}
}

@article{nkeze,
    author = {Yang, H. and Nkeze, J. and Zhao, R.Y.},
    title = {Effects of HIV-1 protease on cellular functions and their potential applications in antiretroviral therapy.},
    journal = {Cell \& Bioscience.},
    volume = {2},
    number = {32},
    pages = {1--8},
    year = {2012},
    doi = {10.1186/2045-3701-2-32}
}

@article{miaohuang,
    author = {Miao, Y. and Huang, Y.-M.M. and Walker, R.C. and McCammon, J.A. and Chang, C.-e. A.},
    title = {Ligand binding pathways and conformational transitions of the HIV protease.},
    journal = {Biochemistry.},
    volume = {57},
    number = {9},
    pages = {1533--1541},
    year = {2018},
    doi = {10.1021/acs.biochem.7b01248}
}

@article{shoeman,
    author = {Shoeman, R.L. and Höner, B. and Stoller, T.J. and Kesselmaier, C. and Miedel, M.C. and Traub, P. and Graves, M.C.},
    title = {Human immunodeficiency virus type 1 protease cleaves the intermediate filament proteins vimentin, desmin, and glial fibrillary acidic protein.},
    journal = {Proc. Natl. Acad. Sci. USA.},
    volume = {87},
    number = {16},
    pages = {6336--40},
    year = {1990},
    doi = {10.1073/pnas.87.16.6336}
}

@article{tomasselli,
    author = {Tomasselli, A.G. and Hui, J.O. and Adams, L. and Chosay, J. and Lowery, D. and Greenberg, B. and Yem, A. and Deibel, M.R. and Zürcher-Neely, H. and Heinrikson, R.L.},
    title = {Actin, troponin C, Alzheimer amyloid precursor protein and pro-interleukin 1 beta as substrates of the protease from human immunodeficiency virus.},
    journal = {J. Biol. Chem.},
    volume = {266},
    number = {22},
    pages = {14548--53},
    year = {1991},
    pmid = {1907279}
}

@article{nieminen,
    author = {Nieminen, M. and Henttinen, T. and Merinen, M. and Marttila-Ichihara, F. and Eriksson, J.E. and Jalkanen, S.},
    title = {Vimentin function in lymphocyte adhesion and transcellular migration.},
    journal = {Nature Cell Biology.},
    volume = {8},
    pages = {156--162},
    year = {2006},
    doi = {10.1038/ncb1355}
}

@article{billadeau,
    author = {Billadeau,D. and Nolz, J. and Gomez, T.},
    title = {Regulation of T-cell activation by the cytoskeleton.},
    journal = {Nature | Reviews Immunology.},
    volume = {7},
    pages = {131--143},
    year = {2007},
    doi = {10.1038/nri2021}
}

@article{parajo,
    author = {Garcia-Parajo, M.F. and Cambi, A. and Torreno-Pina, J.A. and Thompson, N. and Jacobson, K.},
    title = {Nanoclustering as a dominant feature of plasma membrane organization.},
    journal = {Journal of Cell Science.},
    volume = {127},
    pages = {4995--5005},
    year = {2014},
    doi = {10.1242/jcs.146340}
}

@article{renyao,
    author = {Ren, Z. and Yao, Q. and Chen, C.},
    title = {HIV-1 envelope glycoprotein 120 increases intercellular adhesion molecule-1 expression by human endothelial cells.},
    journal = {Laboratory Investigation.},
    volume = {82},
    number = {2},
    pages = {245--255},
    year = {2002},
    doi = {10.1038/labinvest.3780418}
}

@article{tardif,
    author = {Tardif, M.R. and Tremblay, M.J.},
    title = {Presence of host ICAM-1 in human immunodeficiency virus type 1 virions increases productive infection of $CD4^{+}$ T lymphocytes by favoring cytosolic delivery of viral material.},
    journal = {Journal of Virology.},
    volume = {77},
    number = {22},
    pages = {12299--12309},
    year = {2003},
    doi = {10.1128/JVI.77.22.1229912309.2003}
}

@article{vasiliver,
    author = {Vasiliver-Shamis, G. and Cho, M.W. and Hioe, C.E. and Dustin, M.L.},
    title = {Human immunodeficiency virus type 1 envelope gp120-induced partial T-cell receptor signaling creates an F-actin-depleted zone in the virological synapse.},
    journal = {Journal of Virology.},
    volume = {83},
    number = {21},
    pages = {11341--55},
    year = {2009},
    doi = {10.1128/JVI.01440-09}
}

@article{niebren,
    author = {Nie, Z. and Bren, G.D. and Vlahakis, S.R. and Algeciras Schimnich, A. and Brenchley, J.M. and Trushin, S.A. and Warren, S. and Schnepple, D.J. and Kovacs, C.M. and Loutfy, M.R. and Douek, D.C. and Badley, A.D.},
    title = {Human immunodefficiency virus type 1 protease cleaves procaspase 8 in vivo.},
    journal = {Journal of Virology.},
    volume = {81},
    number = {13},
    pages = {6947--6956},
    year = {2007},
    doi = {10.1128/JVI.022798-06}
}

@article{leibowitz,
    author = {Leibowitz, B. and Yu, J.},
    title = {Mitochondrial signaling in cell death via the Bcl-2 family.},
    journal = {Cancer Biology \& Therapy.},
    volume = {9},
    number = {6},
    pages = {417--422},
    year = {2010},
    doi = {10.4161/cbt9.6.11392}
}

@article{martin,
    author = {Martin, N. and Welsch, S. and Jolly, C. and Briggs, J.A.G. and Vaux, D. and Sattentau, Q.J.},
    title = {Virological synapse-mediated spread of human immunodefficiency virus type 1 between T cells is sensitive to entry inhibition. },
    journal = {Journal of Virology.},
    volume = {84},
    number = {7},
    pages = {3516--3527},
    year = {2010},
    doi = {10.1128/JVI.02651-09}
}

@article{ashkenazi,
    author = {Ashkenazi, A. and Faingold, O. and Shai, Y.},
    title = {HIV-1 fusion protein exerts complex immunosuppressive effects.},
    journal = {Trends in biochemical sciences.},
    volume = {38},
    number = {7},
    pages = {345--349},
    year = {2013},
    doi = {10.1016/j.tibs.2013.04.003}
}

@article{filarsky,
    author = {Filarsky, M. and Zillner, K. and Araya, I. and Villar-Garea, A. and Merkl, R. and Längst, G. and Németh, A.},
    title = {The extended AT-hook is a novel RNA binding motif.},
    journal = {RNA Biology.},
    volume = {12},
    number = {8},
    pages = {864--876},
    year = {2015},
    doi = {10.1080/15476286.2015.1060394}
}

@article{kogan,
    author = {Kogan, M. and Rappaport, J.},
    title = {HIV-1 accessory protein Vpr: relevance in the pathogenesis of HIV and potential for therapeutic intervention.},
    journal = {Retrovirology.},
    volume = {8},
    number = {25},
    pages = {1--20},
    year = {2011},
    doi = {10.1186/1742-4690-8-25}
}

@article{gonzalez,
    author = {González, M.E.},
    title = {The HIV-1 Vpr protein: a multifaceted target for therapeutic intervention.},
    journal = {Int. J. Mol. Sci.},
    volume = {18},
    number = {1},
    pages = {126},
    year = {2017},
    doi = {10.3390/ijms18010126}
}

@article{zhaowang,
    author = {Zhao, L.J. and Wang, L. and Mukherjee, S. and Narayan, O.},
    title = {Biochemical mechanism of HIV-1 Vpr function. Oligomerization mediated by the N-terminal domain.},
    journal = {J. Biol. Chem.},
    volume = {269},
    number = {51},
    pages = {32131--7},
    year = {1994},
    pmid = {7798208}
}

@article{venkatachari,
    author = {Venkatachari, N.J. and Walker, L.A. and Tastan, O. and Le, T. and Dempsey, T.M. and Li, Y. and Yanamala, N. and Srinivasan, A. and Klein-Seetharaman, J. and Montelaro, R.C. and Ayyavoo, V.},
    title = {Human immunodefficiency virus type 1 Vpr: oligomerization is an essential feature for its incorporation into virus particles.},
    journal = {Virology Journal.},
    volume = {7},
    number = {119},
    pages = {1--11},
    year = {2010},
    doi = {10.1186/1743-422X-7-119}
}

@article{tsurutani,
    author = {Tsurutani, N. and Yasuda, J. and Yamamoto, N. and Choi, B.I. and Kadoki, M. and Ikawura, Y.},
    title = {Nuclear import of the preintegration complex is blocked upon infection by human immunodefficiency virus type 1 in mouse cells.},
    journal = {Journal of Virology.},
    volume = {81},
    number = {2},
    pages = {677--688},
    year = {2007},
    doi = {10.1128/JVI.00870-06}
}

@article{mahalingam,
    author = {Mahalingam, S. and Ayyavoo, V. and Patel, M. and Kieber-Emmons, T. and Weiner, D.B.},
    title = {Nuclear import, virion incorporation, and cell cycle arrest/dfifferentiation are mediated by distinct functional domains of human immunodefficiency virus type 1 vpr.},
    journal = {Journal of Virology.},
    volume = {71},
    number = {9},
    pages = {6339--47},
    year = {1997},
    doi = {10.1128/JVI.71.9.6339-6347.1997}
}

@article{yingwu,
    author = {Wu, Y. and Zhou, X. and O'Barnes, C. and DeLucia, M. and Cohen, A.E. and Gronenborn, A.M. and Ahn, J. and Calero, G.},
    title = {The DDB1-DCAF1-Vpr-UNG2 crystal structure reveals how HI-1 Vpr steers human UNG2 toward destruction.},
    journal = {Nature Structural \& Molecular Biology.},
    volume = {23},
    pages = {933--940},
    year = {2016},
    doi = {10.1038/nsmb.3284}
}

@article{schabla,
    author = {Schabla, N.M. and Mondal, K. and Swanson, P.C.},
    title = {DCAF1 (VprBP): emerging physiological roles for a unique dual-service E3 ubiquitin ligase substrate receptor.},
    journal = {Journal of Molecular Cell Biology.},
    volume = {11},
    number = {9},
    pages = {725--735},
    year = {2019},
    doi = {10.1093/jmcb/mjy085}
}

@article{haber,
    author = {Haber, J.E. and Hayer, W.-D.},
    title = {The fuss about Mus81.},
    journal = {Cell.},
    volume = {107},
    number = {5},
    pages = {551--554},
    year = {2001},
    doi = {10.1016/S0092-8674(01)00593-1}
}

@article{pu,
    author = {Pu, J. and Wang, Q. and Xu, W. and Lu, L. and Jiang, S.},
    title = {Development of Protein- and Peptide-Based HIV Entry Inhibitors Targeting gp120 or gp41.},
    journal = {Viruses.},
    volume = {11},
    number = {8},
    pages = {705},
    year = {2019},
    doi = {10.3390/v11080705}
}

@article{watase,
    author = {Watase, G. and Takisawa, H. and Kanemaki, M.T.},
    title = {Mcm110 plays a role in functioning of the eukaryotic replicative DNA helicase, Cdc45-Mcm-GINS.},
    journal = {Current Biology.},
    volume = {22},
    pages = {343--349},
    year = {2012},
    doi = {10.1016/j.cub.2012.01.023}
}

@article{brosh,
    author = {Brosh Jr. R.M. and Matson, S.W.},
    title = {History of DNA Helicases.},
    journal = {Genes},
    volume = {11},
    number = {255},
    pages = {1--45},
    year = {2020},
    doi = {10.3390/genes11030255}
}

@article{romani,
    author = {Romani, B. and Baygloo, N.S. and Hamidi-Fard, M. and Aghasadeghi, M.R. and Allahbakhshi, E.},
    title = {HIV-1 Vpr protein induces proteasomal degradation of chromatin-associated class I HDACs to overcome latent infection of macrophages.},
    journal = {J. Biol. Chem.},
    volume = {291},
    number = {6},
    pages = {2696--711},
    year = {2016},
    doi = {10.1074/jbc.M115.689018}
}

@article{godet,
    author = {Godet, J. and Mély, Y.},
    title = {Biophysical studies of the nucleic acid chaperone properties of the HIV-1 nucleocapsid protein.},
    journal = {RNA Biol.},
    volume = {7},
    number = {6},
    pages = {687--99},
    year = {2010},
    doi = {10.4161/rna.7.6.13616}
}

@article{klinger,
    author = {Klingler, J. and Anton, H. and Réal, E. and Zeiger, M. and Moog, C. and Mély, Y. and Boutant, E.},
    title = {How HIV-1 Gag Manipulates Its Host Cell Proteins: A Focus on Interactors of the Nucleocapsid Domain.},
    journal = {Viruses.},
    volume = {12},
    number = {8},
    pages = {888},
    year = {2020},
    doi = {10.3390/v12080888}
}

@article{levin,
    author = {Levin, J.G. and Mitra, M. and Mascarenhas, A. and Musier-Forsyth, K.},
    title = {Role of HIV-1 nucleocapsid protein in HIV-1 reverse transcription.},
    journal = {RNA Biol.},
    volume = {7},
    number = {6},
    pages = {754--774},
    year = {2010},
    doi = {10.4161/rna.7.6.14115}
}

@article{turpin,
    author = {Turpin, J.A.},
    title = {The next generation of HIV/AIDS drugs: novel and developmental antiHIV drugs and targets.},
    journal = {Expert Review of Anti-infective Therapy.},
    volume = {1},
    number = {1},
    pages = {97--128},
    year = {2003},
    doi = {10.1586/14787210.1.1.97}
}

@article{sancineto,
    author = {Sancineto, L. and Iraci, N. and Tabarrini, O. and Santi, C.},
    title = {NCp7: targeting a multitasking protein for next-generation anti-HIV drug development part 1: covalent inhibitors.},
    journal = {Drug Discov. Today.},
    volume = {23},
    number = {2},
    pages = {260--71},
    year = {2018},
    doi = {10.1016/j.drudis.2017.10.017}
}

@article{iraci,
    author = {Iraci, N. and Tabarrini, O. and Santi, C. and Sancineto, L.},
    title = {NCp7: targeting a multitask protein for next-generation anti-HIV drug development part 2. Noncovalent inhibitors and nucleic acid binders.},
    journal = {Drug Discov. Today.},
    volume = {23},
    number = {3},
    pages = {687--95},
    year = {2018},
    doi = {10.1016/j.drudis.2018.01.022}
}

@article{eugenia,
    author = {González, M.E.},
    title = {Vpu protein: the viroporin encoded by HIV-1.},
    journal = {Viruses.},
    volume = {7},
    pages = {4352--4368},
    year = {2015},
    doi = {10.3390/v7082824}
}

@article{dube,
    author = {Dubé, M. and Bego, M.G. and Pauqay, C. and Cohen, É.A.},
    title = {Modulation of HIV-1-host interaction: role of the Vpu accessory protein.},
    journal = {Retrovirology.},
    volume = {7},
    number = {114},
    pages = {1--19},
    year = {2010},
    doi = {10.1186/1742-4690-7-114}
}

@article{guatelli,
    author = {Ruiz, A. and Guatelli, J.C. and Stephens, E.B.},
    title = {The Vpu protein: new concepts in virus release and CD4 downmodulation.},
    journal = {Curr. HIV. Res.},
    volume = {8},
    number = {3},
    pages = {240--252},
    year = {2010},
    pmid = {20201792}
}

@article{pacini,
    author = {Roy, N. and Pacini, G. and Berlioz-Torrent, C. and Janvier, K.},
    title = {Mechanisms underlying HIV-1 Vpu-mediated viral egress.},
    journal = {Frontiers in Microbiology.},
    volume = {5},
    number = {177},
    pages = {1--9},
    year = {2014},
    doi = {10.3389/fmicb.2014.00177}
}

@article{sato,
    author = {Sato, Y. and Tsuchiya, H. and Yamagata, A. and Okatsu, K. and Tanaka, K. and Saeki, Y. and Fukai, S.},
    title = {Structural insights into ubiquitin recognition and Ufd1 interaction of Npl4.},
    journal = {Nature Communications.},
    volume = {10},
    number = {5708},
    pages = {1--13},
    year = {2019},
    doi = {10.1038/s41467-019-13697-y}
}

@online{vcp,
    author = {NIH},
    title = {VCP valosin containing protein [ Homo sapiens (human) ]},
    url  = {https://www.ncbi.nlm.nih.gov/gene/7415},
    addendum = {(accessed: 03.12.2020)},
}

@article{wildum,
    author = {Wildum, S. and Schindler, M. and Münch, J. and Kirchhoff, F.},
    title = {Contribution of Vpu, Env, and Nef to CD4 down-modulation and resistance of Human Immunodefficiency Virus Type 1-infected T cells to superinfection.},
    journal = {Journal of Virology.},
    volume = {80},
    number = {16},
    pages = {8047--8059},
    year = {2006},
    doi = {10.1128/JVI.00252-06}
}

@article{benson,
    author = {Benson, R.E. and Sanfridson, A. and Ottinger, J.S. and Doyle, C. and Cullen, B.R.},
    title = {Downregulation of cell-surface CD4 expression by simian immunodefficiency virus Nef prevents viral super infection.},
    journal = {J. Exp. Med.},
    volume = {177},
    number = {6},
    pages = {1561--1566},
    year = {1993},
    doi = {10.1084/jem.177.6.1561}
}

@article{ueno,
    author = {Tanaka, M. and Ueno, T. and Nakahara, T. and Sasaki, K. and Ishimoto, A. and Sakai, H.},
    title = {Downregulation of CD4 is required for maintenance of viral infectivity of HIV-1.},
    journal = {Virology.},
    volume = {311},
    pages = {316--325},
    year = {2003},
    doi = {10.1016/S0042-6822(03)00126-0}
}

@article{prevost,
    author = {Prévost, J. and Pickering, S. and Mumby, M.J. and Medjahed, H. and Gendron-Lepage, G. and Delgado, G.G. and Dirk, B.S. and Dikeakos, J.D. and Stürzel, C.M. and Sauter, D. and Kirchhoff, F. and Bibollet-Ruche, F. and Hahn, B.H. and Dubé, M. and Kaufmann, D.E. and Neil, S.J.D. and Finzi, A. and Richard, J.},
    title = {Upregulation of BST-2 by type I interferons reduces the capacity of vpu to protect HIV-1-infected cells from NK cell responses.},
    journal = {mBio.},
    volume = {10},
    number = {3},
    pages = {e01113--19},
    year = {2019},
    doi = {10.1128/mBio.01113-19}
}

@article{kuhl,
    author = {Kuhl, B.D. and Cheng, V. and Wainberg, M.A. and Liang, C.},
    title = {Tetherin and its viral antagonists.},
    journal = {J. Neuroimmune Pharmacol.},
    volume = {6},
    number = {2},
    pages = {188--201},
    year = {2011},
    doi = {10.1007/s11481-010-9256-1}
}

@online{kcnk3,
    author = {NIH},
    title = {KCNK3 potassium two pore domain channel subfamily K member 3 [ Homo sapiens (human) ]},
    url  = {https://www.ncbi.nlm.nih.gov/gene/3777},
    addendum = {(accessed: 03.12.2020)},
}

@article{geyer,
    author = {Geyer, M. and Fackler, O.T. and Peterlin, B.M.},
    title = {Structure-function relationships in HIV-1 Nef.},
    journal = {EMBO Reports.},
    volume = {2},
    number = {7},
    pages = {580--585},
    year = {2001},
    doi = {10.1093/embo-reports/kve141}
}

@article{xiaofei,
    author = {Jia, X. and Singh, R. and Homann, S. and Yang, H. and Guatelli, J. and Xiong, Y.},
    title = {Structural basis of evasion of cellular adaptive immunity by HIV-1 Nef.},
    journal = {Nature Structural \& Molecular Biology.},
    volume = {19},
    number = {7},
    pages = {701--708},
    year = {2012},
    doi = {10.1038/nsmb.2328}
}

@article{buffalo,
    author = {Buffalo, C.Z. and Iwamoto, Y. and Hurley, J.H. and Ren, X.},
    title = {How HIV Nef proteins hijack membrane traffic to promote infection.},
    journal = {Journal of Virology.},
    volume = {93},
    number = {24},
    pages = {1--18},
    year = {2019},
    doi = {10.1128/JVI.01322-19}
}

@online{ap1b1,
    author = {NIH},
    title = {AP1B1 adaptor related protein complex 1 subunit beta 1 [ Homo sapiens (human) ]},
    url  = {https://www.ncbi.nlm.nih.gov/gene/162},
    addendum = {(accessed: 03.12.2020)},
}

@online{cd28,
    author = {NIH},
    title = {CD28 CD28 molecule [ Homo sapiens (human) ]},
    url  = {https://www.ncbi.nlm.nih.gov/gene/940},
    addendum = {(accessed: 03.12.2020)},
}

@article{pugliese,
    author = {Pugliese, A. and Vidoot, V. and Beltramo, T. and Petrini, S. and Torre, D.},
    title = {A review of HIV-1 Tat protein biological effects.},
    journal = {Cell Biochemistry and Function.},
    volume = {23},
    pages = {223--227},
    year = {2005},
    doi = {10.1027/cbf.1147}
}

@article{sertznig,
    author = {Sertznig, H. and Hillebrand, F. and Erkelenz, S. and Schaal, H. and Widera, M.},
    title = {Behind the scenes of HIV-1 replication: alternative splicing as the dependency factor on the quiet.},
    journal = {Virology.},
    volume = {516},
    pages = {176--188},
    year = {2018},
    doi = {10.1016/j.virol.2018.01.011}
}

@article{joseph,
    author = {Joseph, A.M. and Ladha, J.S. and Mojamdar, M. and Mitra, D.},
    title = {Human immunodefficiency virus-1 nef protein interacts with Tat and enhances HIV-1 gene expression.},
    journal = {FEBS Letters.},
    volume = {548},
    number = {27462},
    pages = {37--42},
    year = {2003},
    doi = {10.1016/S0014-5793(03)00725-7}
}

@article{dandekar,
    author = {Dandekar, D.H. and Ganesh, K.N. and Mitra, D.},
    title = {HIV-1 Tat directly binds to NFKB enhancer sequence: role in viral and cellular gene expression.},
    journal = {Nucleic Acid Research.},
    volume = {32},
    number = {4},
    pages = {1270--1278},
    year = {2004},
    doi = {10.1093/nar/gkh289}
}

@article{clark,
    author = {Clark, E. and Nava, B. and Caputi, M.},
    title = {Tat is a multifunctional viral protein that modulates cellular gene expression and functions.},
    journal = {Oncotarget.},
    volume = {8},
    number = {16},
    pages = {27569--27581},
    year = {2017},
    doi = {10.18632/oncotarget.15174}
}

@article{pollard,
    author = {Pollard, V.W. and Malim, M.H.},
    title = {The HIV-1 Rev protein.},
    journal = {Annu. Rev. Microbiol.},
    volume = {52},
    pages = {491--532},
    year = {1998},
    doi = {10.1146/annurev.micro.52.1.491}
}

@article{meggio,
    author = {Meggio, F. and D'Agostino, D.M. and Ciminale, V. and Chiecco-Bianchi, L. and Pinna, L.A.},
    title = {Phosphorylation of HIV-1 Rev protein: implication of protein kinase CK2 and pro-directed kinases.},
    journal = {Biochemical and biophysical research communications.},
    volume = {226},
    number = {1392},
    pages = {547--554},
    year = {1996},
    doi = {10.1006/bbrc.1996.1392}
}

@article{wingfield,
    author = {Wingfield, P.T. and Stahl, S.J. and Payton, M.A. and Venkatesan, S. and Misra, M. and Steven, A.C.},
    title = {HIV-1 Rev expressed in recombinant \textit{Escherichia coli}: purification, polymerization, and conformational properties.},
    journal = {Biochemistry.},
    volume = {30},
    number = {7527},
    pages = {7527--7534},
    year = {1991},
    doi = {10.1021/bi00244a023}
}

@article{fernandes,
    author = {Fernandes, J. and Jayaraman, B. and Frankel, A.},
    title = {The HIV-1 Rev response element. An RNA scaffold that directs the cooperative assembly of a homo-oligomeric ribonucleoprotein complex.},
    journal = {RNA Biology.},
    volume = {9},
    number = {1},
    pages = {6--11},
    year = {2012},
    doi = {10.4161/rna.9.1.18178}
}

@article{behrens,
    author = {Behrens, R.T. and Aligeti, M. and Pocock, G.M. and Higgins, C.A. and Sherer, N.M.},
    title = {Nuclear export signal masking regulates HIV-1 Rev trafficking and viral RNA nuclear export.},
    journal = {Journal of Virology.},
    volume = {91},
    number = {3},
    pages = {e02107--16},
    year = {2017},
    doi = {10.1128/JVI.02107-16}
}

@article{hutten,
    author = {Hutten, S. and Kehlenbach, R.H.},
    title = {CRM1-mediated nuclear export: to the pore and beyond.},
    journal = {Cell.},
    volume = {17},
    number = {4},
    pages = {193--202},
    year = {2007},
    doi = {10.1016/j.tcb.2007.02.003}
}

@article{fica,
    author = {Fica, S.M. and Tuttle, N. and Novak, T. and Li, N.-S. and Lu, J. and Koodathingal, P. and Dai, Q. and Staley, J.P. and Piccirilli, J.A.},
    title = {RNA catalyses nuclear pre-mRNA splicing},
    journal = {Nature},
    volume = {0},
    pages = {1--18},
    year = {2013},
    doi = {10.1038/nature12734}
}

@article{dick,
    author = {Dick, A. and Cocklin, S.},
    title = {Recent Advances in HIV-1 Gag Inhibitor Design and Development.},
    journal = {Molecules.},
    volume = {25},
    number = {7},
    pages = {1687},
    year = {2020},
    doi = {10.3390/molecules25071687}
}

@article{casolari,
    author = {Casolari, J.M. and Silver, P.A.},
    title = {Guardian at the gate: preventing unspliced pre-mRNA export.},
    journal = {Trends in Cell Biology.},
    volume = {14},
    number = {5},
    pages = {222--226},
    year = {2004},
    doi = {10.1016/j.tcb.2004.03.007}
}

@article{talhouarne,
    author = {Talhouarne, G.J.S. and Gall, J.G.},
    title = {Lariat intronic RNAs in the cytoplasm of vertebrate cells.},
    journal = {Proc. Natl. Acad. Sci. USA.},
    volume = {115},
    number = {34},
    pages = {E7970--E7977},
    year = {2018},
    doi = {10.1073/pnas.1808816115}
}

@article{buckley,
    author = {Buckley, P.T. and Khaladkar, M. and Kim, J. and Eberwine, J.},
    title = {Cytoplasmic intron retention, function, splicing, and the sentil RNA hypothesis.},
    journal = {Wiley Interdiscip. Rev. RNA.},
    volume = {5},
    number = {2},
    pages = {223--230},
    year = {2014},
    doi = {10.1002/wrna.1203}
}

@article{favaro,
    author = {Favaro, J.P. and Borg, K.T. and Arrigo, S.J. and Schmidt, M.G.},
    title = {Effect of Rev on the intranuclear localization of HIV-1 unspliced RNA.},
    journal = {Virology.},
    volume = {249},
    number = {VY989312},
    pages = {286--296},
    year = {1998},
    doi = {10.1006/viro.1998.9312}
}

@article{kozak,
    author = {Rose, K.M. and Marin, M. and Kozak, S.L. and Kabat, D.},
    title = {The viral infectivity factor (Vif) of HIV-1 unveiled.},
    journal = {Trends in Molecular Medicine.},
    volume = {10},
    number = {6},
    pages = {291--298},
    year = {2004},
    doi = {10.1016/j.molmed.2004.04.008}
}

@article{goncalves,
    author = {Yang, X. and Goncalves, J. and Gabuzda, D.},
    title = {Phosphorylation of Vif and its role in HIV-1 replication.},
    journal = {The Journal of Biological Chemistry.},
    volume = {271},
    number = {17},
    pages = {10121--10129},
    year = {1996},
    doi = {10.1074/jbc.271.17.10121}
}

@article{koning,
    author = {Koning, F.A. and Newman, E.N.C. and Kim, E.-Y. and Kunstman, K.J. and Wolinsky, S.M. and Malim, M.H.},
    title = {Defining APOBEC3 expression patterns in human tissues and hematopoietic cell subsets.},
    journal = {Journal of Virology.},
    volume = {83},
    number = {18},
    pages = {9474--9485},
    year = {2009},
    doi = {10.1128/JVI.01089-09}
}

@article{covino,
    author = {Covino, D.A. and Gauzzi, M.C. and Fantuzzi, L.},
    title = {Understanding the regulation of APOBEC3 expression: current evidence and much to learn.},
    journal = {Journal of Leukocyte Biology.},
    volume = {103},
    number = {3},
    pages = {433--444},
    year = {2018},
    doi = {10.1002/JLB.2MR0717-310R}
}

@article{iwatani,
    author = {Okada, A. and Iwatani, Y.},
    title = {APOBEC3G-mediated G-to-A hypermutation of the HIV-1 genome: the missing link in antiviral molecular mechanisms.},
    journal = {Frontiers in Microbiology.},
    volume = {7},
    number = {2027},
    pages = {1--8},
    year = {2016},
    doi = {10.3389/fmicb.2016.02027}
}

@article{xianghui,
    author = {Yu, X. and Yu, Y. and Liu, B. and Luo, K. and Kong, W. and Mao, P. and Yu, X.-F.},
    title = {Induction of APOBEC3G ubiquitination and degradation by an HIV-1 Vif-Cul5-SCF complex.},
    journal = {Science.},
    volume = {302},
    number = {5647},
    pages = {1056--60},
    year = {2003},
    doi = {10.1126/science.1089591}
}

@article{buckman,
    author = {Buckman, J.S. and Bosche, W.J. and Gorelick, R.J.},
    title = {Human immunodeficiency virus type 1 nucleocapsid zn(2+) fingers are required for efficient reverse transcription, initial integration processes, and protection of newly synthesized viral DNA.},
    journal = {Journal of Virology.},
    volume = {77},
    number = {2},
    pages = {1469--80},
    year = {2003},
    doi = {10.1128/jvi.77.2.1469-1480.2003}
}

@article{langer,
    author = {Langer, S. and Sauter, D.},
    title = {Unusual fusion proteins of HIV-1.},
    journal = {Frontiers in Microbiology.},
    volume = {7},
    pages = {2152},
    year = {2017},
    doi = {10.3389/fmicb.2016.02152}
}

@article{sharma,
    author = {Sharma, S. and Arunachalam, P.S. and Menon, M. and Ragupathy, V. and Satya, R.V. and Jebaraj, J. and Aralaguppe, S.G. and Rao, C. and Pal, S. and Saravanan, S. and Murugavel, K.G. and Balakrishnan, P. and Solomon, S. and Hewlett, I. and Ranga, U.},
    title = {PTAP motif duplication in the p6 Gag protein confers a replication advantage on HIV-1 subtype C.},
    journal = {The Journal of biological chemistry.},
    volume = {293},
    number = {30},
    pages = {11687--11708},
    year = {2018},
    doi = {10.1074/jbc.M117.815829}
}

@article{vanDomselaar,
    author = {van Domselaar, R. and Njenda, D.T. and Rao, R. and Sönnerborg, A. and Singh, K. and Neogi, U.},
    title = {HIV-1 Subtype C with PYxE Insertion Has Enhanced Binding of Gag-p6 to Host Cell Protein ALIX and Increased Replication Fitness.},
    journal = {Journal of Virology.},
    volume = {93},
    number = {9},
    pages = {e00077-19},
    year = {2019},
    doi = {10.1128/JVI.00077-19}
}

@article{spearman,
    author = {Spearman, P.},
    title = {HIV-1 Gag as an Antiviral Target: Development of Assembly and Maturation Inhibitors.},
    journal = {Current topics in medicinal chemistry.},
    volume = {16},
    number = {10},
    pages = {1154--66},
    year = {2016},
    doi = {10.2174/1568026615666150902102143}
}

@article{scarsi,
    author = {Scarsi, K.K. and Havens, J.P. and Podany, A.T. and Avedissian, S.N. and Fletcher, C.V.},
    title = {HIV-1 integrase inhibitors: a comparative review of efficacy and safety.},
    journal = {Drugs.},
    volume = {80},
    pages = {1649--1676},
    year = {2020},
    doi = {10.1007/s40265-020-01379-9}
}

@article{deeks,
    author = {Deeks, E.D.},
    title = {Bictegravir/Emtricitabine/Tenofovir alafenamide: a review in HIV-1 infection.},
    journal = {Drugs.},
    volume = {78},
    pages = {1817--1828},
    year = {2018},
    doi = {10.1007/s40625-018-1010-7}
}

@article{morales-ramirez,
    author = {Morales-Ramirez, J. and Bogner, J.R. and Molina, J.M. and Lombaard, J. and Dicker, I.B. and Stock, D.A. and DeGrosky, M. and Gartland, M. and Pene Dumitrescu, T. and Min, S. and Llamoso, C. and Joshi, S.R. and Lataillade, M.},
    title = {Safety, efficacy, and dose response of the maturation inhibitor GSK3532795 (formerly known as BMS-955176) plus tenofovir/emtricitabine once daily in treatment-naive HIV-1-infected adults: Week 24 primary analysis from a randomized Phase IIb trial.},
    journal = {PloS One.},
    volume = {13},
    number = {10},
    pages = {e0205368},
    year = {2018},
    doi = {10.1371/journal.pone.0205368}
}

@article{tsiang,
    author = {Tsiang, M. and Jones, G.S. and Goldsmith, J. and Mulato, A. and Hansen, D. and Kan, E. and Tsai, L. and Bam, R.A. and Stepan, G. and Stray, K.M. and Niedziela-Majka, A. and Yant, S.R. and Yu, H. and Kukolj, G. and Cihlar, T. and Lazerwith, S.E. and White, K.L. and Jin, H.},
    title = {Antiviral activity of bictegravir (GS-9883), a novel potent HIV-1 integrase strand transfer inhibitor with an improved resistance profile.},
    journal = {Antimicrobial agents and Chemotherapy.},
    volume = {60},
    number = {12},
    pages = {7086--8000},
    year = {2016},
    doi = {10.1128/AAC.01474-16}
}

@article{neogi,
    author = {Neogi, U. and Singh, K. and Aralaguppe, S.G. and Rogers, L.C. and Njenda, D.T. and Sarafianos, S.G. and Hejdman, B. and Söonerborg, A.},
    title = {\textit{Ex vivo} antiretroviral potency of newer integrase strand transfer inhibitors cabotegravir and bictegravir in HIV-1 non-B subtypes.},
    journal = {AIDS},
    volume = {32},
    number = {4},
    pages = {469--476},
    year = {2018},
    doi = {10.1097/QAD.0000000000001726}
}

@article{nwetwin,
    author = {Win, N.N. and Ngwe, H. and Abe, I. and Morita, H.},
    title = {Naturally occurring Vpr inhibitors from medicinal plants of Myanmar.},
    journal = {J. Nat. Med.},
    volume = {71},
    number = {4},
    pages = {579--589},
    year = {2017},
    doi = {10.1007/s1148-017-1104-7}
}

@article{choi,
    author = {Choi, S.B. and Choong, Y.S. and Saito, A. and Wahab, H.A. and Najimudin, N. and Watanabe, N. and Osada, H. and Ong, E.B.B.},
    title = {In silico investigation of a HIV-1 Vpr inhibitor binding site: potential for virtual screening and anti-HIV drug design.},
    journal = {Mol. Inform.},
    volume = {33},
    number = {11--12},
    pages = {742--748},
    year = {2014},
    doi = {10.1002/minf.201400080}
}

@article{hagiwara,
    author = {Hagiwara, K. and Ishii, H. and Murakami, T. and Takeshima, S-n. and Chutiwitoonchai, N. and Kodama, E.N.},
    title = {Synthesis of a Vpr-binding derivative for use as a novel HIV-1 inhibitor.},
    journal = {PLoS One.},
    volume = {10},
    number = {12},
    pages = {e0145573},
    year = {2015},
    doi = {10.1371/journal.pone.0145573}
}

@article{kuhl2,
    author = {Kuhl, B.D. and Cheng, V. and Donahue, D.A. and Sloan, R.D. and Liang, C. and Wilkinson, J. and Wainberg, M.A.},
    title = {The HIV-1 Vpu viroporin inhibitor BIT225 does not affect vpu-mediated tetherin antagonism.},
    journal = {PLoS One.},
    volume = {6},
    number = {11},
    pages = {e27660},
    year = {2011},
    doi = {10.1371/journal.pone.0027660}
}

@article{mwimanzi,
    author = {Mwimanzi, P. and Tietjen, I. and Miller, S.C. and Shahid, A. and Cobarrubias, K. and Kinloch, N.N. and Baraki, B. and Richard, J. and Finzi, A. and Fedida, D. and Brumme, Z.L. and Brockman, M.A.},
    title = {Novel acylguanidine-based inhibitor of HIV-1.},
    journal = {Journal of Virology.},
    volume = {90},
    number = {20},
    pages = {9495--9509},
    year = {2016},
    doi = {10.1128/JVI.01107-16}
}

@article{langarizadeh,
    author = {Langarizadeh, M.A. and Abiri, A. and Ghasemshirazi, S. and Foroutan, N. and Khodadadi, A. and Faghih-Mirzaei, E.},
    title = {Phlorotannins as HIV Vpu inhibitors, an \textit{in silico} virtual screening study of marine natural products.},
    journal = {Biotechnology and Applied Biochemistry.},
    year = {2020},
    doi = {10.1002/bab.2014}
}

@article{venkatesan,
    author = {Venkatesan, J. and Keekan, K.K. and Anil, S. and Bhatnagar, I. and Kim, S.-K.},
    title = {Phlorotannins.},
    journal = {Encyclopedia of Food Chemistry.},
    pages = {515--527},
    year = {2019},
    doi = {10.1016/B978-0-08-100596-5.22360-3}
}

@article{dikeakos,
    author = {Dikeakos, J.D. and Atkins, K.M. and Thomas, L. and Emert-Sedlak, L. and Byeon, I.-J.L. and Jung, J. and Ahn, J. and Wortman, M.D. and Kukull, B. and Saito, M. and Koizumi, H. and Williamson, D.M. and Hiyoshi, M. and Barklis, E. and Takiguchi, M. and Suzu, S. and Gronenborn, A.M. and Smithgall, T.E. and Thomas, G.},
    title = {Small molecule inhibition of HIV-1 induced MHC-1 down-regulation identifies a temporally regulated switch in Nef action.},
    journal = {Molecular Biology of the Cell.},
    volume =  {21},
    pages = {3279--3292},
    year = {2010},
    doi = {10.1091/mbc.E10-05-0470}
}

@article{dekaban,
    author = {Dekaban, G.A. and Dikeakos, J.D.},
    title = {HIV-1 Nef inhibitors: a novel class of HIV-specific immune adjuvants in support of a cure.},
    journal = {AIDS Res. Ther.},
    volume =  {14},
    number = {53},
    pages = {1--4},
    year = {2017},
    doi = {10.1186/s12981-017-0175-6}
}

@article{emert,
    author = {Emert-Sedlak, L. and Kodama, T. and Lerner, E.C. and Dai, W. and Foster, C. and Day, B.W. and Lazo, J.S. and Smithgall, T.E.},
    title = {Chemical library screens targeting an HIV-1 accessory factor/host cell kinase complex identify novel antiretroviral compounds.},
    journal = {ACS Chem. Biol.},
    volume =  {4},
    number = {11},
    pages = {939--947},
    year = {2009},
    doi = {10.1021/cb900195c}
}

@article{richter,
    author = {Richter, S.N. and Palù, G.},
    title = {Inhibitors of HIV-1 Tat-mediated transactivation.},
    journal = {Curr. Med. Chem.},
    volume =  {13},
    number = {11},
    pages = {1305--15},
    year = {2006},
    doi = {10.2174/092986706776872989}
}

@article{daelemans,
    author = {Daelemans, D. and Afonina, E. and Nilsson, J. and Werner, G. and Kjems, J. and De Clercq, E. and Pavlakis, G.N. and Vandamme, A.-M.},
    title = {A synthetic HIV-1 Rev inhibitor interfering with the CRM1-mediated nuclear export.},
    journal = {Proc. Natl. Acad. Sci. USA.},
    volume =  {99},
    number = {22},
    pages = {14440--5},
    year = {2002},
    doi = {10.1073/pnas.212285299}
}

@article{balachandran,
    author = {Balachandran, A. and Wong, R. and Stoilov, P. and Pan, S. and Blencowe, B. and Cheung, P. and Harrigan, P.R. and Cochrane, A.},
    title = {Identification of small molecule modulators of HIV-1 Tat and Rev protein accumulation.},
    journal = {Retrovirology.},
    volume =  {14},
    number = {7},
    year = {2017},
    doi = {10.1186/s12977-017-0330-0}
}

@article{nathans,
    author = {Nathans, R. and Cao, H. and Sharova, N. and Ali, A. and Sharkey, M. and Stranska, R. and Stevenson, M. and Rana, T.M.},
    title = {Small-molecule inhibition of HIV-1 Vif.},
    journal = {Nat. Biotechnol.},
    volume =  {26},
    number = {10},
    pages = {1187--1192},
    year = {2009},
    doi = {10.1038/nbt.1496}
}

@article{mazhang,
    author = {Ma, L. and Zhang, Z. and Liu, Z. and et al.},
    title = {Identification of small molecule compounds targeting the interaction of HIV-1 Vif and human APOBEC3G by virtual screening and biological evaluation.},
    journal = {Sci. Rep.},
    volume =  {8},
    number = {8067},
    pages = {1--12},
    year = {2018},
    doi = {10.1038/s41598-018-26318-3}
}

@article{pery,
    author = {Pery, E. and Sheehy, A. and Nebane, N.M. and Misra, V. and Mankowski, M.K. and Rasmussen, L. and White, E.L. and Ptak, R.G. and Gabuzda, D.},
    title = {Redoxal, an inhibitor of de novo pyrimidine biosynthesis, augments APOBEC3G antiviral activity against human immunodeficiency virus type 1.},
    journal = {Virology.},
    volume =  {484},
    pages = {276--287},
    year = {2015},
    doi = {10.1016/j.virol.2015.06.014}
}

@article{ghosh,
    author = {Ghosh, A.K. and Osswald, H.L. and Prato, G.},
    title = {Recent Progress in the Development of HIV-1 Protease Inhibitors for the Treatment of HIV/AIDS.},
    journal = {J. Med. Chem.},
    volume =  {59},
    number = {11},
    pages = {5172--5208},
    year = {2016},
    doi = {10.1021/acs.jmedchem.5b01697}
}

@article{kumar,
    author = {Kumar, G.N. and Rodrigues, A.D. and Buko, A.M. and Denissen, J.F.},
    title = {Cytochrome P450-mediated metabolism of the HIV-1 protease inhibitor ritonavir (ABT-538) in human liver microsomes.},
    journal = {J. Pharmacol. Exp. Ther.},
    volume =  {277},
    pages = {423--31},
    year = {1996},
    PMID = {8613951}
}

@article{mitsuya,
    author = {Mitsuya, H. and Maeda, K. and Das, D. and Ghosh, A.K.},
    title = {Development of protease inhibitors and the fight with drug-resistant HIV-1 variants.},
    journal = {Adv. Pharmacol.},
    volume =  {56},
    pages = {169--197},
    year = {2008},
    doi = {10.1016/S1054-3589(07)56006-0}
}

@article{perno,
    author = {Perno, C.F. and Cozzi-Lepri, A. and Balotta, C. and Forbici, F. and Violin, M. and Bertoli, A. and Facchi, G. and Pezzotti, P. and Cadeo, G. and Tositti, G. and Pasquinucci, S. and Pauluzzi, S. and Scalzini, A. and Salassa, B. and Vincenti, A. and Phillips, A.N. and Dianzani, F. and Appice, A. and Angarano, G. and Monno, L. and Ippolito, G. and Moroni, M. and d' Arminio Monforte, A. and Italian Cohort Naive Antiretroviral (I.CO.N.A.)},
    title = {Secondary mutations in the protease region of human immunodeficiency virus and virologic failure in drug-naive patients treated with protease inhibitor-based therapy.},
    journal = {J. Infect. Dis.},
    volume =  {184},
    number = {8},
    pages = {983--91},
    year = {2001},
    doi = {10.1086/323604}
}

@article{ghosh2,
    author = {Ghosh, A.K. and Anderson, D.D. and Weber, I.T. and Mitsuya, H.},
    title = {Enhancing protein backbone binding--a fruitful concept for combating drug-resistant HIV.},
    journal = {Angew. Chem. Int. Ed. Engl.},
    volume =  {51},
    number = {8},
    pages = {1778--1802},
    year = {2012},
    doi = {10.1002/anie.201102762}
}

@article{wu,
    author = {Xu, L. and Liu, H. and Murray, B.P. and Callebaut, C. and Lee, M.S. and Hong, A. and Strickley, R.G. and Tsai, L.K. and Stray, K.M. and Wang, Y. and Rhodes, G.R. and Desai, M.C.},
    title = {Cobicistat (GS-9350): A Potent and Selective Inhibitor of Human CYP3A as a Novel Pharmacoenhancer.},
    journal = {ACS. Med. Chem. Lett.},
    volume =  {1},
    number = {5},
    pages = {209--13},
    year = {2010},
    doi = {10.1021/ml1000257}
}

@article{rai,
    author = {Rai, M.A. and Pannek, S. and Fichtenbaum, C.J.},
    title = {Emerging reverse transcriptase inhibitors for HIV-1 infection.},
    journal = {Expert Opin. Emerg. Drugs.},
    volume =  {23},
    number = {2},
    pages = {149--57},
    year = {2018},
    doi = {10.1080/14728214.2018.1474202}
}

@inbook{marcelin,
    author = "Marcelin, A.G.",
    title = "Resistance to nucleoside reverse transcriptase inhibitors.",
    editor = "Geretti, A.M.",
    booktitle = "Antiretroviral Resistance in Clinical Practice.",
    publisher = "London: Mediscript",
    year = "2006",
    chapter = "1",
    PMID = "21249766"
}

@article{asahchop,
    author = {Asahchop, E.L. and Wainberg, M.A. and Sloan, R.D. and Tremblay, C.L.},
    title = {Antiviral drug resistance and the need for development of new HIV-1 reverse transcriptase inhibitors.},
    journal = {Antimicrob. Agents. Chemother.},
    volume =  {56},
    number = {10},
    pages = {5000--08},
    year = {2012},
    doi = {10.1128/AAC.00591-12}
}

@article{adams,
    author = {Adams, J. and Patel, N. and Mankaryous, N. and Tadros, M. and Miller, C.D.},
    title = {Nonnucleoside reverse transcriptase inhibitor resistance and the role of the second-generation agents.},
    journal = {Ann. Pharmacother.},
    volume =  {44},
    number = {1},
    pages = {157--65},
    year = {2010},
    doi = {10.1345/aph.1M359}
}

@article{das,
    author = {Das, K. and Sarafianos, S.G. and Clark, A.D. and Boyer, P.L. and Hughes, S.H. and Arnold, E.},
    title = {Crystal structures of clinically relevant Lys103Asn/Tyr181Cys double mutant HIV-1 reverse transcriptase in complexes with ATP and non-nucleoside inhibitor HBY 097.},
    journal = {J. Mol. Biol.},
    volume =  {365},
    number = {1},
    pages = {77--89},
    year = {2007},
    doi = {10.1016/j.jmb.2006.08.097}
}

@article{das2,
    author = {Das, K. and Ding, J. and Hsiou, Y. and Clark, A.D. and Moereels, H. and Koymans, L. and Andries, K. and Pauwels, R. and Janssen, P.A. and Boyer, P.L. and Clark, P. and Smith, R.H. and Kroeger Smith, M.B. and Michejda, C.J. and Hughes, S.H. and Arnold, E.},
    title = {Crystal structures of 8-Cl and 9-Cl TIBO complexed with wild-type HIV-1 RT and 8-Cl TIBO complexed with the Tyr181Cys HIV-1 RT drug-resistant mutant.},
    journal = {J. Mol. Biol.},
    volume =  {264},
    number = {5},
    pages = {1085--1100},
    year = {1996},
    doi = {10.1006/jmbi.1996.0698}
}

@article{ren,
    author = {Ren, J. and Nichols, C. and Bird, L. and Chamberlain, P. and Weaver, K. and Short, S. and Stuart, D.I. and Stammers, D.K.},
    title = {Structural mechanisms of drug resistance for mutations at codons 181 and 188 in HIV-1 reverse transcriptase and the improved resilience of second generation non-nucleoside inhibitors.},
    journal = {J. Mol. Biol.},
    volume =  {312},
    number = {4},
    pages = {795--805},
    year = {2001},
    doi = {10.1006/jmbi.2001.4988}
}

@article{gulick,
    author = {Gulick R.M.},
    title = {Investigational Antiretroviral Drugs: What is Coming Down the Pipeline.},
    journal = {Top. Antivir. Med.},
    volume =  {25},
    number = {4},
    pages = {127--32},
    year = {2018},
    PMID = {29689540}
}

@article{markowitz,
    author = {Markowitz, M. and Grobler, J.A.},
    title = {Islatravir for the treatment and prevention of infection with the human immunodeficiency virus type 1.},
    journal = {Curr. Opin. HIV. AIDS.},
    volume =  {15},
    number = {1},
    pages = {27--32},
    year = {2020},
    doi = {10.1097/COH.0000000000000599}
}

@article{blevins,
    author = {Blevins, S.R. and Hester, E.K. and Chastain, D.B. and Cluck, D.B.},
    title = {Doravirine: A Return of the NNRTI Class?.},
    journal = {Ann. Pharmacother.},
    volume =  {54},
    number = {1},
    pages = {64--74},
    year = {2020},
    doi = {10.1177/1060028019869641}
}

@article{al-salama,
    author = {Al-Salama Z.T.},
    title = {Elsulfavirine: First Global Approval.},
    journal = {Drugs.},
    volume =  {77},
    number = {16},
    pages = {1811--16},
    year = {2017},
    doi = {10.1007/s40265-017-0820-3}
}

@inproceedings{murphy,
    author = {Murphy, R.L. and Kravchenko, A. and Orlova-Morozova, E. and Nagimova, F. and Kozirev, O. and Shimonava, T. and Deulina, M. and Vostokova, N. and Zozulya, O. and Bichko, V.},
    title = {Elsulfavirine as compared to efavirenz in combination with TDF/FTC: 48-week study},
    addendum = {Conference on Retroviruses and Opportunistic Infections (CROI).},
    year = {2017}
}

@article{tramontano,
    author = {Tramontano, E. and Corona, A. and Menéndez-Arias, L.},
    title = {Ribonuclease H, an unexploited target for antiviral intervention against HIV and hepatitis B virus.},
    journal = {Antiviral Res.},
    volume =  {171},
    pages = {104613},
    year = {2019},
    doi = {10.1016/j.antiviral.2019.104613}
}

@article{borkow,
    author = {Borkow, G. and Fletcher, R.S. and Barnard, J. and Arion, D. and Motakis, D. and Dmitrienko, G.I. and Parniak, M.A.},
    title = {Inhibition of the ribonuclease H and DNA polymerase activities of HIV-1 reverse transcriptase by N-(4-tert-butylbenzoyl)-2-hydroxy-1-naphthaldehyde hydrazone.},
    journal = {Biochemistry.},
    volume =  {36},
    number = {11},
    pages = {3179--85},
    year = {1997},
    doi = {10.1021/bi9624696}
}

@article{himmel,
    author = {Himmel, D.M. and Sarafianos, S.G. and Dharmasena, S. and Hossain, M.M. and McCoy-Simandle, K. and Ilina, T. and Clark, A.D. and Knight, J.L. and Julias, J.G. and Clark, P.K. and Krogh-Jespersen, K. and Levy, R.M. and Hughes, S.H. and Parniak, M.A. and Arnold, E.},
    title = {HIV-1 reverse transcriptase structure with RNase H inhibitor dihydroxy benzoyl naphthyl hydrazone bound at a novel site.},
    journal = {ACS. Chem. Biol.},
    volume =  {1},
    number = {11},
    pages = {702--12},
    year = {2006},
    doi = {10.1021/cb600303y}
}

@article{wang,
    author = {Wang, X. and Gao, P. and Menendez-Arias, L. and Liu, X. and Zhan, P.},
    title = {Update on Recent Developments in Small Molecular HIV-1 RNase H Inhibitors (2013-2016): Opportunities and Challenges.},
    journal = {Curr. Med. Chem.},
    volume =  {25},
    number = {14},
    pages = {1682--1702},
    year = {2018},
    doi = {10.2174/0929867324666170113110839}
}

@article{llu,
    author = {Lu, L. and et al.},
    title = {Development of small-molecule HIV netry inhibitors specifically targeting gp120 or gp41.},
    journal = {Curr. Top Med. Chem.},
    volume =  {16},
    number = {10},
    pages = {1074--1090},
    year = {2016},
    url = {https://www.ncbi.nlm.nih.gov/pmc/articles/PMC4775441/pdf/nihms735344.pdf}
}

@article{zyang,
    author = {Yang, Z. and et al.},
    title = {Preclinical pharmacokinetics of a novel HIV-1 attachment inhibitor MS-378806 and prediction of its human pharmacokinetics.},
    journal = {Biopharmaceutics and Drug Disposition.},
    volume =  {26},
    pages = {387--402},
    year = {2005},
    doi = {10.1002/bdd.471}
}

@article{ywang,
    author = {Wang, Y. and et al.},
    title = {Discovery of 4-benzoyl-1-[(4-methoxy-1H-pyrrolo[2,3-b]pyridin-3-yl)oxoacetyl]-2-(R)-methypiperazine (BMS-378806): a novel HIV-1 attachment inhibitor that interferes with CD4-gp120 interactions.},
    journal = {J. Med. Chem.},
    volume =  {46},
    pages = {4236--4239},
    year = {2003},
    doi = {10.1021/jm034082o}
}

@article{lzhu,
    author = {Zhu, L. and et al.},
    title = {Pharmacokinetic interactions between BMS-626529, the active moiety of the HIV-1 attachment inhibitor prodrug BMS-663068, and ritonavir or ritonavir-boosted atazanavir in healthy subjects.},
    journal = {Antimicrobial agents and chemotherapy.},
    volume =  {59},
    number = {7},
    pages = {3816--3823},
    year = {2015},
    doi = {10.1128/AAC.04914-14}
}

@article{nettles,
    author = {Nettles, R.E. and et al.},
    title = {Pharmacodynamics, safety, and pharmacokinetics of BMS-663068, an oral HIV-1 attachment inhibitor in HIV-1 infected subjects.},
    journal = {The Journal of Infectious Diseases.},
    volume =  {206},
    pages = {1002--1011},
    year = {2012},
    doi = {10.1093/infdis/jis432}
}

@article{nowicka,
    author = {Nowicka-Sans, B. and et al.},
    title = {In vitro antiviral characteristics of HIV-1 attachment inhibitor BMS-626529, the active compound of the prodrug BMS-663068.},
    journal = {Antimicrobial agents and chemotherapy.},
    volume =  {56},
    number = {7},
    pages = {3498--3507},
    year = {2012},
    doi = {10.1128/AAC.00426-12}
}

@article{kozal,
    author = {Kozal, M. and Aberg, J. and Pialoux, G. and Cahn, P. and Thompson, M. and Molina, J.-M. and Grinsztejn, B. and Diaz, R. and Castagna, A. and Kumar, P. and Latiff, G. and DeJesus, E. and Gummel, M. and Gartland, M. and Pierce, A. and Ackerman, P. and Llamoso, C. and Lataillade, M.},
    title = {Fostemsavir in adults with multidrug-resistant HIV-1 infection.},
    journal = {The New England Journal of Medicine.},
    volume =  {382},
    pages = {1232--43},
    year = {2020},
    doi = {10.1056/NEJMoa1902493}
}

@article{feasey,
    author = {Feasey, N.A. and Healey, P. and Gordon, M.A.},
    title = {Review article: the aetiology, investigation and management of diarrhoea in the HIV-positive patient.},
    journal = {Aliment. Pharmacol. Ther.},
    volume =  {34},
    pages = {587--603},
    year = {2011},
    doi = {10.1111/j.1365-2036.2011.04781.x}
}

@article{mitalipov,
    author = {Mitalipov, S. and Wolf, D.},
    title = {Totipotency, pluripotency and nuclear reprogramming.},
    journal = {Adv. Biochem. Eng. Biotechnol.},
    volume =  {114},
    pages = {185--199},
    year = {2009},
    doi = {10.1007/10_2008_45}
}

@article{mercardouribe,
    author = {Niu, N. and Mercado-Uribe, I. and Liu, J.},
    title = {Dedifferentiation into blastomere-like cancer stem cells via formation of polyploid giant cancer cells.},
    journal = {Oncogene.},
    volume =  {36},
    pages = {4887--4900},
    year = {2017},
    doi = {10.1038/onc.2017.72}
}

@article{jäger,
    author = {Jäger, S. and Cimermancic, P. and Gulbahce, N. and Johnson, J.R. and McGovern, K.E. and Clarke, S.C. and Shales, M. and Mercenne, G. and Pache, L. and Li, K. and Hernandez, H. and Jang, G.M. and Roth, S.L. and Akiva, E. and Marlett, J. and Stephens, M. and D'Orso, I. and Fernandes, J. and Fahey, M. and Mahon, C. and O'Donoghue, A.J. and Todorovic, A. and Morris, J.H. and Maltby, D.A. and Alber, T. and Cagney, G. and Bushman, F.D. and Young, J.A. and Chanda, S.K. and Sundquist, W.I. and Kortemme, T. and Hernandez, R.D. and Craik, C.S. and Burlingame, A. and Sali, A. and Frankel, A.D. and Krogan, N.J.},
    title = {Global landscape of HIV-human protein complexes.},
    journal = {Nature.},
    volume = {481},
    number = {7381},
    pages = {365--370},
    year = {2011},
    doi = {10.1038/nature10719}
}

@article{cai,
    author = {Cai, L. and Gochin, M. and Liu, K.},
    title = {Biochemistry and biophysics of HIV-1 gp41 - membrane interactions and implications for HIV-1 envelope protein mediated viral-cell fusion and fusion inhibitor design.},
    journal = {Current topics in medicinal chemistry.},
    volume = {11},
    number = {24},
    pages = {2959--2984},
    year = {2011},
    doi = {10.2174/156802611798808497}
}

@article{yakovian,
    author = {Yakovian, O. and Schwarzer, R. and Sajman, J. and Neve-Oz, Y. and Razvag, Y. and Herrmann, A. and Sherman, E.},
    title = {Gp41 dynamically interacts with the TCR in the immune synapse and promotes early T cell activation.},
    journal = {Scientific reports.},
    volume = {8},
    number = {1},
    pages = {9747},
    year = {2018},
    doi = {10.1038/s41598-018-28114-5}
}

@article{zhou,
    author = {Zhou, X. and Liu, Z. and Zhang, J. and Adelsberger, J.W. and Yang, J. and Burton, G.F.},
    title = {Alpha-1-antitrypsin interacts with gp41 to block HIV-1 entry into CD4+ T lymphocytes.},
    journal = {BMC microbiology.},
    volume = {16},
    number = {1},
    pages = {172},
    year = {2016},
    doi = {10.1186/s12866-016-0751-2}
}

@article{ivanusic,
    author = {Ivanusic, D.},
    title = {HIV-1 cell-to-cell spread: CD63-gp41 interaction at the virological synapse.},
    journal = {AIDS Res Hum Retroviruses.},
    volume = {30},
    number = {9},
    pages = {844--845},
    year = {2014},
    doi = {10.1089/aid.2014.0116}
}

@article{wyma,
    author = {Wyma, D.J. and Kotov, A. and Aiken, C.},
    title = {Evidence for a stable interaction of gp41 with Pr55(Gag) in immature human immunodeficiency virus type 1 particles.},
    journal = {Journal of virology.},
    volume = {74},
    number = {20},
    pages = {9381--9387},
    year = {2000},
    doi = {10.1128/jvi.74.20.9381-9387.2000}
}

@article{thielens,
    author = {Thielens, N.M. and Bally, I.M. and Ebenbichler, C.F. and Dierich, M.P. and Arlaud, G.J.},
    title = {Further characterization of the interaction between the C1q subcomponent of human C1 and the transmembrane envelope glycoprotein gp41 of HIV-1.},
    journal = {Journal of immunology.},
    volume = {151},
    number = {11},
    pages = {6583--6592},
    year = {1993},
    PMID = {8245486}
}

@article{chen,
    author = {Chen, Y.H. and Xiao, Y. and Wu, W. and Yang, J. and Sui, S. and Dierich, M.P.},
    title = {The C domain of HIV-1 gp41 binds the putative cellular receptor protein P62.},
    journal = {AIDS.},
    volume = {13},
    number = {9},
    pages = {1021--1024},
    year = {1999},
    doi = {10.1097/00002030-199906180-00003}
}

@article{qi,
    author = {Qi, M. and Williams, J.A. and Chu, H. and Chen, X. and Wang, J.J. and Ding, L. and Akhirome, E. and Wen, X. and Lapierre, L.A. and Goldenring, J.R. and Spearman, P.},
    title = {Rab11-FIP1C and Rab14 direct plasma membrane sorting and particle incorporation of the HIV-1 envelope glycoprotein complex.},
    journal = {PLoS pathogens.},
    volume = {9},
    number = {4},
    pages = {e1003278},
    year = {2013},
    doi = {10.1371/journal.ppat.1003278}
}

@article{sanhadji,
    author = {Sanhadji, K. and Tardy, J.C. and Touraine, J.L.},
    title = {HIV-1 infection: functional competition between gp41 and interleukin-2.},
    journal = {Comptes rendus biologies.},
    volume = {333},
    number = {8},
    pages = {608--612},
    year = {2010},
    doi = {10.1016/j.crvi.2009.01.009}
}

@article{checkley,
    author = {Checkley, M.A. and Luttge, B.G. and Freed, E.O.},
    title = {HIV-1 envelope glycoprotein biosynthesis, trafficking, and incorporation.},
    journal = {J. Mol. Biol.},
    volume = {410},
    number = {4},
    pages = {582--608},
    year = {2011},
    doi = {10.1016/j.jmb.2011.04.042}
}

@article{henrick,
    author = {Henrick, B.M. and Yao, X.D. and Zahoor, M.A. and Abimiku, A. and Osawe, S. and Rosenthal, K.L.},
    title = {TLR10 Senses HIV-1 Proteins and Significantly Enhances HIV-1 Infection.},
    journal = {Frontiers in immunology.},
    volume = {10},
    pages = {420},
    year = {2019},
    doi = {10.3389/fimmu.2019.00482}
}

@article{takeshita,
    author = {Takeshita, S. and Breen, E.C. and Ivashchenko, M. and Nishanian, P.G. and Kishimoto, T. and Vredevoe, D.L. and Martinez-Maza, O.},
    title = {Induction of IL-6 and IL-10 production by recombinant HIV-1 envelope glycoprotein 41 (gp41) in the THP-1 human monocytic cell line.},
    journal = {Cell. Immunol.},
    volume = {165},
    number = {2},
    pages = {234--242},
    year = {1995},
    doi = {10.1006/cimm.1995.1210}
}

@article{stoiber,
    author = {Stoiber, H. and Ebenbichler, C. and Schneider, R. and Janatova, J. and Dierich, M.P.},
    title = {Interaction of several complement proteins with gp120 and gp41, the two envelope glycoproteins of HIV-1.},
    journal = {AIDS},
    volume = {9},
    number = {1},
    pages = {19--26},
    year = {1995},
    doi = {10.1097/00002030-199501000-00003}
}

@article{planelles,
    author = {Planelles, V. and Benichou, S.},
    title = {Vpr and its interactions with cellular proteins.},
    journal = {Curr. Top. Microbiol. Immunol.},
    volume = {339},
    pages = {177--200},
    year = {2009},
    doi = {10.1007/978-3-642-02175-6_9}
}

@article{zhao,
    author = {Zhao, R.Y. and Li, G. and Bukrinsky, M.I.},
    title = {Vpr-host interactions during HIV-1 viral life cycle.},
    journal = {J. Neuroimmune. Pharmacol.},
    volume = {6},
    number = {2},
    pages = {216--229},
    year = {2011},
    doi = {10.1007/s11481-011-9261-z}
}

@article{fabryova,
    author = {Fabryova, H. and Strebel, K.},
    title = {Vpr and Its Cellular Interaction Partners: R We There Yet?},
    journal = {Cells.},
    volume = {8},
    number = {11},
    pages = {1310},
    year = {2019},
    doi = {10.3390/cells8111310}
}

@article{pereira,
    author = {Pereira, E.A. and daSilva, L.L.},
    title = {HIV-1 Nef: Taking Control of Protein Trafficking.},
    journal = {Traffic.},
    volume = {17},
    number = {9},
    pages = {976--996},
    year = {2016},
    doi = {10.1111/tra.12412}
}

@article{pyeon,
    author = {Pyeon, D. and Rojas, V.K. and Price, L. and Kim, S. and Meharvan, S. and Park, I.W.},
    title = {HIV-1 Impairment via UBE3A and HIV-1 Nef Interactions Utilizing the Ubiquitin Proteasome System.},
    journal = {Viruses.},
    volume = {11},
    number = {12},
    pages = {1098},
    year = {2019},
    doi = {10.3390/v11121098}
}

@article{marrero,
    author = {Marrero-Hernández, S. and Márquez-Arce, D. and Cabrera-Rodríguez, R. and Estévez-Herrera, J. and Pérez-Yanes, S. and Barroso-González, J. and Madrid, R. and Machado, J.D. and Blanco, J. and Valenzuela-Fernández, A.},
    title = {HIV-1 Nef Targets HDAC6 to Assure Viral Production and Virus Infection.},
    journal = {Front. Microbiol.},
    volume = {10},
    pages = {2437},
    year = {2019},
    doi = {10.3389/fmicb.2019.02437}
}

@article{arhel,
    author = {Arhel, N.J. and Kirchhoff, F.},
    title = {Implications of Nef: host cell interactions in viral persistence and progression to AIDS.},
    journal = {Curr. Top. Microbiol. Immunol.},
    volume = {339},
    pages = {147--175},
    year = {2009},
    doi = {10.1007/978-3-642-02175-6_8}
}

@article{hu,
    author = {Hu, Y. and Desimmie, B.A. and Nguyen, H.C. and Ziegler, S.J. and Cheng, T.C. and Chen, J. and Wang, J. and Wang, H. and Zhang, K. and Pathak, V.K. and Xiong, Y.},
    title = {Structural basis of antagonism of human APOBEC3F by HIV-1 Vif.},
    journal = {Nat. Struct. Mol. Biol.},
    volume = {26},
    number = {12},
    pages = {1176--1183},
    year = {2019},
    doi = {10.1038/s41594-019-0343-6}
}

@article{wang2,
    author = {Wang, J. and Becker, J.T. and Shi, K. and Lauer, K.V. and Salamango, D.J. and Aihara, H. and Shaban, N.M. and Harris, R.S.},
    title = {The Role of RNA in HIV-1 Vif-Mediated Degradation of APOBEC3H.},
    journal = {J. Mol. Biol.},
    volume = {431},
    number = {24},
    pages = {5019--5031},
    year = {2019},
    doi = {10.1016/j.jmb.2019.09.014}
}

@article{salamango,
    author = {Salamango, D.J. and Ikeda, T. and Moghadasi, S.A. and Wang, J. and McCann, J.L. and Serebrenik, A.A. and Ebrahimi, D. and Jarvis, M.C. and Brown, W.L. and Harris, R.S.},
    title = {HIV-1 Vif Triggers Cell Cycle Arrest by Degrading Cellular PPP2R5 Phospho-regulators.},
    journal = {Cell. Rep.},
    volume = {29},
    number = {5},
    pages = {1057--1065.e4},
    year = {2019},
    doi = {10.1016/j.celrep.2019.09.057}
}

@article{wang3,
    author = {Wang, X. and Wang, X. and Wang, W. and Zhang, J. and Wang, J. and Wang, C. and Lv, M. and Zuo, T. and Liu, D. and Zhang, H. and Wu, J. and Yu, B. and Kong, W. and Wu, H. and Yu, X.},
    title = {Both Rbx1 and Rbx2 exhibit a functional role in the HIV-1 Vif-Cullin5 E3 ligase complex in vitro.},
    journal = {Biochem. Biophys. Res. Commun.},
    volume = {461},
    number = {4},
    pages = {624--629},
    year = {2015},
    doi = {10.1016/j.bbrc.2015.04.077}
}

@article{huttenhain,
    author = {Hüttenhain, R. and Xu, J. and Burton, L.A. and Gordon, D.E. and Hultquist, J.F. and Johnson, J.R. and Satkamp, L. and Hiatt, J. and Rhee, D.Y. and Baek, K. and Crosby, D.C. and Frankel, A.D. and Marson, A. and Harper, J.W. and Alpi, A.F. and Schulman, B.A. and Gross, J.D. and Krogan, N.J.},
    title = {ARIH2 Is a Vif-Dependent Regulator of CUL5-Mediated APOBEC3G Degradation in HIV Infection.},
    journal = {Cell. Host. Microbe.},
    volume = {26},
    number = {1},
    pages = {86--99.e7},
    year = {2019},
    doi = {10.1016/j.chom.2019.05.008}
}

@article{azimi,
    author = {Azimi, F.C. and Lee, J.E.},
    title = {Structural perspectives on HIV-1 Vif and APOBEC3 restriction factor interactions.},
    journal = {Protein. Sci.},
    volume = {29},
    number = {2},
    pages = {391--406},
    year = {2020},
    doi = {doi.org/10.1002/pro.3729}
}

@article{spector,
    author = {Spector, C. and Mele, A.R. and Wigdahl, B. and Nonnemacher, M.R.},
    title = {Genetic variation and function of the HIV-1 Tat protein.},
    journal = {Med. Microbiol. Immunol.},
    volume = {208},
    number = {2},
    pages = {131--169},
    year = {2019},
    doi = {10.1007/s00430-019-00583-z}
}

@article{jean,
    author = {Jean, M.J. and Power, D. and Kong, W. and Huang, H. and Santoso, N. and Zhu, J.},
    title = {Identification of HIV-1 Tat-Associated Proteins Contributing to HIV-1 Transcription and Latency.},
    journal = {Viruses.},
    volume = {9},
    number = {4},
    pages = {67},
    year = {2017},
    doi = {10.3390/v9040067}
}

@article{mahmoudi,
    author = {Mahmoudi, T. and Parra, M. and Vries, R.G. and Kauder, S.E. and Verrijzer, C.P. and Ott, M. and Verdin, E.},
    title = {The SWI/SNF chromatin-remodeling complex is a cofactor for Tat transactivation of the HIV promoter.},
    journal = {J. Biol. Chem.},
    volume = {281},
    number = {29},
    pages = {19960--19968},
    year = {2006},
    doi = {10.1074/jbc.M603336200}
}

@article{gautier,
    author = {Gautier, V.W. and Gu, L. and O'Donoghue, N. and Pennington, S. and Sheehy, N. and Hall, W.W.},
    title = {In vitro nuclear interactome of the HIV-1 Tat protein.},
    journal = {Retrovirology.},
    volume = {6},
    pages = {47},
    year = {2009},
    doi = {10.1186/1742-4690-6-47}
}

@article{remoli,
    author = {Remoli, A.L. and Marsili, G. and Perrotti, E. and Acchioni, C. and Sgarbanti, M. and Borsetti, A. and Hiscott, J. and Battistini, A.},
    title = {HIV-1 Tat Recruits HDM2 E3 Ligase To Target IRF-1 for Ubiquitination and Proteasomal Degradation.},
    journal = {mBio.},
    volume = {7},
    number = {5},
    pages = {e01528-16},
    year = {2016},
    doi = {10.1128/mBio.01528-16}
}

@article{cujec,
    author = {Cujec, T.P. and Okamoto, H. and Fujinaga, K. and Meyer, J. and Chamberlin, H. and Morgan, D.O. and Peterlin, B.M.},
    title = {The HIV transactivator TAT binds to the CDK-activating kinase and activates the phosphorylation of the carboxy-terminal domain of RNA polymerase II.},
    journal = {Genes. Dev.},
    volume = {11},
    number = {20},
    pages = {2645--2657},
    year = {1997},
    doi = {10.1101/gad.11.20.2645}
}

@article{li,
    author = {Li, X.Y. and Green, M.R.},
    title = {The HIV-1 Tat cellular coactivator Tat-SF1 is a general transcription elongation factor.},
    journal = {Genes. Dev.},
    volume = {12},
    number = {19},
    pages = {2992--2996},
    year = {1998},
    doi = {10.1101/gad.12.19.2992.}
}

@article{connors,
    author = {Connors, M. and et al.},
    title = {HIV infection induces changes in CD4(+) T-cell phenotype and depletions within the CD4(+) T-cell repertoire that are not immediately restored by antiviral or immune-based therapies.},
    journal = {Nature Medicine.},
    volume = {3},
    number = {5},
    pages = {533--541},
    year = {1997},
    doi = {10.1038/nm0597-533}
}

@conference{fontanelli2,
    author = {Fontanelli, O.},
    title = {Modelo de venta de garage y la inevitavilidad de la desigualdad.},
    series = {Seminario de Economía y Ciencias de la Complejidad},
    year = {2020}
}

@article{boghosian,
    author = {Boghosian, B.M. and Devitt-Lee, A. and Wnag, H.},
    title = {The growth of oligarchy in a yard-sale model of asset exhchange: a logistic equation for wealth condensation.},
    journal = {arXiv.},
    pages = {1--15},
    year = {2016},
    url = {https://arxiv.org/pdf/1608.05851.pdf}
}

@article{chorro,
    author = {Chorro, C.},
    title = {A simple probabilistic approach of the Yard-Sale model.},
    journal = {Statistics and Probability Letters.},
    pages = {1--6},
    year = {2016},
    doi = {10.1016/j.spl.2016.01.012}
}

@article{egghe,
    author = {Egghe, L.},
    title = {Mathematical derivation of the impact factor distribution.},
    journal = {Journal of Infometrics.},
    volume = {3},
    number = {4},
    pages = {290--295},
    year = {2009}
}

@article{miramontes,
    author = {Li, W. and Miramontes, P.},
    title = {Fitting ranked english and spanish letter frequency distribution in US and Mexican presidential speeches.},
    journal = {Journal of Quantitative Linguistics.},
    volume = {18},
    number = {4},
    pages = {359--380},
    year = {2011}
}

@article{freudenberg,
    author = {Li, W. and Freudenberg, J. and Miramontes, P.},
    title = {Diminishing return for increased mappability with longer sequencing reads: implications of the k-mer distributions in the human genome.},
    journal = {BMC Bioinformatics.},
    volume = {15},
    number = {1},
    pages = {2},
    year = {2014}
}

@article{nguyen,
    author = {Nguyen, V. and Markelov, O. and Serdyuk, A. and Vasenev, A. and Bogachev, M.},
    title = {Universal sank-size statistics in network traffic: modeling collective access patterns by Zipf's law with long-term correlations.},
    journal = {Europhysics Letters.},
    volume = {123},
    number = {5},
    pages = {50001},
    year = {2018}
}

@article{solbak,
    author = {Solbak, S.M. and Reksten, T.R. and Hahn, F. and Wray, V. and Henklein, P. and Henklein, P. and Halskau, Ø. and Schubert, U. and Fossen, T.},
    title = {HIV-1 p6 - a structured to flexible multifunctional membrane-interacting protein.},
    journal = {Biochim. Biophys. Acta.},
    volume = {1828},
    number = {2},
    pages = {816--823},
    year = {2013},
    doi = {10.1016/j.bbamem.2012.11.010}
}

@article{sharma2,
    author = {Sharma, S. and Arunachalam, P.S. and Menon, M. and Ragupathy, V. and Satya, R.V. and Jebaraj, J. and Aralaguppe, S.G. and Rao, C. and Pal, S. and Saravanan, S. and Murugavel, K.G. and Balakrishnan, P. and Solomon, S. and Hewlett, I. and Ranga, U.},
    title = {PTAP motif duplication in the p6 Gag protein confers a replication advantage on HIV-1 subtype C.},
    journal = {J. Biol. Chem.},
    volume = {293},
    number = {30},
    pages = {11687--11708},
    year = {2018},
    doi = {10.1074/jbc.M117.815829}
}

@article{solbak2,
    author = {Solbak, S.M. and Reksten, T.R. and Röder, R. and Wray, V. and Horvli, O. and Raae, A.J. and Henklein, P. and Henklein, P. and Fossen, T.},
    title = {HIV-1 p6-Another viral interaction partner to the host cellular protein cyclophilin A.},
    journal = {Biochim. Biophys. Acta.},
    volume = {1824},
    number = {4},
    pages = {667--678},
    year = {2012},
    doi = {10.1016/j.bbapap.2012.02.002}
}

@article{popov2,
    author = {Popov, S. and Popova, E. and Inoue, M. and Wu, Y. and Göttlinger, H.},
    title = {HIV-1 gag recruits PACSIN2 to promote virus spreading.},
    journal = {Proc. Natl. Acad. Sci. USA.},
    volume = {115},
    number = {27},
    pages = {7093--7098},
    year = {2018},
    doi = {10.1073/pnas.1801849115}
}

@article{müller,
    author = {Müller, B. and Patschinsky, T. and Kräusslich, H.G.},
    title = {The late-domain-containing protein p6 is the predominant phosphoprotein of human immunodeficiency virus type 1 particles.},
    journal = {J. Virol.},
    volume = {76},
    number = {3},
    pages = {1015--1024},
    year = {2002},
    doi = {10.1128/jvi.76.3.1015-1024.2002}
}

@article{ott,
    author = {Ott, D.E. and Coren, L.V. and Copeland, T.D. and Kane, B.P. and Johnson, D.G. and Sowder, R.C. and Yoshinaka, Y. and Oroszlan, S. and Arthur, L.O. and Henderson, L.E.},
    title = {Ubiquitin is covalently attached to the p6Gag proteins of human immunodeficiency virus type 1 and simian immunodeficiency virus and to the p12Gag protein of Moloney murine leukemia virus.},
    journal = {J. Virol.},
    volume = {72},
    number = {4},
    pages = {2962--2968},
    year = {1998},
    doi = {10.1128/JVI.72.4.2962-2968.1998}
}

@article{yu,
    author = {Yu, K.L. and Lee, S.H. and Lee, E.S. and You, J.C.},
    title = {HIV-1 nucleocapsid protein localizes efficiently to the nucleus and nucleolus.},
    journal = {Virology.},
    volume = {492},
    pages = {204--212},
    year = {2016},
    doi = {10.1016/j.virol.2016.03.002}
}

@article{anton,
    author = {Anton, H. and Taha, N. and Boutant, E. and Richert, L. and Khatter, H. and Klaholz, B. and Rondé, P. and Réal, E. and de Rocquigny, H. and Mély, Y.},
    title = {Investigating the cellular distribution and interactions of HIV-1 nucleocapsid protein by quantitative fluorescence microscopy.},
    journal = {PloS. One.},
    volume = {10},
    number = {2},
    pages = {e0116921},
    year = {2015},
    doi = {10.1371/journal.pone.0116921}
}

@article{karnati,
    author = {Karnati, K.R. and Wang, Y.},
    title = {Structural and binding insights into HIV-1 protease and P2-ligand interactions through molecular dynamics simulations, binding free energy and principal component analysis.},
    journal = {J. Mol. Graph. Model.},
    volume = {92},
    pages = {112--122},
    year = {2019},
    doi = {10.1016/j.jmgm.2019.07.008}
}

@article{rumlova,
    author = {Rumlová, M. and Křížová, I. and Keprová, A. and Hadravová, R. and Doležal, M. and Strohalmová, K. and Pichová, I. and Hájek, M. and Ruml, T.},
    title = {HIV-1 protease-induced apoptosis.},
    journal = {Retrovirology.},
    volume = {11},
    pages = {37},
    year = {2014},
    doi = {10.1186/1742-4690-11-37}
}

@article{wagner,
    author = {Wagner, R.N. and Reed, J.C. and Chanda, S.K.},
    title = {HIV-1 protease cleaves the serine-threonine kinases RIPK1 and RIPK2.},
    journal = {Retrovirology.},
    volume = {12},
    pages = {74},
    year = {2015},
    doi = {10.1186/s12977-015-0200-6}
}

@article{impens,
    author = {Impens, F. and Timmerman, E. and Staes, A. and Moens, K. and Ariën, K.K. and Verhasselt, B. and Vandekerckhove, J. and Gevaert, K.},
    title = {A catalogue of putative HIV-1 protease host cell substrates.},
    journal = {Biol. Chem.},
    volume = {393},
    number = {9},
    pages = {915--931},
    year = {2012},
    doi = {10.1515/hsz-2012-0168}
}

@article{sheng,
    author = {Sheng, N. and Erickson-Viitanen, S.},
    title = {Cleavage of p15 protein in vitro by human immunodeficiency virus type 1 protease is RNA dependent.},
    journal = {J. Virol.},
    volume = {68},
    number = {10},
    pages = {6207--6214},
    year = {1994},
    doi = {10.1128/JVI.68.10.6207-6214.1994}
}

@article{wangw,
    author = {Wang, W. and Naiyer, N. and Mitra, M. and Li, J. and Williams, M.C. and Rouzina, I. and Gorelick, R.J. and Wu, Z. and Musier-Forsyth, K.},
    title = {Distinct nucleic acid interaction properties of HIV-1 nucleocapsid protein precursor NCp15 explain reduced viral infectivity.},
    journal = {Nucleic. Acids. Res.},
    volume = {42},
    number = {11},
    pages = {7145--7159},
    year = {2014},
    doi = {10.1093/nar/gku335}
}

@article{yuXG,
    author = {Yu, X.G. and Shang, H. and Addo, M.M. and Eldridge, R.L. and Phillips, M.N. and Feeney, M.E. and Strick, D. and Brander, C. and Goulder, P.J. and Rosenberg, E.S. and Walker, B.D. and Altfeld, M. and HIV Study Collaboration},
    title = {Important contribution of p15 Gag-specific responses to the total Gag-specific CTL responses.},
    journal = {AIDS.},
    volume = {16},
    number = {3},
    pages = {321--328},
    year = {2002},
    doi = {10.1097/00002030-200202150-00002}
}

@article{caccuri,
    author = {Caccuri, F. and Marsico, S. and Fiorentini, S. and Caruso, A. and Giagulli, C.},
    title = {HIV-1 Matrix Protein p17 and its Receptors.},
    journal = {Curr. Drug. Targets.},
    volume = {17},
    number = {1},
    pages = {23--32},
    year = {2016},
    doi = {10.2174/1389450116666150825110840}
}

@article{bugatti,
    author = {Bugatti, A. and Paiardi, G. and Urbinati, C. and Chiodelli, P. and Orro, A. and Uggeri, M. and Milanesi, L. and Caruso, A. and Caccuri, F. and D’Ursi, P. and Rusnati, M.},
    title = {Heparin and heparan sulfate proteoglycans promote HIV-1 p17 matrix protein oligomerization: computational, biochemical and biological implications.},
    journal = {Sci. Rep.},
    volume = {9},
    number = {1},
    pages = {15768},
    year = {2019},
    doi = {10.1038/s41598-019-52201-w}
}

@article{lu,
    author = {Lu, J. and Jia, J. and Zhang, J. and Liu, X.},
    title = {HIV p17 enhances T cell proliferation by suppressing autophagy through the p17-OLA1-GSK3$\beta$ axis under nutrient starvation [published online ahead of print, 2020 Aug 13]},
    journal = {J. Med. Virol.},
    year = {2020},
    doi = {10.1002/jmv.26423}
}

@online{biogrid,
    author = "The BioGrid",
    title = "BioGrid",
    url  = "https://thebiogrid.org/",
    addendum = "(accessed: 04.02.2021)"
}

@online{string,
    author = "STRING Consortium 2020.",
    title = "STRING",
    url  = "https://string-db.org/cgi/input?sessionId=bDawDjRl9gu9&input_page_show_search=on",
    addendum = "(accessed: 04.02.2021)"
}

@article{tang,
    author = {Tang, S. and Zhao, J. and Wang, A. and Viswanath, R. and Harma, H. and Little, R.F. and Yarchoan, R. and Stramer, S.L. and Nyambi, P.N. and Lee, S. and Wood, O. and Wong, E.Y. and Wang, X. and Hewlett, I.K.},
    title = {Characterization of immune responses to capsid protein p24 of human immunodeficiency virus type 1 and implications for detection.},
    journal = {Clin. Vaccine. Immunol.},
    volume = {17},
    number = {8},
    pages = {1244--1251},
    year = {2010},
    doi = {10.1128/CVI.00066-10}
}

@article{couturier,
    author = {Couturier, J. and Orozco, A.F. and Liu, H. and Budhiraja, S. and Siwak, E.B. and Nehete, P.N. and Sastry, K.J. and Rice, A.P. and Lewis, D.E.},
    title = {Regulation of cyclin T1 during HIV replication and latency establishment in human memory CD4 T cells.},
    journal = {Virol. J.},
    volume = {16},
    number = {1},
    pages = {22},
    year = {2019},
    doi = {10.1186/s12985-019-1128-6}
}

@article{fu,
    author = {Fu, E. and Pan, L. and Xie, Y. and Mu, D. and Liu, W. and Jin, F. and Bai, X.},
    title = {Tetraspanin CD63 is a regulator of HIV-1 replication.},
    journal = {Int. J. Clin. Exp. Pathol.},
    volume = {8},
    number = {2},
    pages = {1184--1198},
    year = {2015},
    PMID = {25973004}
}

@article{bejarano,
    author = {Bejarano, D.A. and Peng, K. and Laketa, V. and Börner, K. and Jost, K.L. and Lucic, B. and Glass, B. and Lusic, M. and Müller, B. and Kräusslich, H.G.},
    title = {HIV-1 nuclear import in macrophages is regulated by CPSF6-capsid interactions at the nuclear pore complex.},
    journal = {eLife.},
    volume = {8},
    pages = {e41800},
    year = {2019},
    doi = {10.7554/eLife.41800}
}

\end{document}